\documentclass[twocolumn,times,twocolappendix]{aastex631}

\usepackage{amsmath}

\newcommand{\Msun}{\ensuremath{M_\odot}}

\def\Angstrom{\textup{\AA}}

\newcommand{\UF}{\affiliation{Department of Astronomy, University of Florida, Bryant Space Science Center, Gainesville, FL 32611-2055 USA}}
\newcommand{\CfA}{\affiliation{Center for Astrophysics \textbar{} Harvard \& Smithsonian, 60 Garden Street, Cambridge, MA 02138-1516, USA}}
\newcommand{\IAIFI}{\affiliation{The NSF AI Institute for Artificial Intelligence and Fundamental Interactions, USA}}

\newcommand{\UT}{\affiliation{Department of Astronomy, The University of Texas at Austin, 2515 Speedway, Stop C1400, Austin, TX 78712, USA}}
\shorttitle{Type IIn Supernovae. I}
\shortauthors{Hiramatsu et al.}

\begin{document}
%\begin{CJK*}{UTF8}{gbsn}

\title{\bf \large Type IIn Supernovae. I. \\ Uniform Light-curve Characterization and a Bimodality in the Radiated Energy Distribution}

\author[0000-0002-1125-9187]{Daichi Hiramatsu}
%(平松大地)}
\UF\CfA\IAIFI

\correspondingauthor{Daichi~Hiramatsu}
\email{dhiramatsu@ufl.edu}

\author[0000-0002-9392-9681]{Edo~Berger}
\CfA\IAIFI

\author[0000-0001-6395-6702]{Sebastian~Gomez}
\UT\CfA

\author[0000-0003-0526-2248]{Peter~K.~Blanchard}
\CfA\IAIFI

\author[0000-0003-0871-4641]{Harsh~Kumar}
\CfA\IAIFI

\author[0000-0002-2866-6416]{Wasundara~Athukoralalage}
\CfA

\begin{abstract}
We present the largest uniform study to date of Type~IIn supernovae (SNe~IIn), focusing in this first paper on the multiband optical light curves of 490 SNe~IIn. The sample, constructed from multiple surveys, extends to $z\approx0.8$, with the majority of events at $z\lesssim0.3$.  We construct uniform multiband and bolometric light curves using Gaussian process regression, and determine key observed properties in the rest frame (e.g., peak luminosity, timescales, radiated energy). We find that SNe~IIn span broad ranges in peak luminosity ($\sim 10^{42}-10^{44}$\,erg\,s$^{-1}$) and timescales ($\sim 20-300$ days above 50\% of peak luminosity), but the sample divides into two clear groups in the luminosity-timescale phase space around the median peak luminosity ($\approx 10^{43}$\,erg\,s$^{-1}$): faint--fast and luminous--slow groups. This leads to a strong bimodality in the radiated energy distribution, with peaks at $\sim 10^{49}$ and $\sim 2\times10^{50}$\,erg, with the latter events having a characteristic timescale of $\sim 100$ days, and the former appearing to bifurcate into two branches with timescales of $\sim 40$ and $\sim 70$ days. Therefore, SNe~IIn exhibit at least two dominant groupings, and perhaps three, which are likely reflective of different progenitor and/or circumstellar medium (CSM) formation pathways. We do not find any obvious transition in SN~IIn properties at the arbitrary cutoff of $\approx -20$ mag used for the designation ``Type~IIn~Superluminous~Supernovae'' or ``SLSN-IIn,'' and we argue that this classification should be abandoned. The absence of SNe~IIn with timescales of $\lesssim 15$ days defines the region occupied by fast transients with evidence for interaction with a hydrogen-poor CSM. 
\end{abstract}

% from https://astrothesaurus.org/concept-select/
\keywords{
\href{https://vocabs.ardc.edu.au/repository/api/lda/aas/the-unified-astronomy-thesaurus/current/resource.html?uri=http://astrothesaurus.org/uat/1668}{Supernovae (1668)}; 
\href{https://vocabs.ardc.edu.au/repository/api/lda/aas/the-unified-astronomy-thesaurus/current/resource.html?uri=http://astrothesaurus.org/uat/304}{Core-collapse supernovae (304)}; 
\href{https://vocabs.ardc.edu.au/repository/api/lda/aas/the-unified-astronomy-thesaurus/current/resource.html?uri=http://astrothesaurus.org/uat/1731}{Type II supernovae (1731)}; 
\href{https://vocabs.ardc.edu.au/repository/api/lda/aas/the-unified-astronomy-thesaurus/current/resource.html?uri=http://astrothesaurus.org/uat/732}{Massive stars (732)};
\href{https://vocabs.ardc.edu.au/repository/api/lda/aas/the-unified-astronomy-thesaurus/current/resource.html?uri=http://astrothesaurus.org/uat/1613}{Stellar mass loss (1613)};
\href{https://vocabs.ardc.edu.au/repository/api/lda/aas/the-unified-astronomy-thesaurus/current/resource.html?uri=http://astrothesaurus.org/uat/241}{Circumstellar matter (241)}
}

\section{Introduction} 
\label{sec:intro}

Among the hydrogen-rich, Type II supernovae (SNe~II), Type IIn SNe (SNe~IIn) are classified based on the presence of ``narrow'' Balmer series emission lines in their spectra \citep{Schlegel1990MNRAS.244..269S,Filippenko1997ARA&A..35..309F}. The main power source of SNe IIn is thought to be the shock interaction between the supernova (SN) ejecta and pre-existing circumstellar medium (CSM; see \citealt{Smith2017hsn..book..403S} for a review), resulting in the most heterogeneous SN class in terms of observed properties (see \citealt{Kiewe2012ApJ...744...10K,Taddia2013,Nyholm2020} for small sample studies). The observed heterogeneity likely reflects the diversity in pre-explosion mass loss responsible for the CSM formation. However, the nature of the underlying SNe and their progenitors remains elusive, as their observational signatures are mostly hidden below the photosphere formed at the CSM interaction layer.  It is therefore possible that a diversity of progenitor channels and explosion properties also contributes to the observed diversity.

The direct progenitor identification of SN~IIn 2005gl in a pre-explosion image pointed to a massive ($>50\,M_\odot$) luminous blue variable as a potential progenitor for that event \citep{Gal-Yam2007ApJ...656..372G,Gal-Yam2009Natur.458..865G}. Given the observed heterogeneous properties of SNe~IIn, however, various other progenitor channels are also possible, including for example, super-asymptotic giant branch stars ($\sim8-10\,M_\odot$; e.g., \citealt{Kankare2012MNRAS.424..855K,Mauerhan2013MNRAS.431.2599M,Smith2013MNRAS.434..102S,Moriya2014A&A...569A..57M,Hiramatsu2021NatAs...5..903H}), extreme red supergiants ($\sim17-25\,M_\odot$; e.g., \citealt{Fullerton2006ApJ...637.1025F,Smith2009AJ....137.3558S,Moriya2011MNRAS.415..199M,Mauerhan2012MNRAS.424.2659M,Hiramatsu2021ApJ...913...55H}), interacting massive binaries ($>20\,M_\odot$; e.g., \citealt{Chevalier2012ApJ...752L...2C,Soker2013ApJ...764L...6S,Kashi2013MNRAS.436.2484K,Schroder2020ApJ...892...13S,Metzger2022ApJ...932...84M,Hiramatsu2024ApJ...964..181H}), and even pulsational pair-instability SNe ($\sim110-140\,\Msun$; e.g.,  \citealt{Woosley2007Natur.450..390W,Blinnikov2010PAN....73..604B,Moriya2013MNRAS.428.1020M,Woosley2017ApJ...836..244W}).

In an effort to explore and map the diversity of progenitors, CSM, and explosion properties, here we present the first large-scale study of nearly 500 SNe IIn with well-sampled light curves and confirmed spectroscopic classification.  In this first paper, we focus on a uniform analysis of the optical light-curve properties to extract observed features (e.g., peak luminosity, timescales, radiated energy) and study their distributions and correlations.  Subsequent papers will present a similar population-level analysis in relation to numerical simulations; detailed semi-analytical modeling of each SN to determine the explosion and CSM properties; a spectroscopic study of the CSM and CSM-SN shock interaction regions; and studies of the host galaxies and SN IIn locations within their hosts.

This paper (Paper I) is structured as follows.  In \S\ref{sec:sample}, we present our methods to construct an unprecedentedly large sample of SNe IIn. In \S\ref{sec:phot}, we present the data sources for light-curve construction. In \S\ref{sec:gp}, we present a uniform Gaussian process (GP) regression method to create uniform rest-frame and time-sampled light curves, and describe the measurements extracted from these GP light curves. In \S\ref{sec:ana}, we analyze the light-curve properties and explore various correlations between measured and/or calculated properties.  We summarize our key findings and discuss their implications in \S\ref{sec:sum} and conclude in \S\ref{sec:conc}.

\section{Sample Construction} 
\label{sec:sample}

We construct a sample of SNe~IIn from transients discovered between 2016-01-01 and 2023-12-31 and classified as ``IIn," ``IIn-pec," or ``SLSN-II" by our last query date on 2025 July 31 on the Transient Name Server\footnote{\url{https://www.wis-tns.org/}} (TNS) and/or the Weizmann Interactive Supernova Data Repository\footnote{\url{https://wiserep.weizmann.ac.il}} (WISeREP; \citealt{Yaron2012PASP..124..668Y}). The discovery start date is chosen to match the official roll-out of TNS, while the end date is chosen to allow enough temporal coverage after discovery by the last query date ($\gtrsim580$ days). The classification categories are selected not to miss any SN IIn candidates given the ambiguous and somewhat arbitrary categorization used by the community, usually based on photometric, rather than spectroscopic, properties (e.g., peak brightness $\lesssim -20$ mag sometimes used for ``SLSN-II'').

To ensure SN IIn classification with narrow (${\rm FWHM}\lesssim3$,$000$\,km\,s$^{-1}$) Balmer series emission lines, we require that at least one public spectrum at any phase is available for each event in order to perform visual inspection and/or cross correlation with the SN spectral libraries \texttt{Superfit} \citep{Howell2005ApJ...634.1190H} (and its new Python implementation, \texttt{NGSF}; \citealt{Goldwasser2022TNSAN.191....1G}), \texttt{SNID} \citep{Blondin2007ApJ...666.1024B}, and \texttt{GELATO} \citep{Harutyunyan2008A&A...488..383H}.
There are 544 events classified as IIn, IIn-pec, or SLSN-II on TNS and/or WISeREP (as of 2025 July 31). We exclude 16 events without public spectra and an additional 74 contaminants with spectra inconsistent with the SN~IIn classification\footnote{In particular, we exclude eight misclassified SNe~II with flash features (e.g., \citealt{Khazov2016ApJ...818....3K,Bruch2021ApJ...912...46B,Jacobson2024ApJ...970..189J}) by identifying high ionization lines (e.g., He~{\sc ii}, C~{\sc iii}, C~{\sc iv}, N~{\sc iii}, N~{\sc iv}) and characteristic light-curve morphology (i.e., $\sim$100 day optically thick phase) that are atypical of SNe~IIn. Even with any residual contamination at a similar level of a few events, it is at the $\sim$1\% level, not affecting the population-level analysis of this work.} (see Appendix~\ref{sec:extra} for the details).
We then apply a light-curve quality cut based on coverage of the peak\footnote{This quality cut could bias against events with faster rise times than the typical ATLAS and ZTF survey cadence of 2--3 days; however, its effect on the population-level analysis is negligible given the low volumetric rate of such fast transients, particularly those with SN~IIn-like spectral features ($\lesssim$0.1\% of core-collapse SNe; \citealt{Drout2014ApJ...794...23D,Pursiainen2018MNRAS.481..894P,Ho2023ApJ...949..120H}). The gap we find at short timescales (rise time $\lesssim5$ days; see \S\ref{sec:ana}) is longer than the survey cadence, and thus unlikely caused by an observational bias.} to enable subsequent measurements (see \S\ref{sec:gp}), which leads to the final sample size of 427 SNe~IIn.
Our sample selection criteria are summarized in Table~\ref{tab:cut}. 

Of the 74 contaminants, whose spectra are inconsistent with SN IIn, 13 events are classified as SLSN-II with broad (${\rm FWHM}>5$,$000$\,km\,s$^{-1}$) H$\alpha$ emission lines, as similarly seen in normal (i.e., fainter) SNe~II; 11 of these events have been published in the sample study by \cite{Kangas2022MNRAS.516.1193K},\footnote{Excluding 3 ``SLSNe~I.5'' (SNe~2019pud, 2019zcr, and 2020jhm) with early spectroscopic similarity to hydrogen-poor SLSNe~I.} and two additional events (SNe~2023gpw and 2023acft) have been more recently discovered and classified (\citealt{2023TNSCR1334....1K,2024TNSCR.672....1C}; see \citealt{Kangas2026A&A...705A..52K} for a detailed study on SN~2023gpw). Given their different spectroscopic properties (i.e., possibly different emission mechanisms), we only use these 13 ``broad-lined SLSNe-II'' as a comparison sample\footnote{The broad-lined SLSNe-II are shown in the distribution histograms/plots in \S\ref{sec:ana}, but excluded in the calculations of medians, percentiles, and light-curve templates.} in this work. 

In addition to the large SN~IIn sample constructed above, we also include small literature samples from: the Carnegie Supernova Project (CSP; \citealt{Stritzinger2012ApJ...756..173S,Taddia2013}), Caltech Core-Collapse Project (CCCP; \citealt{Kiewe2012ApJ...744...10K}), Sloan Digital Sky Survey (SDSS; \citealt{Sako2018PASP..130f4002S}), Panoramic Survey Telescope and Rapid Response System (Pan-STARRS) 1 Medium Deep Survey (PS1; \citealt{Villar2020ApJ...905...94V}), and Palomar Transient Facility (PTF; \citealt{Nyholm2020}). The sample sizes are given in Table~\ref{tab:tot}, resulting in the largest total sample to date of 490 SNe IIn.

In Appendix~\ref{sec:extra}, we provide additional sample descriptions; the whole SN~IIn sample and the broad-lined SLSNe-II sample are summarized in Tables~\ref{tab:sample} and \ref{tab:broad}, respectively, while the excluded events are summarized in Table~\ref{tab:exc} with the reasoning.

\begin{deluxetable}{cc}
%\tabletypesize{\scriptsize}
\tablecaption{Sample Selection Criteria\label{tab:cut}}
\tablehead{
\colhead{Selection Criterion} & \colhead{Sample Size}
}
\startdata
IIn/IIn-pec/SLSN-II on TNS/WISeREP & 544 \\
At least one publicly available spectrum & 528 \\
Verified SN~IIn classification & 454 \\
Coverage of the light-curve peak in the $g$/$r$ bands & 427 \\
\enddata
\end{deluxetable}

\begin{deluxetable}{cc}
%\tabletypesize{\scriptsize}
\tablecaption{Total Sample Size\label{tab:tot}}
\tablehead{
\colhead{Data Source} & \colhead{Sample Size\tablenotemark{a}}
}
\startdata
This work & 427 (454) \\
CSP & 2 (7) \\
CCCP & 3 (4) \\
SDSS & 8 (11) \\
PS1 & 20 (24) \\
PTF & 30 (42) \\
\hline
Total & 490 (542) \\
\enddata
\tablenotetext{a}{Numbers in parentheses are before the light-curve quality cut.}
\end{deluxetable}

\section{Photometry}
\label{sec:phot}

We obtained public photometry up to 2025 July 31 (${\rm MJD}=60888$) from the Asteroid Terrestrial-impact Last Alert System (ATLAS; \citealt{Tonry2018PASP..130f4505T,Smith2020PASP..132h5002S}), Zwicky Transient Facility (ZTF; \citealt{Bellm2019PASP..131a8002B,Graham2019PASP..131g8001G}), and \textit{Gaia} Spacecraft \citep{Gaia2016A&A...595A...1G}. For the archival IIn events, we use their published photometry from the references listed in \S\ref{sec:sample}.

\subsection{ATLAS}
\label{sec:atlas}

ATLAS photometry was retrieved from the ATLAS forced photometry server\footnote{\url{https://fallingstar-data.com/forcedphot/}} \citep{Shingles2021TNSAN...7....1S} in the $c$ and $o$ bands. Following \citet{Young2024zndo..10978968Y}, quality cuts and sigma clipping at the $5\sigma$ level were applied to filter outlier data points. For each SN in each band, a baseline was measured as the median flux of at least 30 data points in the time window well before ($>500$ days) or after ($>1000$ days) the SN discovery. In cases with a step discontinuity in the baseline due to a change in the difference template image during the SN, two baseline measurements were made both before and after the SN. The flux measurements were then corrected for the baseline.

\subsection{ZTF}
\label{sec:ztf}

ZTF photometry was retrieved from the ZTF forced photometry service\footnote{\url{https://ztfweb.ipac.caltech.edu/batchfp.html}} \citep{Masci2019PASP..131a8003M,Masci2023arXiv230516279M} in the $g$, $r$, and $i$ bands. Following \citet{Masci2023arXiv230516279M}, quality cuts were applied to filter outlier data points. Additional sigma clipping at the $5\sigma$ level was applied to further reduce the scatter. For each SN in each combination of band, ZTF field, and CCD quadrant, flux measurements were corrected for the baseline, as similarly done for the ATLAS photometry (\S\ref{sec:atlas}). Flux uncertainties were corrected for the reduced $\chi^2$ distribution of point-spread function fits \citep{Masci2023arXiv230516279M}.

\subsection{Gaia}
\label{sec:gaia}

\textit{Gaia} photometry was retrieved from the \textit{Gaia} Science Alerts\footnote{\url{http://gsaweb.ast.cam.ac.uk/alerts/home}} \citep{Hodgkin2021A&A...652A..76H} in the $G$ band. The epochs with ``null'' or ``untrusted'' magnitude measurements were discarded. For each magnitude measurement, an uncertainty was estimated from the median standard deviation at the given magnitude (see Figure~15 of \citealt{Hodgkin2021A&A...652A..76H}).

\subsection{Additional Data Sources}
\label{sec:add}

In addition to the ATLAS, ZTF, and \textit{Gaia} photometry, we include available Pan-STARRS photometry in the $g$, $r$, $i$, and $z$ bands for a subset of our SN~IIn sample (13 events;\footnote{Out of 15 events classified as SNe~IIn or SLSN-II in YSE DR1, excluding AT~2020rdu without a public spectrum and SN~2020tan as a non-IIn contaminant.} see Table~\ref{tab:sample}) from the Young Supernova Experiment (YSE; \citealt{Jones2021ApJ...908..143J}) Data Release 1 (DR1; \citealt{Aleo2023ApJS..266....9A}). For the highly energetic SN~IIn 2016aps \citep{Nicholl2020NatAs...4..893N}, we also include the published photometry\footnote{Via WISeREP: \url{https://www.wiserep.org/object/14528}; it is the only object in our sample whose published photometry is available on WISeREP.} and use the SN as a comparison object in this work.

\subsection{Postprocessing}
\label{sec:post}

Throughout this work, we report the flux measurements in $\mu$Jy. All the photometry (including the literature samples from CSP, CCCP, SDSS, PS1, and PTF) is weighted by uncertainty and stacked every 0.5 day in flux space to account for negative flux as well (if available). The detection significance ($\sigma$) is determined from the ratio of measured flux ($F_\nu$) to its error ($F_{\nu,{\rm err}}$). For the measurements above and below $3\sigma$, we report their magnitudes ($-2.5\,{\rm log}_{10}(F_\nu)+{\rm ZP}$) and $3\sigma$ upper limits ($-2.5\,{\rm log}_{10}(3\times F_{\nu,{\rm err}})+{\rm ZP}$), respectively, where ``ZP'' is the zero-point in the AB magnitude system, i.e., $-2.5\,{\rm log}_{10}(3631\,{\rm Jy}/\mu{\rm Jy})=23.9$ mag.

We assume a standard $\Lambda$CDM cosmology with $H_0=71.0$\, km\,s$^{-1}$\,Mpc$^{-1}$, $\Omega_{\Lambda}=0.7$, and $\Omega_m=0.3$ and convert the redshift to a luminosity distance. Due to the lack of spectra with resolved host galaxy Na~{\sc i}~D absorption for the majority of the sample, we only account and correct for the Milky Way (MW) extinction \citep{Schlafly2011ApJ...737..103S}, assuming the \cite{Fitzpatrick1999PASP..111...63F} reddening law with $R_V=3.1$.

\section{Gaussian Process}
\label{sec:gp}

The main objective of this large sample study is to perform uniform measurements in a model-independent way for such measurements to be meaningful and comparable. Moreover, given the heterogeneous nature of the sample, it is also required to overcome the different survey systematics (i.e., cadences, filters, depths). 
Therefore, we perform a two-dimensional GP to interpolate the sample light curves between epochs and frequencies.

\begin{figure}
    \centering
    \includegraphics[width=0.49\textwidth]{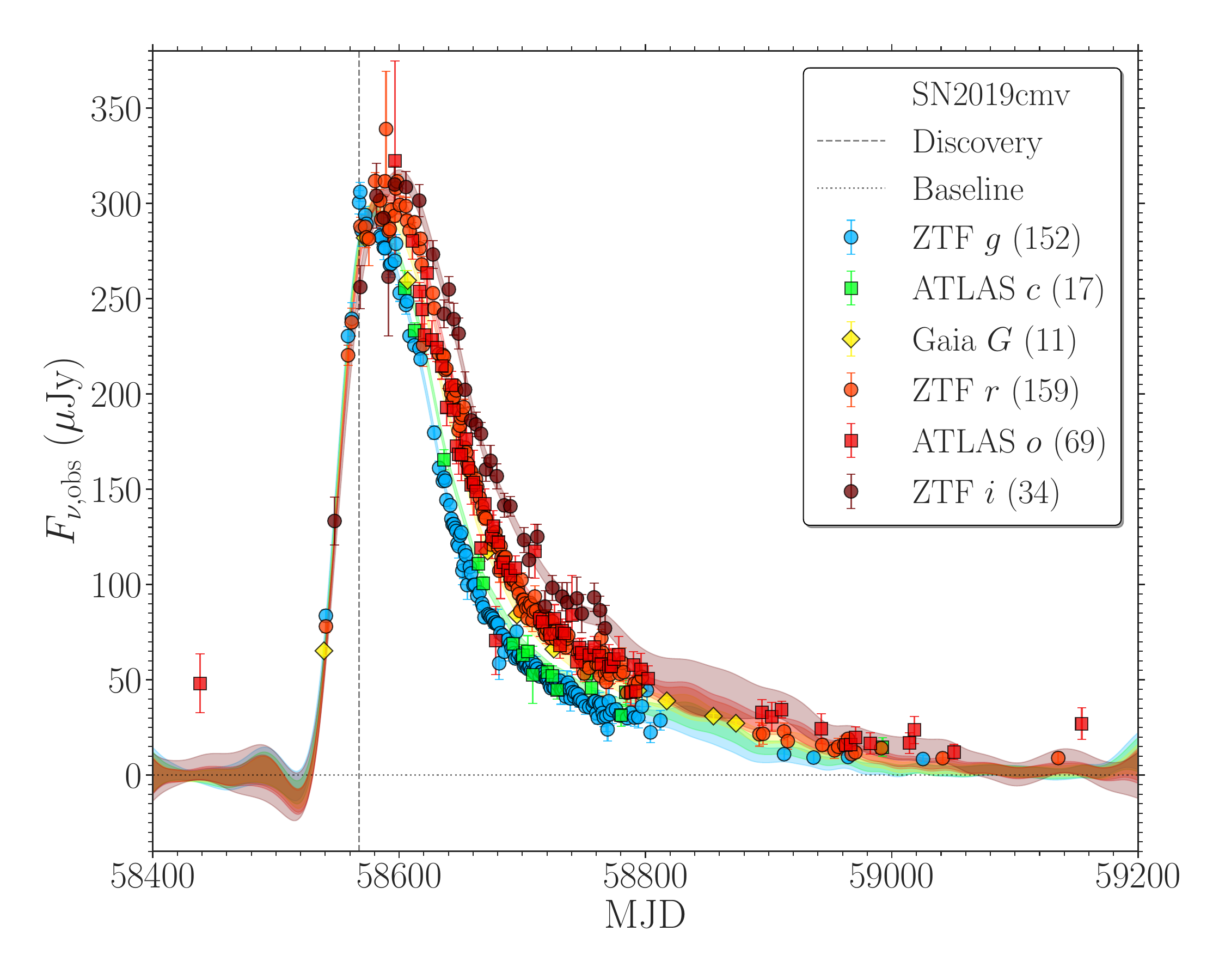}
\caption{An example of the multiband observed (points) and 2D (time and frequency) GP light curves (lines and shaded regions). The error bars and shaded regions represent $1\sigma$ uncertainty in the observed and GP light curves, respectively. Only the $\ge 3\sigma$ detections (with the number in the parentheses for each band) are shown for clarity, but the nondetections are also used to anchor the GP light curves to zero flux (i.e., baseline) before and after the SN.
}
    \label{fig:GP}
\end{figure}

We do so by using \texttt{GaussianProcessRegressor} from \texttt{scikit-learn} \citep{Pedregosa2011JMLR...12.2825P} with a combination of Mat\'{e}rn, Constant, and White (i.e., noise) kernels (Constant $\times$ Mat\'{e}rn ($\nu=2.5$) $+$ White). For each SN, the input dataset is made of observed light curves shifted to the rest frame, consisting of a 2D array of time and effective filter frequency ($\nu_{\rm eff,obs}(1+z)$) and 1D arrays of flux ($F_{\nu,{\rm obs}}/(1+z)$) and its variance ($(F_{\rm err,obs}/(1+z))^2$) where $z$ is the redshift. The kernels are then optimized in the GP regression fit and used to predict the rest-frame flux ($F_{\nu,{\rm rest}}$) with associated uncertainty ($F_{\nu,\rm err,rest}$) as a function of $\nu_{\rm eff,rest}$, i.e., the spectral energy distribution (SED), at any given time. Thus, the SED information is incorporated in the flux frame transformation, i.e., a $K$-correction, in addition to the cosmological correction of $1/(1+z)$. For a given filter with $\nu_{\rm eff,obs}$, $F_{\nu,{\rm rest}}(1+z)$ at $\nu_{\rm eff,rest}=\nu_{\rm eff,obs}(1+z)$ gives the observer-frame flux. An example multiband GP light curve is shown in Figure~\ref{fig:GP}.

% (e.g., $5\times10^{14}$\,Hz in the $r$-band)

\subsection{Measurements}
\label{sec:mea}

\begin{figure}
    \centering
    \includegraphics[width=0.49\textwidth]{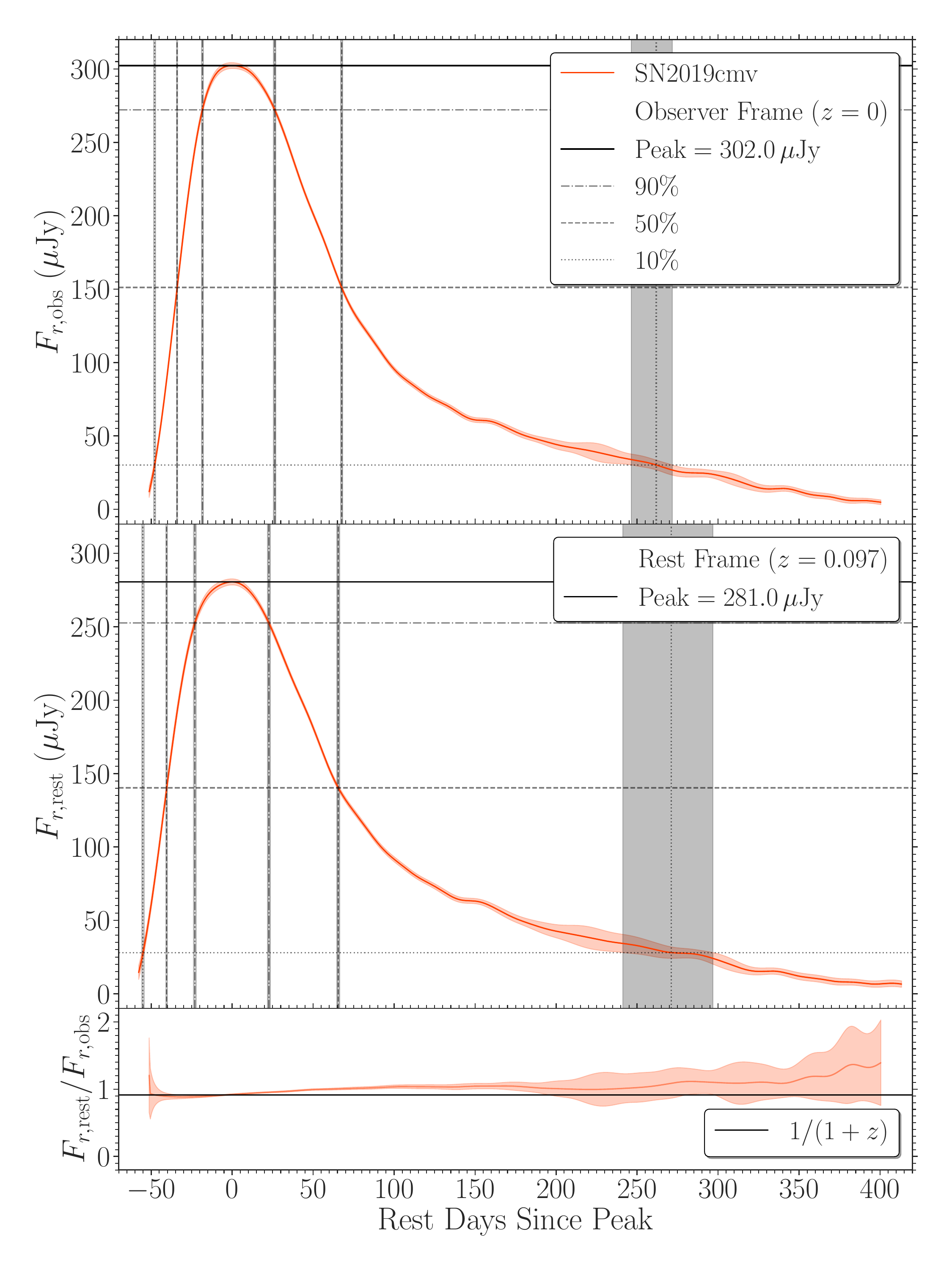}
\caption{An example of the peak brightness and timescale measurements of the $r$-band GP light curves in the observer (\textit{top}) and rest (\textit{middle}) frames, along with the flux ratio in the two frames, i.e., the $K$-correction (\textit{bottom}). The color-shaded region corresponds to the $1\sigma$ flux uncertainty in the GP light curve ($F_{\nu, {\rm err}}$), which is propagated to the timescale uncertainty at each available 10\% flux measurement with respect to the peak, i.e., the duration between $F_{\nu}+F_{\nu, {\rm err}}\leq F_{\nu,X\%} \leq F_{\nu}-F_{\nu, {\rm err}}$ on the rise or $F_{\nu}-F_{\nu, {\rm err}}\leq F_{\nu,X\%} \leq F_{\nu}+F_{\nu, {\rm err}}$ on the decline (gray shaded regions). The $K$-correction incorporates the SED information from the 2D GP (Figure~\ref{fig:GP}) and differs from the cosmological-only correction of $1/(1+z)$. 
}
    \label{fig:KC}
\end{figure}

\begin{figure}
    \centering
    \includegraphics[width=0.49\textwidth]{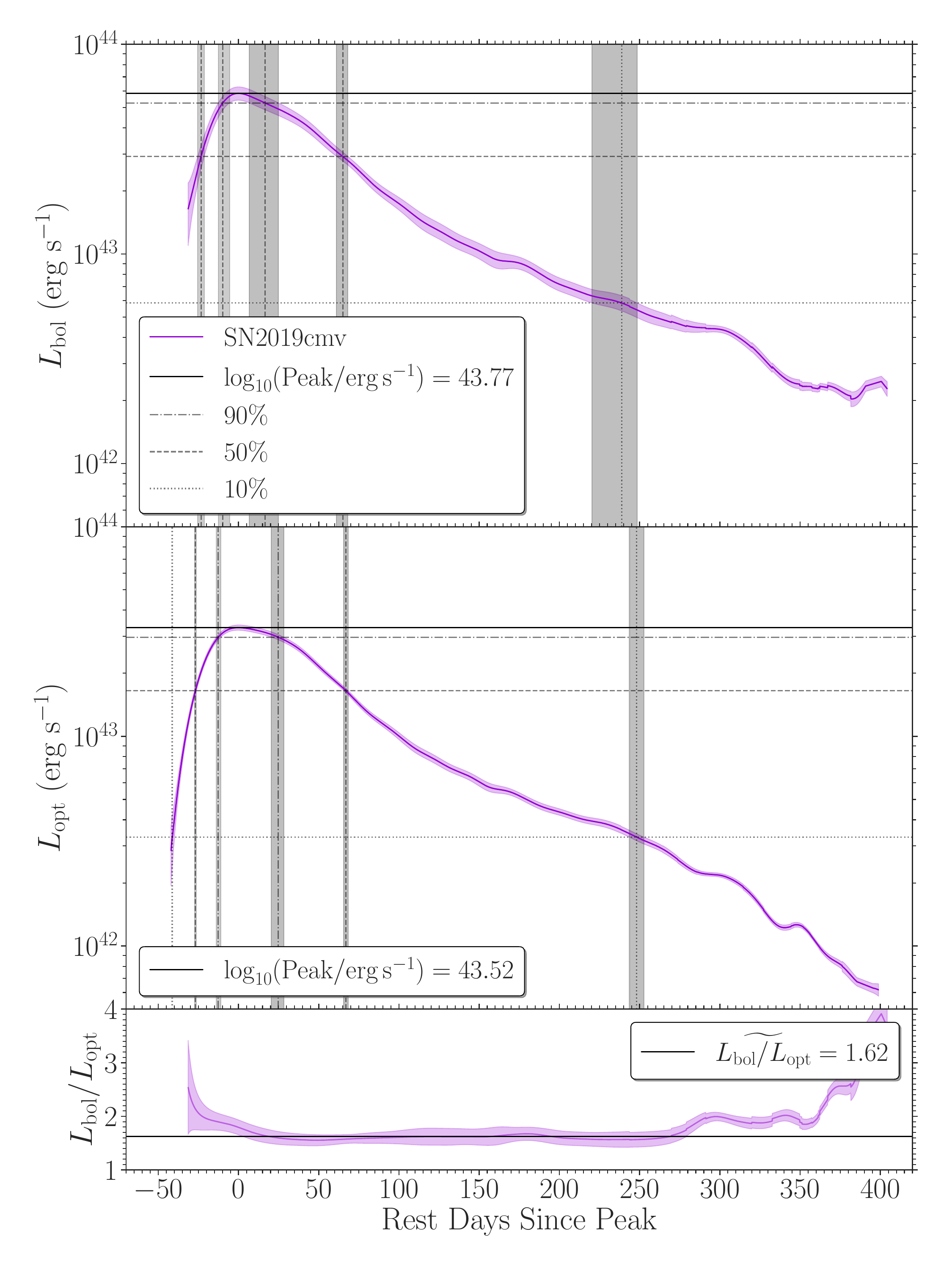}
\caption{Similar to Figure~\ref{fig:KC}, but for the bolometric (\textit{top}) and pseudo-bolometric (i.e., optical: 3250--8900$\,\Angstrom$; \textit{middle}) luminosity from blackbody SED fits, along with their ratio, i.e., the optical-bolometric correction (\textit{bottom}). Given only the optical coverage, the blackbody temperature and hence the bolometric luminosity are not well constrained in the early phase when the SED peaks in ultraviolet ($\gtrsim10$,$000$\,K). Nevertheless, the optical-bolometric correction is roughly constant around the median over the bulk of the light curve, from roughly the peak to about 250 days (in this case).
}
    \label{fig:BC}
\end{figure}

For a uniform light-curve characterization, we perform simple measurements directly on the 2D GP light curves in the well-constrained phase above $3\sigma$ ($F_\nu/F_{\nu, {\rm err}}\geq3$). For each SN in each band, we first find a maximum flux and then measure the timescale at each available 10\% flux level on the rise and decline with respect to the maximum value, as illustrated in Figure~\ref{fig:KC}. For example, $t_{50,{\rm rise}}$ corresponds to the duration between $50\%$ flux to the maximum on the rise, while $t_{50,{\rm decl}}$ corresponds to the duration between the maximum to 50\% flux on the decline. $F_{\nu, {\rm err}}$ is propagated to the lower and upper timescale uncertainties (Figure~\ref{fig:KC}), and we discard measurements without proper uncertainty quantification due to insufficient light-curve coverage/depth. To ensure that the maximum flux is indeed the peak flux, we require that both $t_{90,{\rm rise}}$ and $t_{90,{\rm decl}}$ are properly measured with uncertainty. With this requirement for the rest-frame peak coverage in the ZTF $g$ and/or $r$ band, we exclude 27 objects from the 454 with verified IIn classification in our sample (Tables~\ref{tab:cut} and \ref{tab:exc}). We perform the same for the literature IIn sample (Tables~\ref{tab:tot} and \ref{tab:exc}).

Using the 2D GP light curves, we construct bolometric and pseudo-bolometric (i.e., optical: 3250--8900$\,\Angstrom$) light curves by fitting and integrating a blackbody SED\footnote{This is reasonable given the typical SN~IIn SED dominated by a blue continuum with a narrow H$\alpha$ emission line, except in the late nebular phase ($\gtrsim300$ days) when H$\alpha$ starts to dominate the SED.} at every epoch containing GP interpolations above $3\sigma$ in at least three bands.\footnote{Excluding \textit{Gaia} $G$ band in the fits, given its broad wavelength coverage.} We then perform the same set of measurements on the bolometric and optical light curves, as illustrated in Figure~\ref{fig:BC}. 
We note that the SEDs generally peak blueward of the optical range in the early phase (i.e., $\gtrsim10$,$000$\,K), and the fitted blackbody temperatures may be underestimated by up to $\sim10$,$000$\,K along with the radius overestimations, which translates to a factor of up to $\sim5$ in the bolometric luminosity  (e.g., \citealt{Valenti2016MNRAS.459.3939V,Arcavi2022ApJ...937...75}). Given this potential uncertainty in the bolometric luminosity at early times, we center our analysis on the optical luminosity.

\section{Analysis} 
\label{sec:ana}

In the following analysis, we use the rest-frame measurements in the $r$-band and pseudo-bolometric (optical) light curves, unless otherwise noted. We note that the overall trends and correlations are similar in the other bands and in the bolometric light curves, and we provide all the measurements as the data behind Figure~\ref{fig:Mr_z}.

\subsection{Peak Brightness versus Redshift}
\label{sec:z}

\begin{figure}
    \centering
    \includegraphics[width=0.49\textwidth]{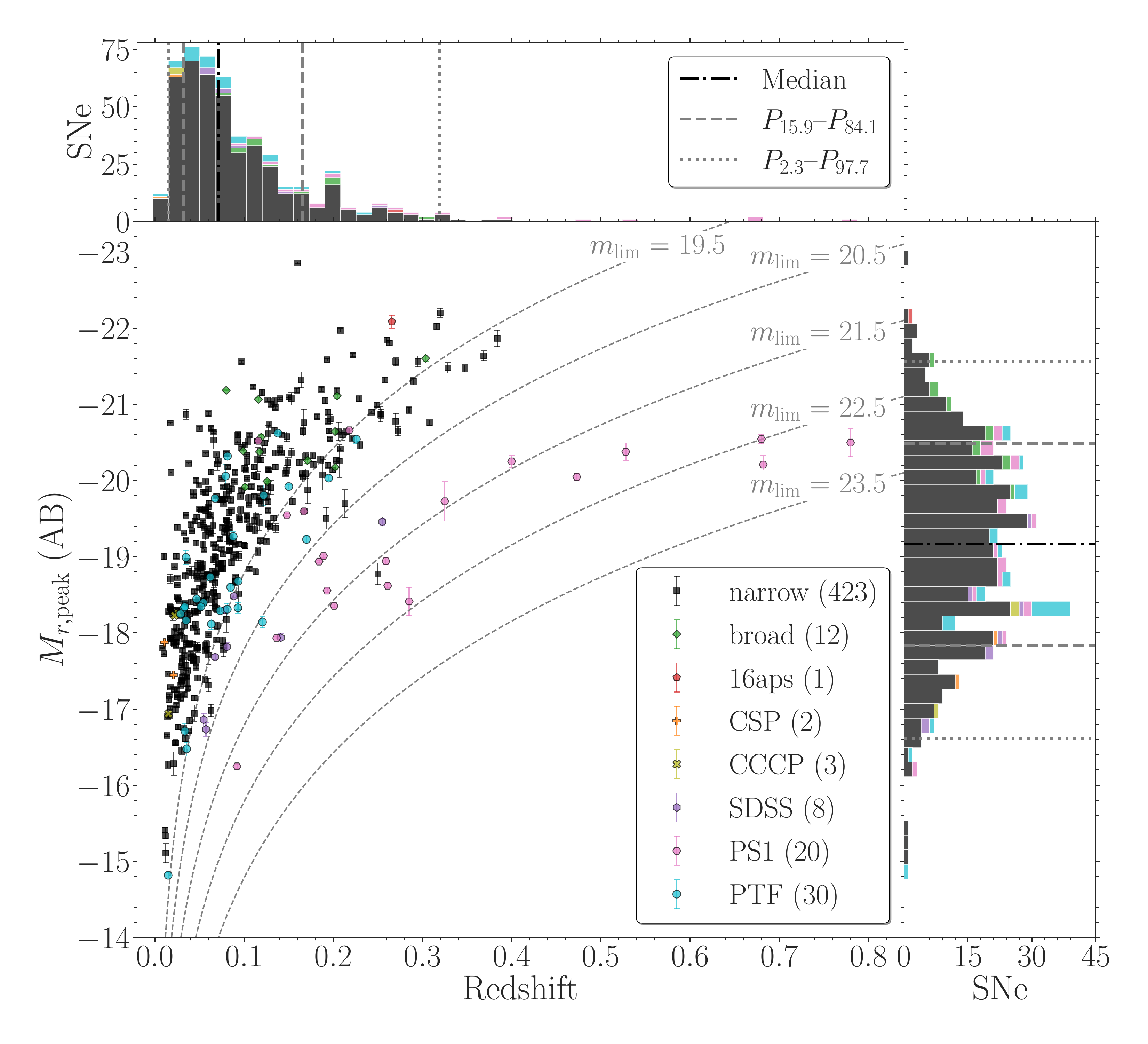}
\caption{Peak $r$-band absolute magnitude vs. redshift, with projected histograms. The magnitude errors denote $1\sigma$ uncertainty. The dashed curves correspond to a range of survey limiting magnitudes. For presentation purposes, the sample is divided to ``narrow-lined'' SN IIn (this work), ``broad-lined'' SLSN-II (\citealt{Kangas2022MNRAS.516.1193K} and this work), the highly energetic SN IIn 2016aps \citep{Nicholl2020NatAs...4..893N}, and literature SN IIn samples from previous surveys, analyzed uniformly here: CSP \citep{Stritzinger2012ApJ...756..173S,Taddia2013}, CCCP \citep{Kiewe2012ApJ...744...10K}, SDSS \citep{Sako2018PASP..130f4002S}, PS1 \citep{Villar2020ApJ...905...94V}, and PTF \citep{Nyholm2020}, with their sample size in the parentheses.
(The data used to create this figure and all other measurements will be available in machine-readable format upon publication.)
}
    \label{fig:Mr_z}
\end{figure}

% SDSS ~ 22.2
% PTF ~20.5-21.0

\begin{figure}
    \centering
    \includegraphics[width=0.45\textwidth]{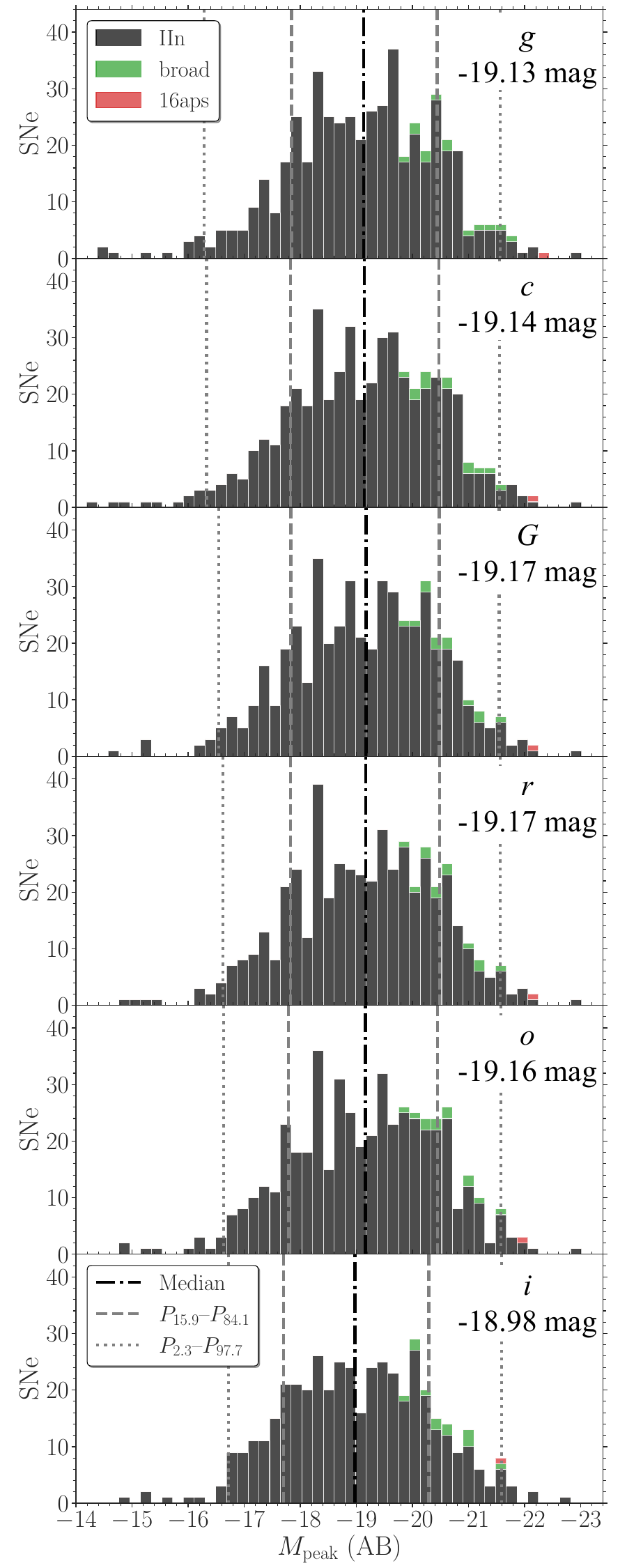}
\caption{Histograms of the peak absolute magnitude in each filter. ``IIn'' includes the entire sample (this work, CSP, CCCP, SDSS, PS1, and PTF), while ``broad'' and ``16aps'' are shown for comparison (as in Figure~\ref{fig:Mr_z}). The vertical lines in each panel mark the median (with its value on the top right) and $P_{15.9}$--$P_{84.1}$ and $P_{2.3}$--$P_{97.7}$ ranges. The peak distributions span $\approx2.6$ and $\approx5.0$ mag in all filters at the $P_{15.9}$--$P_{84.1}$ and $P_{2.3}$--$P_{97.7}$ ranges, respectively.
}
    \label{fig:bandP}
\end{figure}

\begin{figure}
    \centering
    \includegraphics[width=0.45\textwidth]{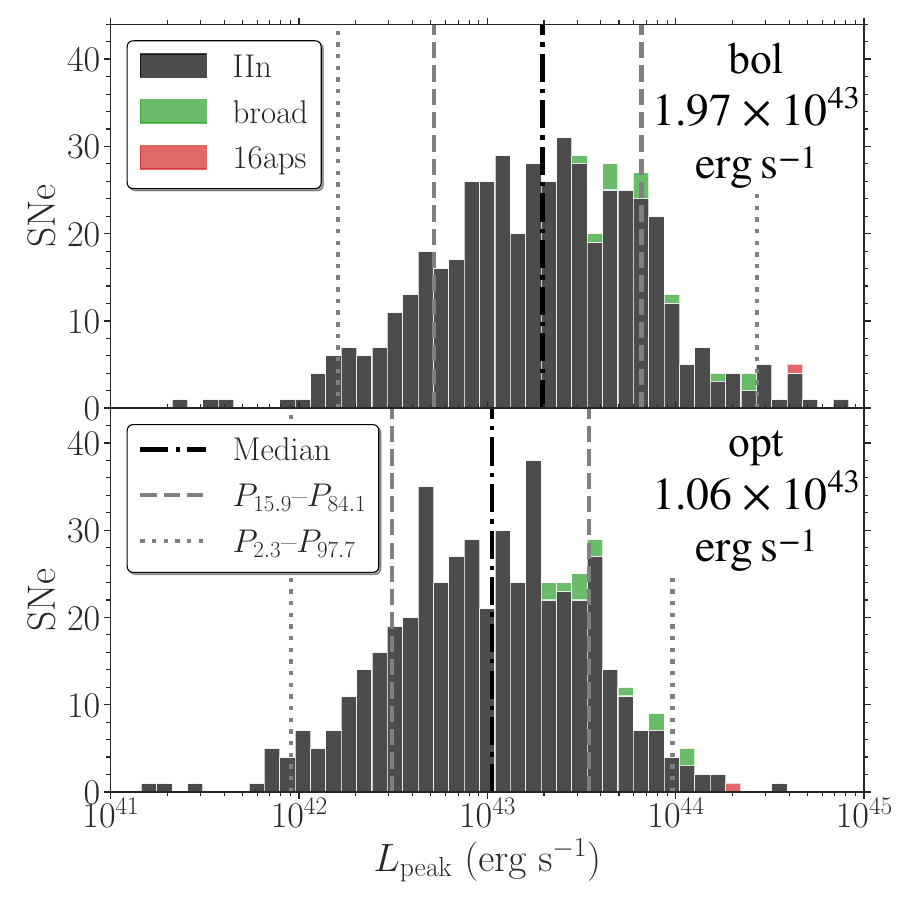}
\caption{Similar to Figure~\ref{fig:bandP}, but for the bolometric (\textit{top}) and optical (\textit{bottom}) luminosities. The $P_{15.9}$--$P_{84.1}$ range is about an order of magnitude, and an overall spread is about 2 orders of magnitude at the $P_{2.3}$--$P_{97.7}$ range. The median offset between the optical and bolometric peaks is about a factor of $1.9$, which can be used as a rough proxy for the overall optical-bolometric correction of the class.
}
    \label{fig:bolP}
\end{figure}

In Figure~\ref{fig:Mr_z}, we plot the peak $r$-band absolute magnitude ($M_{r,{\rm peak}}$) versus redshift for the full SN~IIn sample. The bulk of the population is located at $z\lesssim 0.3$ given the limiting magnitudes of ATLAS ($\approx19.7$ mag) and ZTF ($\approx20.7$ mag), with an extension to $z\approx 0.8$ from PS1, which had a limiting magnitude of $\approx23.3$ mag. The median redshift is $0.071$ with the 15.9th and 84.1th percentiles (which we denote as $P_{15.9}$ and $P_{84.1}$ in this work) at $0.032$ and $0.166$, respectively, and the 2.3th and 97.7th percentiles ($P_{2.3}$ and $P_{97.7}$) at $0.015$ and $0.319$, respectively. The median $\widetilde{M}_{r,{\rm peak}}$ is $-19.16$ mag with the $P_{15.9}$--$P_{84.1}$ and $P_{2.3}$--$P_{97.7}$ spreads of $-17.83$ to $-20.50$ mag and $-16.62$ to $-21.56$ mag, respectively. We do not correct the sample for Malmquist bias, given its heterogeneous nature, but such a correction should marginally lower the median, well within the $P_{15.9}$--$P_{84.1}$ spread (e.g., \citealt{Nyholm2020,Pessi2025A&A...695A.142P,Ransome2025ApJ...987...13R}). We further discuss the effect of Malmquist bias on the radiated energy distribution in \S\ref{sec:energy}.

In addition to the $r$ band, we show histograms of $M_{\rm peak}$ in each available filter in Figure~\ref{fig:bandP}. The median $\widetilde{M}_{\rm peak}$ values are $-19.13^{-1.31}_{+1.28}$, $-19.14^{-1.33}_{+1.31}$, $-19.17^{-1.30}_{+1.34}$, $-19.17^{-1.32}_{+1.34}$, $-19.16^{-1.28}_{+1.37}$, and $-18.98^{-1.31}_{+1.28}$ mag in the $g$, $c$, $G$, $r$, $o$, and $i$ bands, respectively, with the upper/lower bounds corresponding to the $P_{15.9}$--$P_{84.1}$ range. 
The sample spans $\approx 5.0$ mag in all filters at the $P_{2.3}$--$P_{97.7}$ range.

In Figure~\ref{fig:bolP}, we show histograms of the peak bolometric and optical luminosities ($L_{\rm peak}$). The median $\widetilde{L}_{\rm peak}$ values are $(1.97^{+4.63}_{-1.44})\times10^{43}$ and $(1.06^{+2.40}_{-0.76})\times10^{43}$\,erg\,s$^{-1}$ for the bolometric and optical luminosities, respectively, with a median offset of $\approx 1.9$ between them. This offset can be used as a rough proxy for the overall optical-bolometric correction of the sample (although we reiterate the potential underestimation of the bolometric peak luminosity as noted in \S\ref{sec:mea}). 
As similarly seen in the filter peaks (Figure~\ref{fig:bandP}), the sample spans about 2 orders of magnitude in both bolometric and optical peaks at the $P_{2.3}$--$P_{97.7}$ range.

\subsection{Light-curve Morphologies}
\label{sec:lc}

\begin{figure*}
    \centering
    \includegraphics[width=0.49\textwidth]{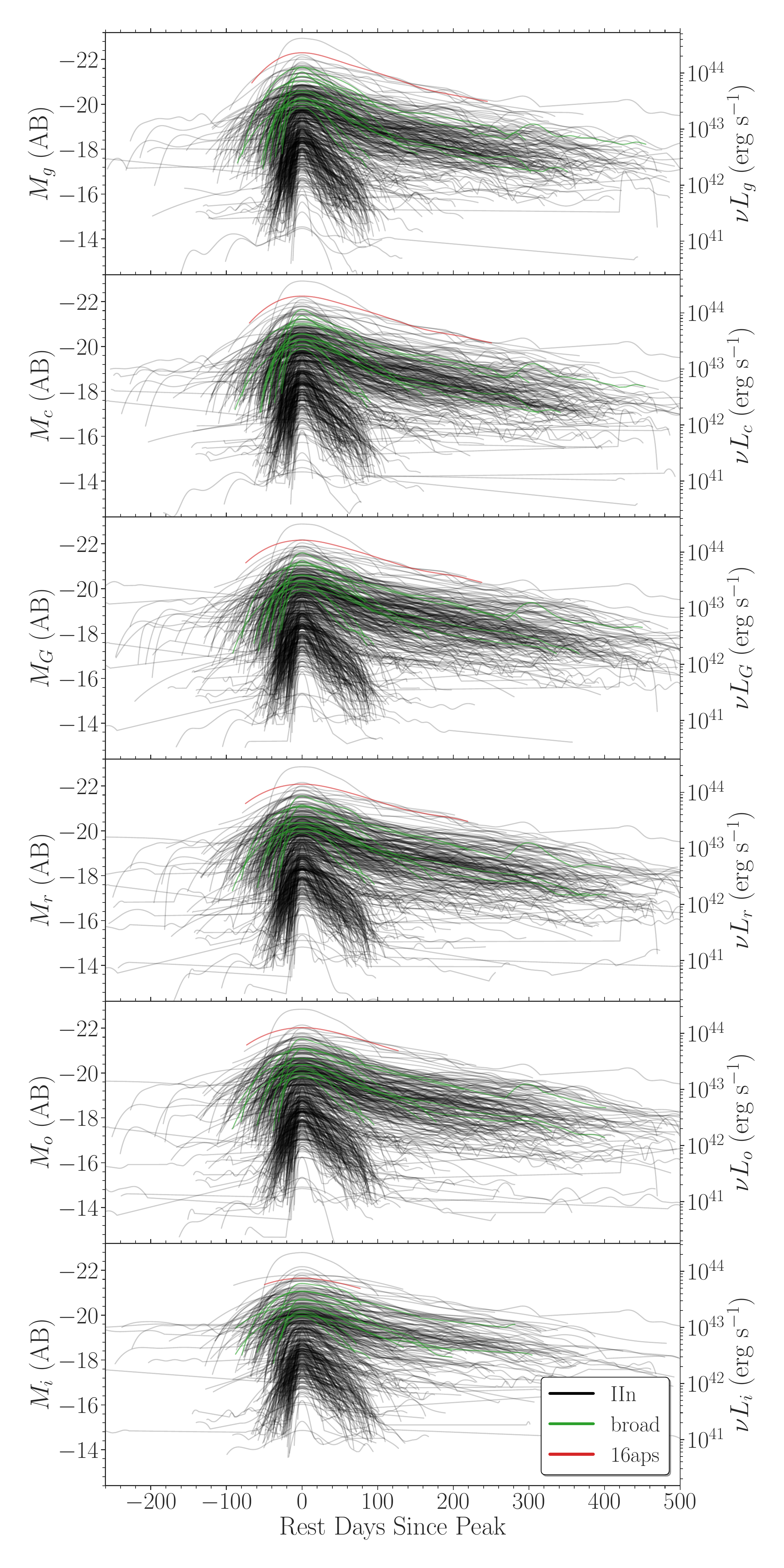}
    \includegraphics[width=0.49\textwidth]{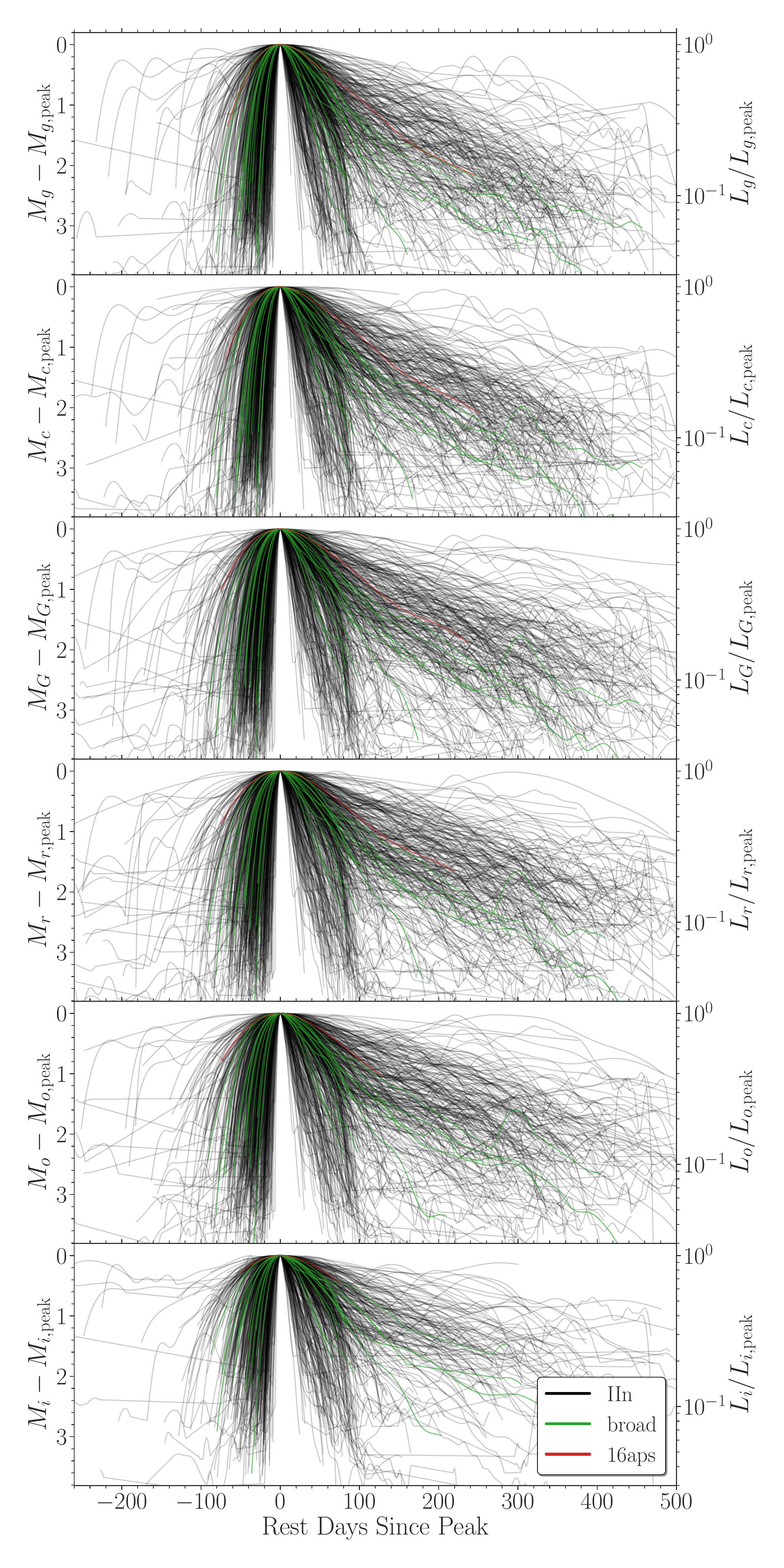}
\caption{{\it Left:} GP light curves in absolute magnitude in each filter ($g$, $c$, $G$, $r$, $o$, and $i$ from \textit{top} to \textit{bottom}) for the IIn sample (black) and comparison broad-lined (green) and SN~2016aps (red) events.  {\it Right:} same light curves, but normalized at peak. While the IIn sample spans wide ranges of brightness and timescale, there are two clear light-curve morphologies: ``faint--fast" and ``luminous--slow" (see also Figure~\ref{fig:LCtemp}). 
(The data used to create this figure will be available in machine-readable format upon publication.)
}
    \label{fig:bandLC}
\end{figure*}

\begin{figure*}
    \centering
    \includegraphics[width=0.49\textwidth]{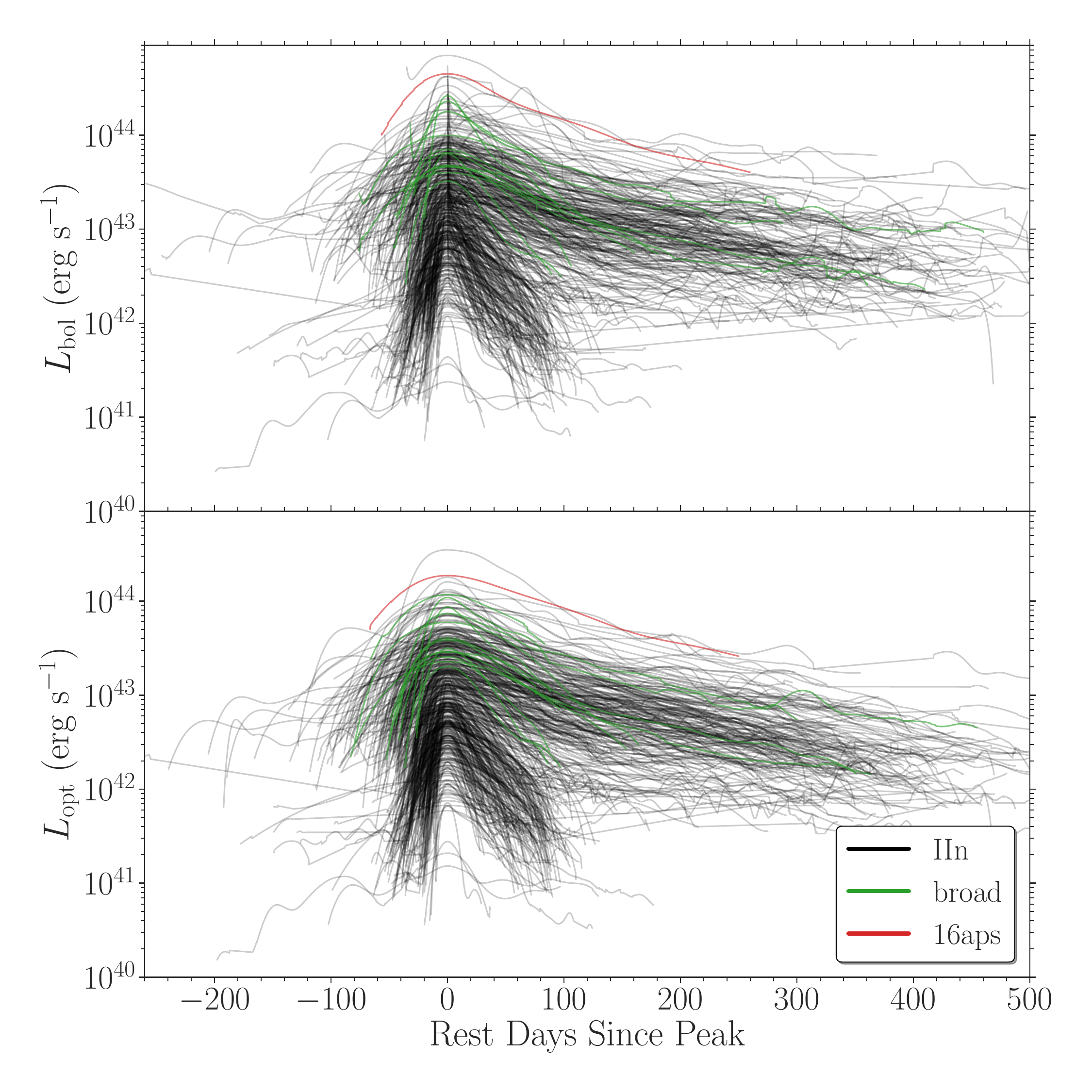}
    \includegraphics[width=0.49\textwidth]{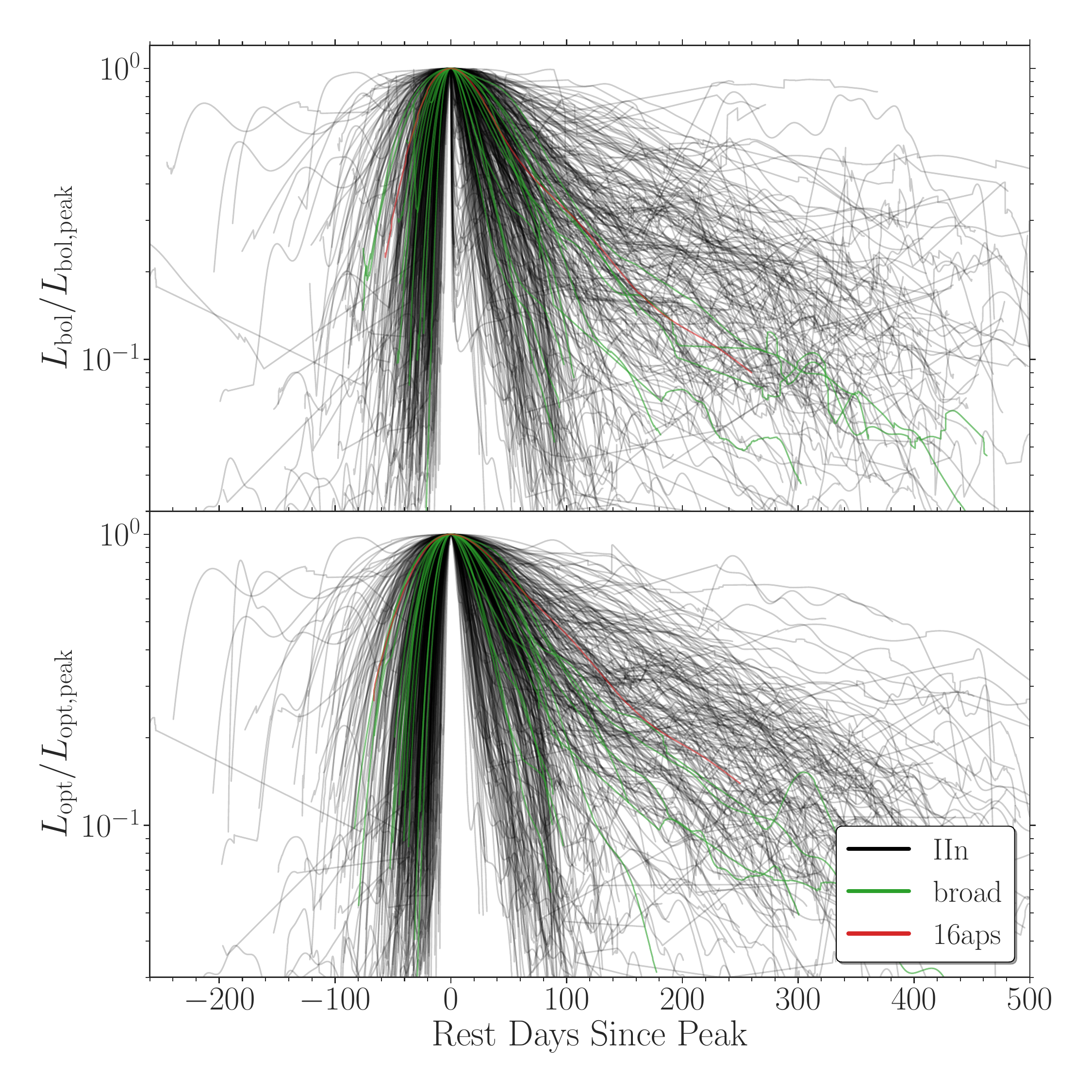}
\caption{Similar to Figure~\ref{fig:bandLC}, but in luminosity space, where the same two light-curve morphologies can be seen (see also Figure~\ref{fig:Lboltemp}).
(The data used to create this figure will be available in machine-readable format upon publication.)
}
    \label{fig:bolLC}
\end{figure*}

In Figure~\ref{fig:bandLC}, we plot all the ensemble of GP light curves in each filter. In addition to the wide spread in peak brightness (\S\ref{sec:z}), the sample also spans a wide range of timescales. Despite the broad range in peak brightness and timescale, however, the population bifurcates into two clear light-curve morphological groups: luminous events with slow temporal evolution, and fainter events with faster evolution, with the two groups separated around the median peak brightness in each filter (Figure~\ref{fig:bandP}). Hereafter, we refer to these two subpopulations as ``luminous--slow'' and ``faint--fast,'' and we note that few events reside between them.

When normalized at peak (right panels of Figure~\ref{fig:bandLC}), the timescale is better visualized, revealing a mostly continuous distribution on the rise of the light curve, and a wide, bimodal spread on the decline, exhibiting the fast and slow groups noted above. We note that the broad-lined SLSNe-II show a light-curve morphology similar to that of the luminous--slow SNe~IIn. In addition, there is an unoccupied region in duration below the shortest timescale of the faint--fast SNe~IIn of $\lesssim15$ days (above the 50\% of peak flux; see \S\ref{sec:time}). 

In Figure~\ref{fig:bolLC}, we plot the bolometric and optical light curves. As in the filter light curves (Figure~\ref{fig:bandLC}), the same two light-curve morphologies are evident, roughly separated around the median peak luminosities of $2\times10^{43}$\,erg\,s$^{-1}$ (bolometric) and $10^{43}$\,erg\,s$^{-1}$ (optical), respectively (Figure~\ref{fig:bolP}). We expand the analysis of the light-curve morphology in \S\ref{sec:temp}, where we construct a light-curve template for each morphological group.

\subsection{Rise and Decline Timescales} 
\label{sec:time}

\begin{figure*}
    \centering
    \includegraphics[width=0.98\textwidth]{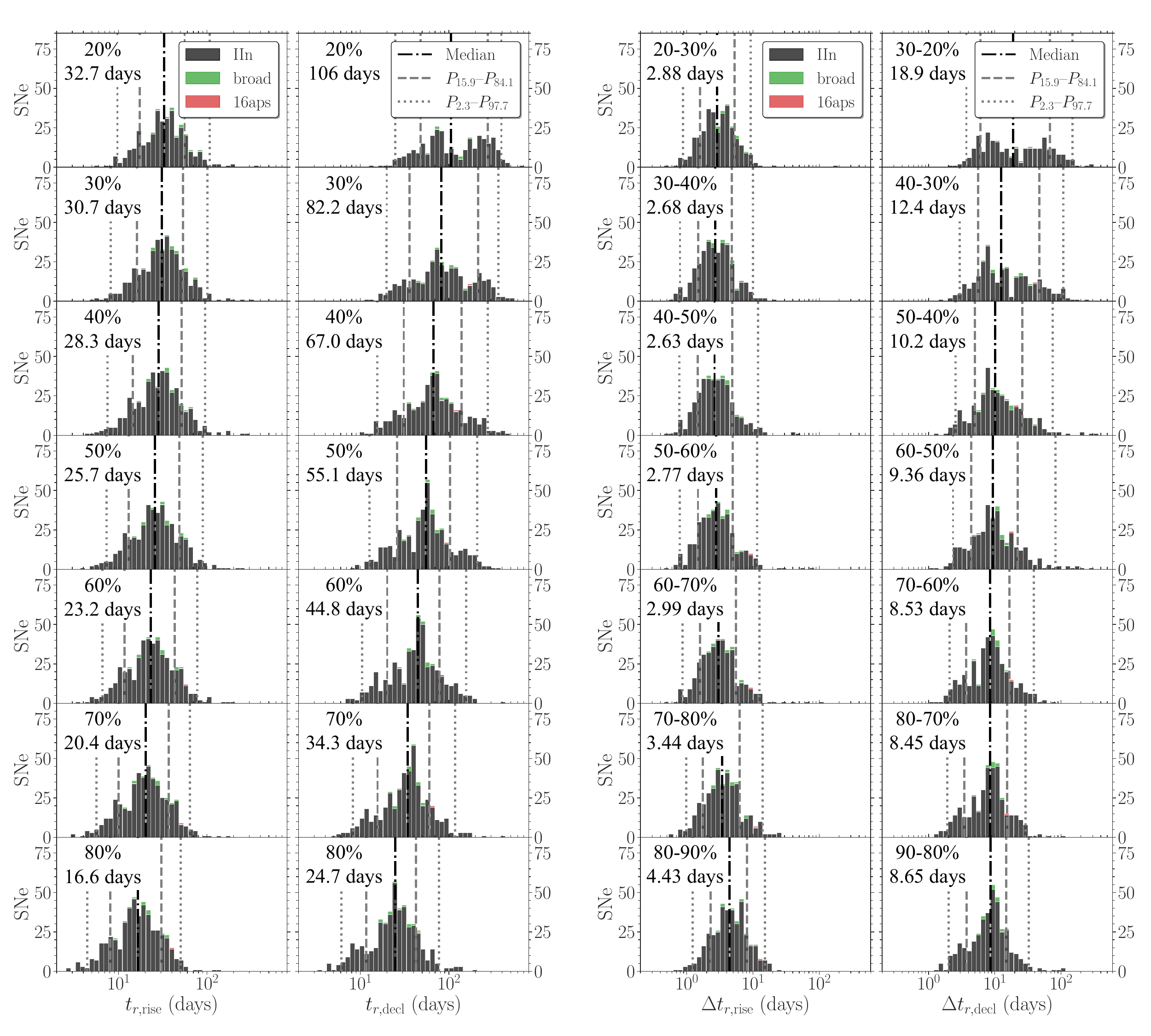}
\caption{{\it Left:} histograms of the $r$-band timescales ($t_r$) above each 10\% flux level relative to peak on the rise and decline. {\it Right:} the difference between successive 10\% flux levels ($\Delta t_r$, namely, the inverted slope) on the rise and decline. The vertical lines in each panel mark the median (with its value on the top left) and $P_{15.9}$--$P_{84.1}$ and $P_{2.3}$--$P_{97.7}$ ranges. For each distribution, the temporal progression along the light curve is counterclockwise from the left-top to right-top panels. The inverted slope is an increasing function, i.e., the light-curve slope becomes shallower with respect to time.
}
    \label{fig:rT}
\end{figure*}

\begin{figure}
    \centering
    \includegraphics[width=0.44\textwidth]{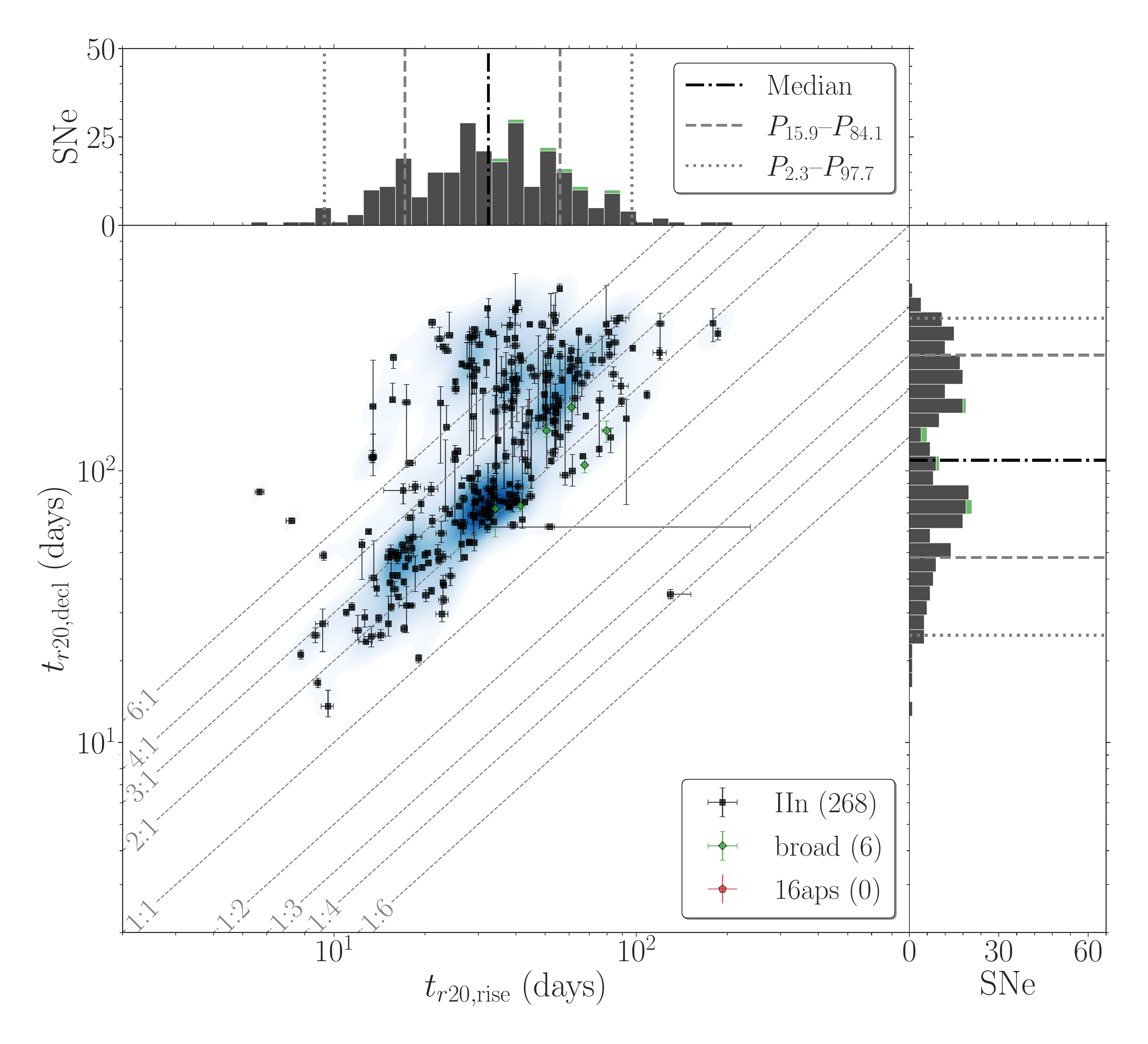}
    \includegraphics[width=0.44\textwidth]{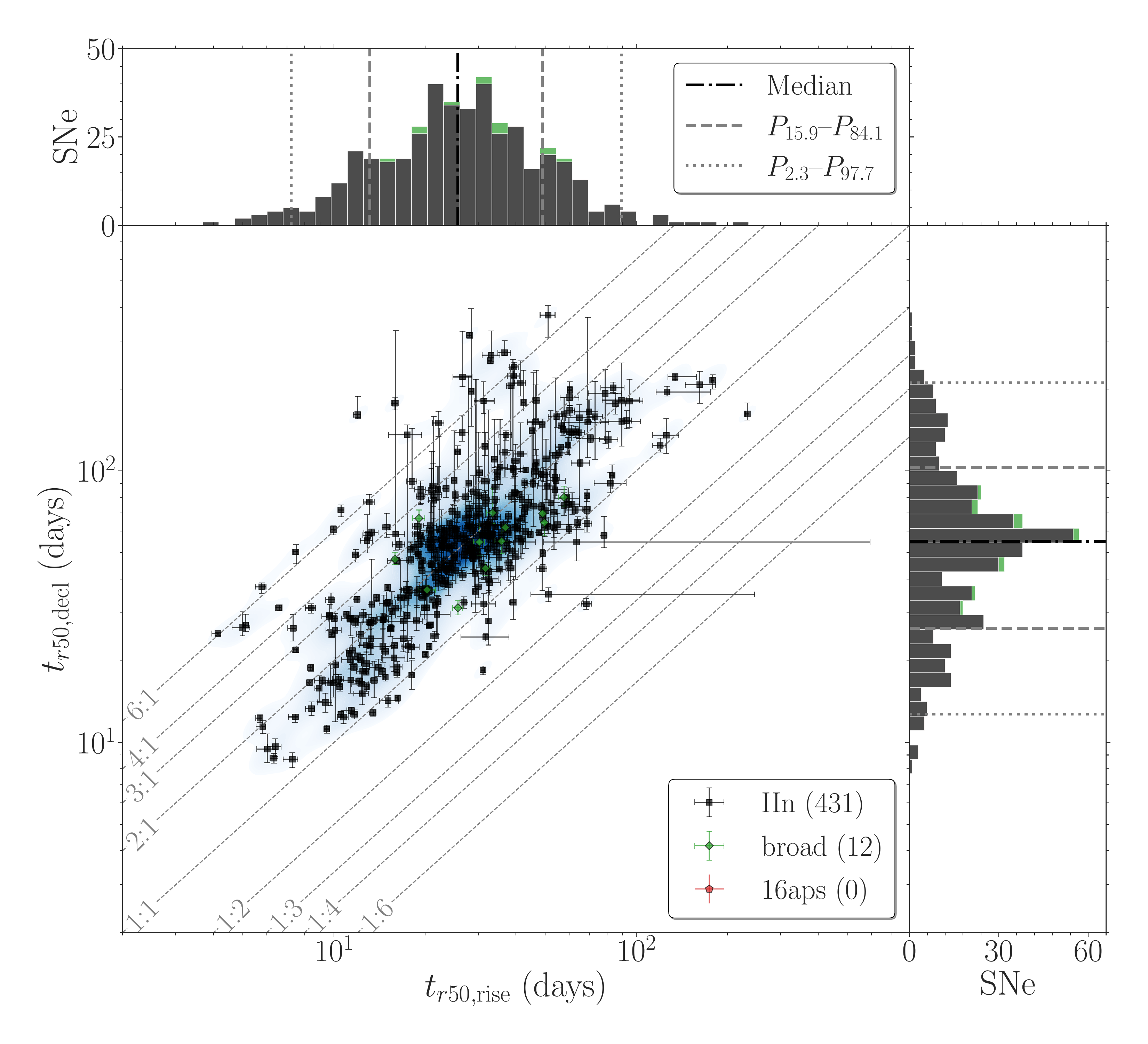}
    \includegraphics[width=0.44\textwidth]{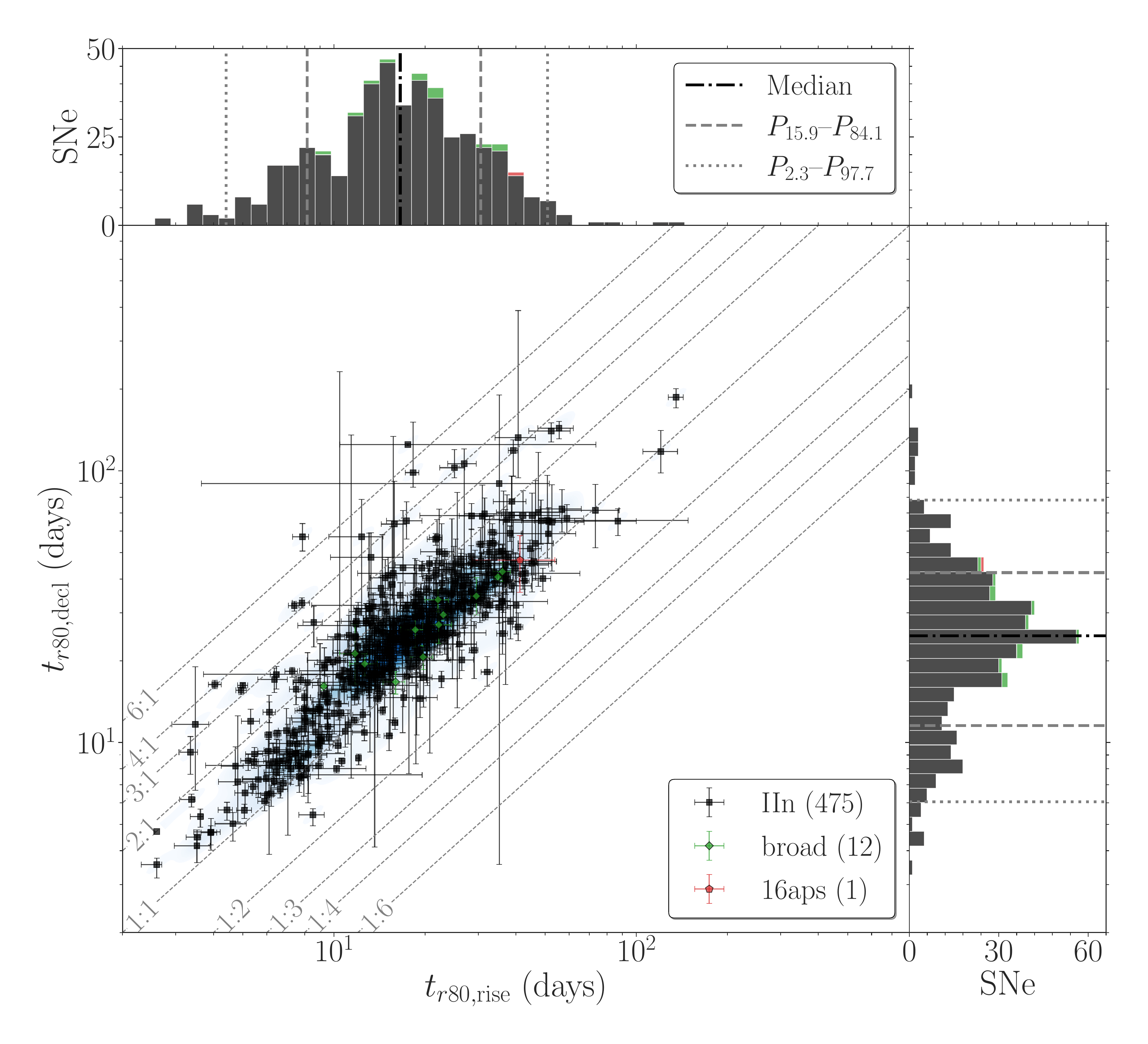} 
\caption{Comparisons of the $r$-band timescales on the rise and decline above three flux levels: 20\% ($t_{r20}$; \textit{top}), 50\% ($t_{r50}$; \textit{middle}), and 80\% ($t_{r80}$; \textit{bottom}). The error bars  denote $1\sigma$ uncertainties, while the underlying blue shaded regions represent the density maps. The light curves become more symmetric around the peak; $t_{r20, {\rm decl}} \approx 3 \times t_{r20, {\rm rise}}$, $t_{r50, {\rm decl}} \approx 2 \times t_{r50, {\rm rise}}$, and $t_{r80, {\rm decl}} \approx 1.5 \times t_{r80, {\rm rise}}$.
 }
    \label{fig:rT205080}
\end{figure}

\begin{figure}
    \centering
    \includegraphics[width=0.49\textwidth]{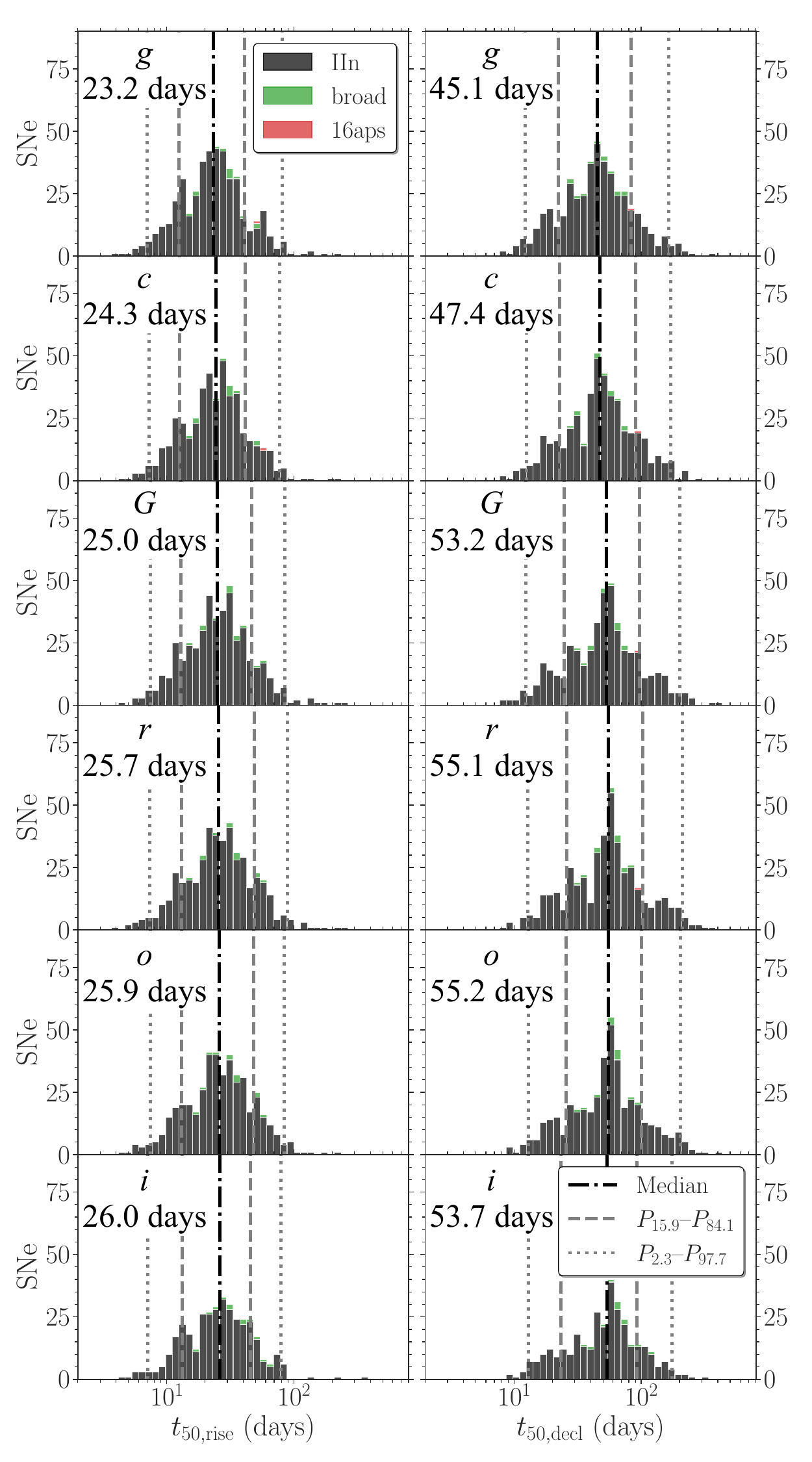}
    \caption{Histograms of the 50\% timescales in each filter on the rise (\textit{left}) and decline (\textit{right}). The vertical lines in each panel mark the median (with its value on top left) and $P_{15.9}$--$P_{84.1}$ and $P_{2.3}$--$P_{97.7}$ ranges. In general, both the rise and decline timescales are faster in bluer filters, while the slower decline timescales in $r$ and $o$ than in $i$ are likely due to the presence of prominent H$\alpha$ emission line.
    }
    \label{fig:bandT50}
\end{figure}

In the left panels of Figure~\ref{fig:rT}, we plot histograms of the $r$-band rise and decline timescales above each 10\% flux level relative to peak. By the measurement definition (Figure~\ref{fig:KC}), the timescales become shorter at higher fractional flux levels. At the 50\% flux level, the median timescales on the rise and decline are $\widetilde{t}_{r50,{\rm rise}}=25.7^{+23.2}_{-12.6}$ and $\widetilde{t}_{r50,{\rm decl}}=55.1^{+47.7}_{-29.1}$ days, respectively ($P_{15.9}$--$P_{84.1}$ ranges). The $P_{2.3}$--$P_{97.7}$ and $P_{0.2}$--$P_{98.8}$ ranges of $t_{r50,{\rm arise}}$ ($t_{r50,{\rm decl}}$) are $7-89$ and $5-234$ days ($13-210$ and $9-322$ days), respectively. At the 50\% flux level, the median total timescale (rise plus decline) is $t_{r50,{\rm rise}}+t_{r50,{\rm decl}}=81.9^{+66.8}_{-41.9}$ days, with the $P_{2.3}$--$P_{97.7}$ and $P_{0.2}$--$P_{98.8}$ ranges of $23-286$ and $15-399$ days, respectively. The lower timescale ranges define a clear unoccupied region in normalized light curves (right panels of Figure~\ref{fig:bandLC}).

In the right panels of Figure~\ref{fig:rT}, we plot histograms of the rise and decline timescale {\it differences} between successive 10\% flux levels. Given the fixed fractional change at each flux level, the differences are effectively the inverted light-curve slopes, and their evolution therefore tracks the light-curve slope evolution from well below to near peak. The overall trend is that the difference increases on the rise to the peak, from $\approx 2.7$ days per 10\% fraction up to the 60\% flux level, and $\approx 3.0-4.4$ days per 10\% fraction in the next 30\% toward the peak. This trend indicates that the light-curve slope becomes shallower closer to the peak. On the decline, the opposite effect is observed as the differences increase, from $\approx 8.5$ days per 10\% fraction down to the 60\% flux level, and $\approx 9.4-18.9$ days per 10\% fraction in the next 30\%.

In Figure~\ref{fig:rT205080}, we plot the $r$-band decline versus rise timescales for the 20\%, 50\%, and 80\% flux levels. The rise and decline timescales are linearly correlated. %, as expected for diffusion-dominated light curves. 
The approximate correlations are $t_{r20, {\rm decl}} \approx 3\times t_{r20, {\rm rise}}$, $t_{r50, {\rm decl}} \approx 2\times t_{r50, {\rm rise}}$, and $t_{r80, {\rm decl}} \approx 1.5\times t_{r80, {\rm rise}}$. This demonstrates that the overall light-curve shape becomes more symmetric around the peak. 
For completeness, we also show the timescale evolution in bolometric and optical luminosity in Figures~\ref{fig:optT}, \ref{fig:bolT}, \ref{fig:optT205080}, and \ref{fig:bolT205080}, where the same trends and similar correlations as in the $r$ band can be seen.

Finally, in Figure~\ref{fig:bandT50}, we plot the rise and decline timescales above the 50\% flux level in all filters. The general trend is that the rise times are comparable in all filters ($\approx 23-26$ days), albeit with slightly faster timescales in the bluer filters.  The decline timescales exhibit a more significant filter dependence, with medians of $45.1^{+37.9}_{-22.7}$ days in the $g$ band to $53.7^{+39.2}_{-30.4}$ days in the $i$ band. This overall trend tracks the declining photospheric temperature near and after the peak.  However, the median decline timescales in the $r$ and $o$ bands ($55.1^{+47.7}_{-29.1}$ and $55.2^{+45.4}_{-29.5}$ days, respectively) are systematically longer than in the $i$ band, which is likely due to the presence of the strong H$\alpha$ emission line in these bands;\footnote{\textit{Gaia} $G$ also covers H$\alpha$; however, the relative contribution from the line emission to the total band flux is lower, given its broader wavelength coverage.} this will be discussed in detail in a follow-up paper in this series focused on spectroscopy.

\subsection{Peak Brightness-Timescale Correlations} 
\label{sec:peaktime}

\begin{figure*}
    \centering
    \includegraphics[width=0.49\textwidth]{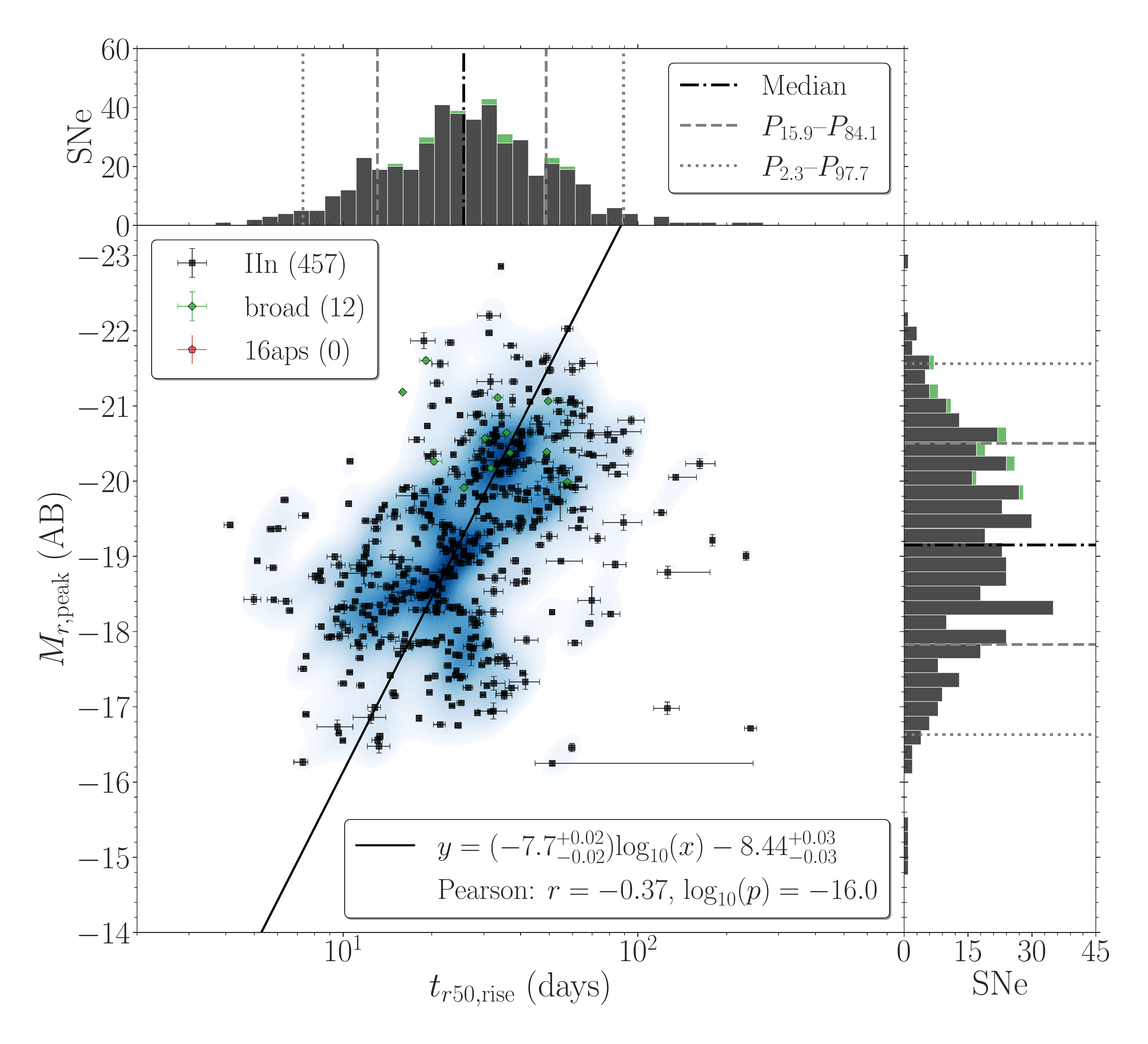}
    \includegraphics[width=0.49\textwidth]{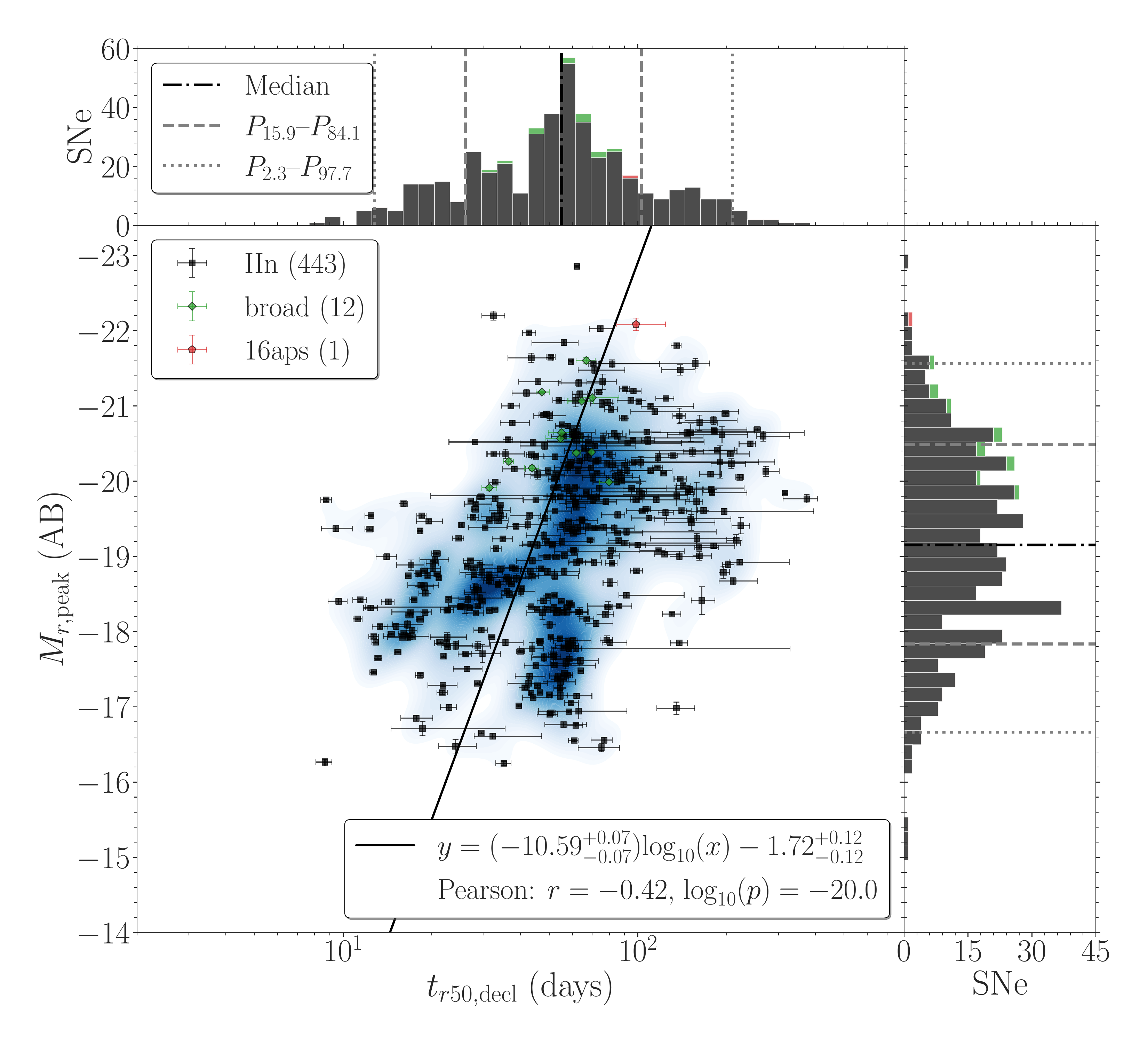}
\caption{Distributions of the peak $r$-band magnitude and 50\% rise time ($t_{r50,{\rm rise}}$; \textit{left}) and decline time ($t_{r50,{\rm decl}}$; \textit{right}). The error bars denote $1\sigma$ uncertainties. In each panel, we show the correlation between peak magnitude and logarithm of the timescale (black line) and provide the best fit and Pearson correlation test values (coefficient $r$ and log p-value). The blue shaded regions represent the density maps, indicating multiple clusters along the overall correlation.  
}
    \label{fig:MrT50}
\end{figure*}

\begin{figure*}
    \centering
    \includegraphics[width=0.49\textwidth]{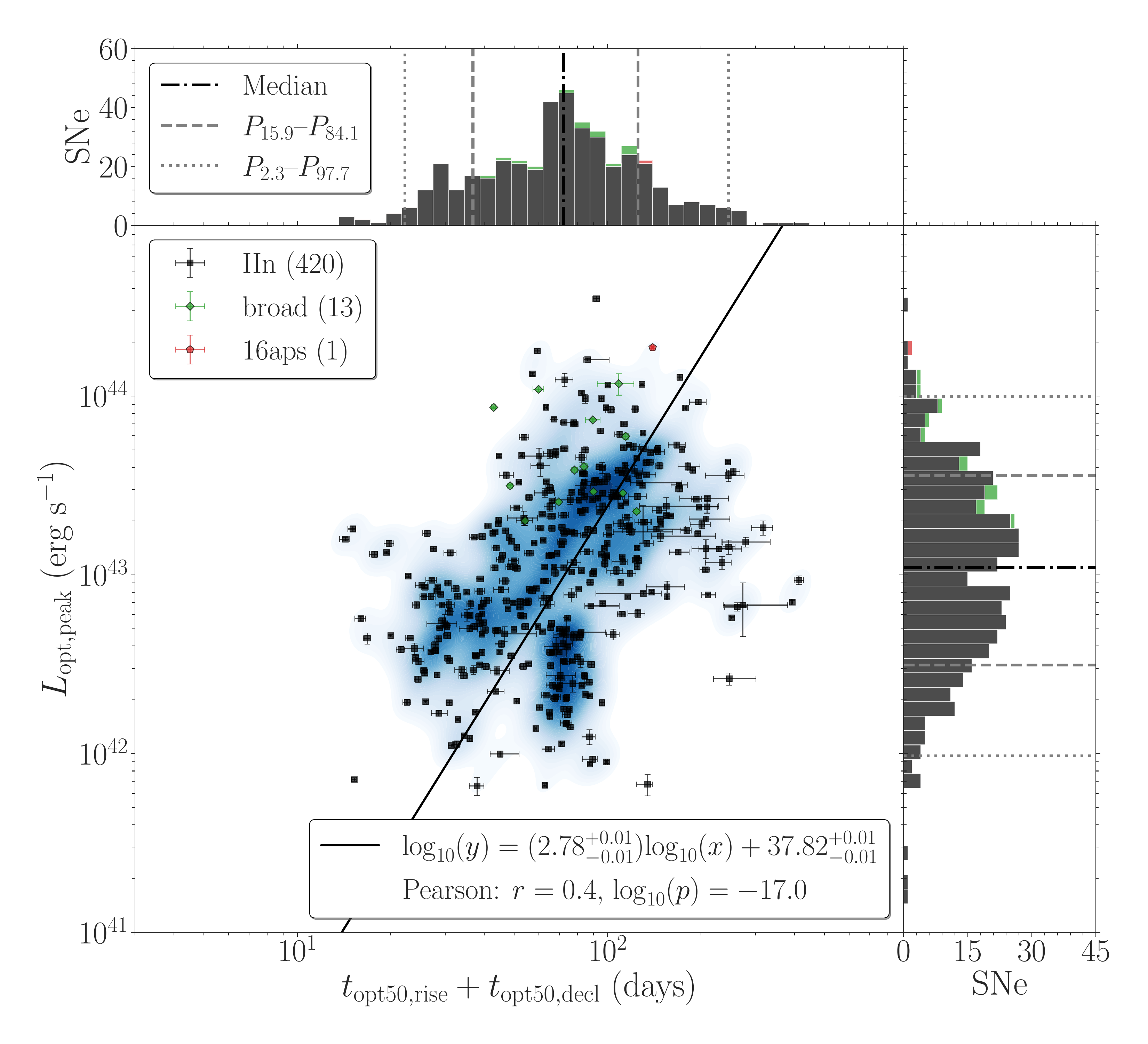}
    \includegraphics[width=0.49\textwidth]{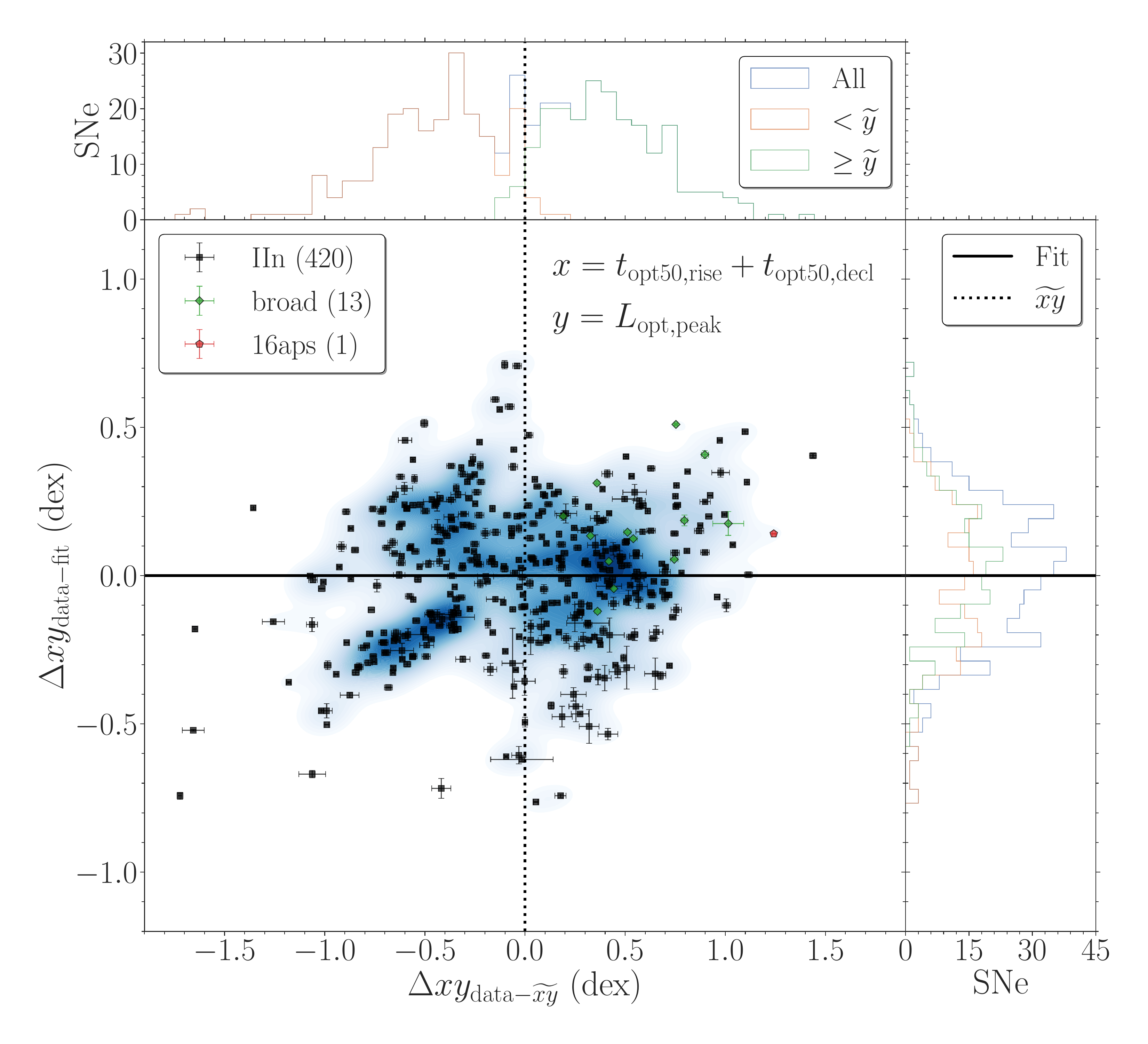}
\caption{\textit{Left}: similar to Figure~\ref{fig:MrT50}, but for the peak optical luminosity and 50\% total timescale ($t_{{\rm opt}50,{\rm rise}}+t_{{\rm opt}50,{\rm decl}}$). \textit{Right}: rotated distribution along the best-fit correlation, centered around the median luminosity ($y$) and timescale ($x$). The sample is clearly bimodal below and above the median luminosity (upper histogram), with two potential branches below the median luminosity.
}
    \label{fig:LTrot}
\end{figure*}

Using the peak brightness (\S\ref{sec:z}) and timescale (\S\ref{sec:time}) measurements, we explore their correlations in the $r$ band in Figure~\ref{fig:MrT50}. We find a clear correlation between peak brightness and both rise and decline timescales, with Pearson $r$ coefficients\footnote{The negative values of the coefficients $r$ are due to the magnitude system.} of $\approx -0.4$, and log-probability (i.e., chance correlation) values of $\approx -16$ (rise time) and $\approx -20$ (decline time).  The correlation is linear between peak absolute magnitude ($M_{r,{\rm peak}}$) and log timescale ($t_{r50,{\rm rise}}$ and $t_{r50,{\rm decl}}$). 
We also show in Figure~\ref{fig:MrT50} the 2D density maps for the sample, which reveal under-densities in the distribution around the medians of both $M_{r,{\rm peak}}$ and the timescales ($t_{r50,{\rm rise}}$, $t_{r50,{\rm decl}}$). Performing a clustering test (Hierarchical Density-Based Spatial Clustering of Applications with Noise\footnote{Via \texttt{HDBSCAN} implemented in \texttt{scikit-learn}.}), we find two robust clusters roughly separated by the medians of $M_{r,{\rm peak}}$ and $t_{r50,{\rm rise}}$ or $t_{r50,{\rm decl}}$; there are also two potential branches\footnote{Sensitive to the choice of minimum sample size of a group for that group to be considered as a cluster in \texttt{HDBSCAN}.} on the below-median ends of $M_{r,{\rm peak}}$ and $t_{r50,{\rm decl}}$ (the two ``leg-like'' structures in the right panel of Figure~\ref{fig:MrT50}).

In the left panel of Figure~~\ref{fig:LTrot}, we plot the peak optical luminosity ($L_{\rm opt,peak}$) versus the total timescale above the 50\% flux level ($t_{{\rm opt}50,{\rm rise}}+t_{{\rm opt}50,{\rm decl}}$). As in the $r$ band, we find a strong correlation and a power-law fit with the best-fit index of $\approx2.78$. %linear fit. 
The clustering analysis again gives rise to the same two robust clusters separated roughly by the medians of both quantities, with two potential branches on the below-median cluster.  

In the right panel of Figure~\ref{fig:LTrot}, we show the same data, but with the axes rotated along the best-fit correlation, centered around the median $\widetilde{L}_{\rm opt,peak}$ and total timescale. The projected histogram along the best-fit correlation shows a clear bimodality below and above the median $\widetilde{L}_{\rm opt,peak}$. The two potential branches below the median $\widetilde{L}_{\rm opt,peak}$ can be seen in the projected histogram, bifurcated by the mean correlation trend (i.e., horizontal line).

Both in the filter and bolometric light curves, the strong correlations observed between the peak brightness and rise timescale are responsible for the mostly continuous distribution on the rise of light curves (\S\ref{sec:lc}; Figures~\ref{fig:bandLC} and \ref{fig:bolLC}). On the decline, however, the more pronounced under-densities around the medians along such correlations are responsible for the clear bifurcation of the light-curve morphology: luminous--slow and faint--fast. Within the faint--fast group, one of the two potential branches with longer timescales (i.e., the right leg in Figure~\ref{fig:MrT50} right panel and Figure~\ref{fig:LTrot} left panel) represents the few transitional events between the two morphological groups (see \S\ref{sec:temp}).

\subsection{Radiated Energy} 
\label{sec:energy}

\begin{figure*}
    \centering
    \includegraphics[width=0.98\textwidth]{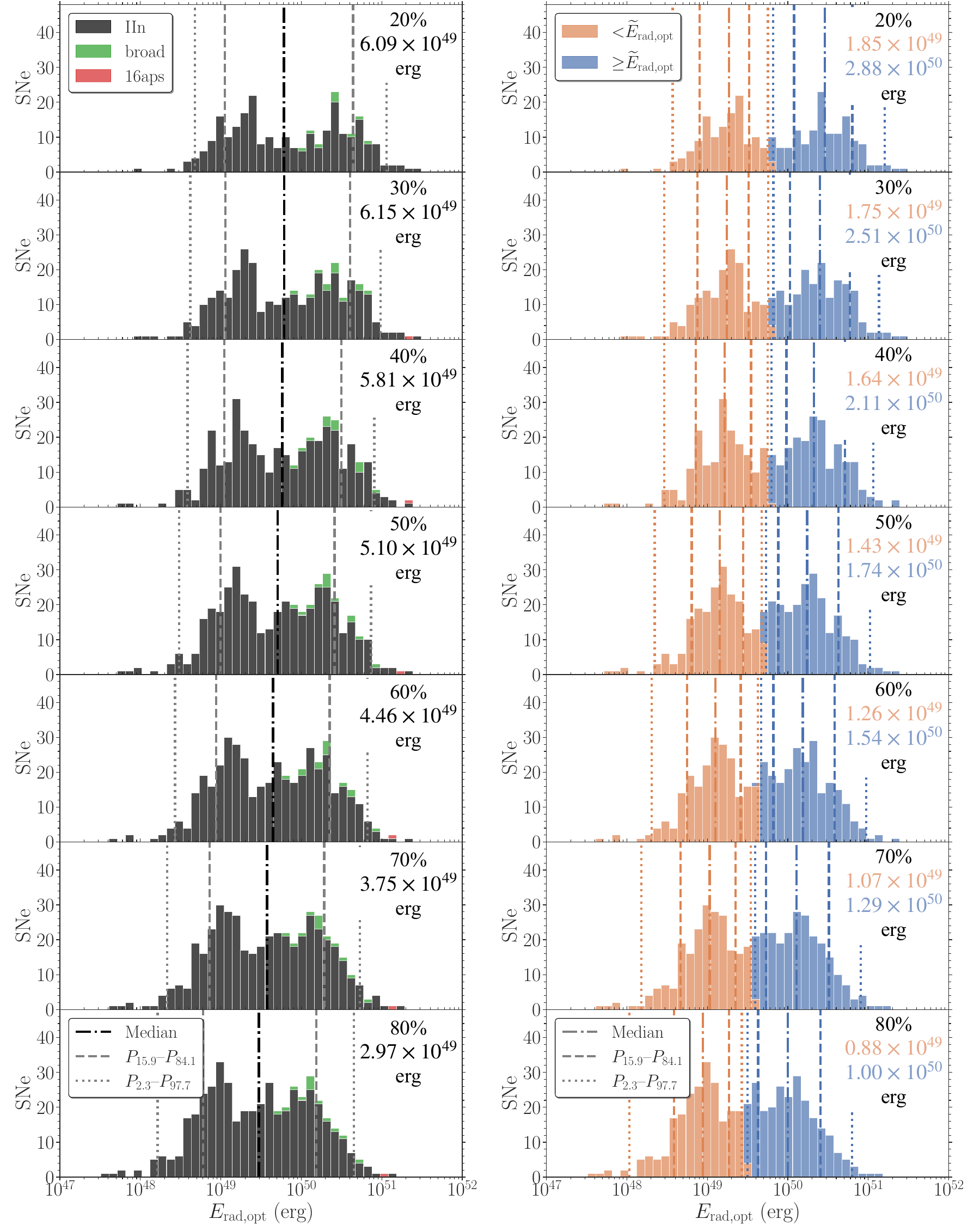}
\caption{\textit{Left}: histograms of the optical radiated energy ($E_{\rm rad, opt}$) above each 10\% flux level. \textit{Right}: same histograms, but separated by the median $\widetilde{E}_{\rm rad, opt}$.  The vertical lines in each panel mark the median (with its value on the top left) and $P_{15.9}$--$P_{84.1}$ and $P_{2.3}$--$P_{97.7}$ ranges. There is a strong bimodality in the distribution, stemming from the correlation between the luminosity and timescale and their bimodal distribution around the median (Figure~\ref{fig:LTrot}). 
}
    \label{fig:Eradopt}
\end{figure*}

\begin{figure}
    \centering
    \includegraphics[width=0.49\textwidth]{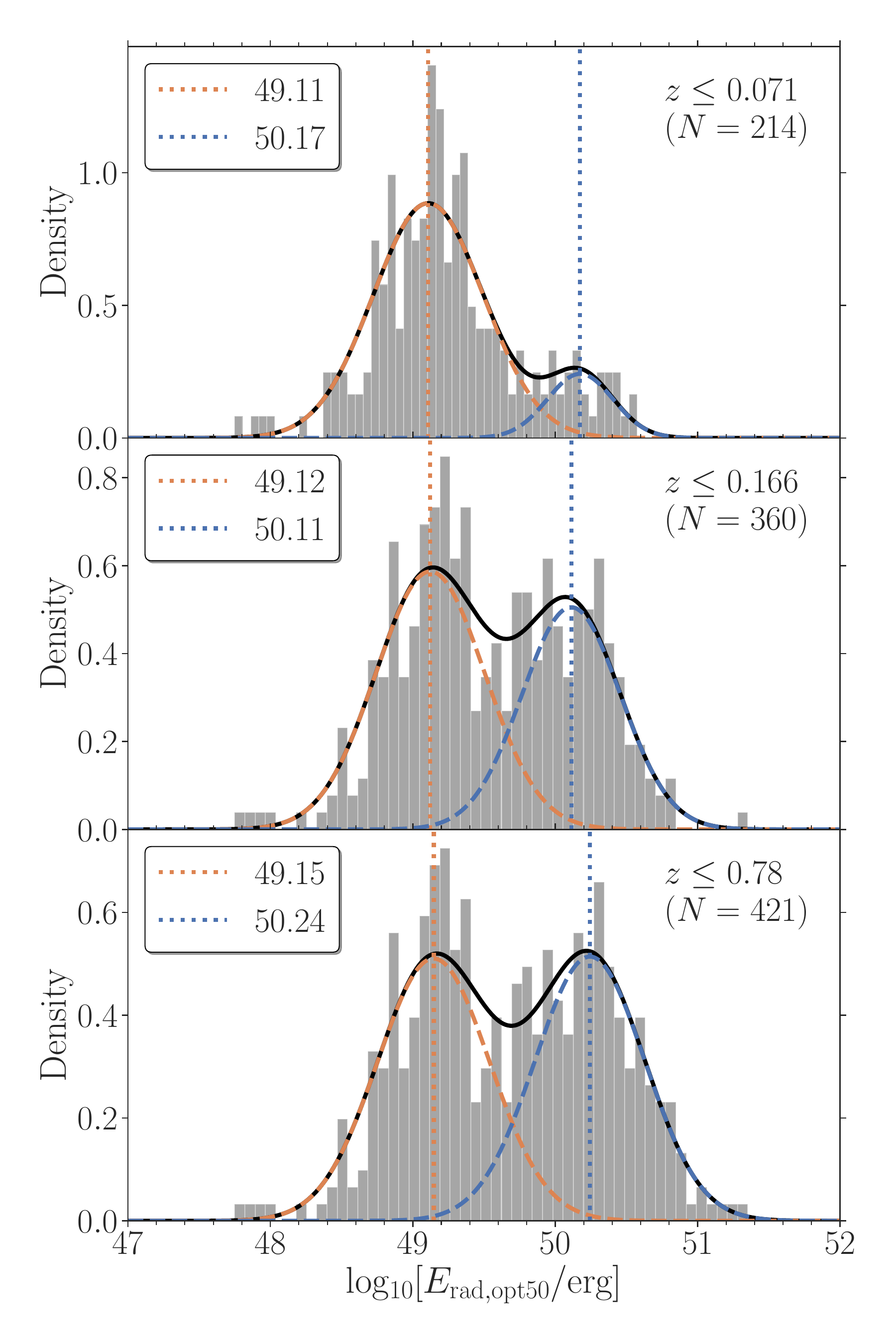}
\caption{Histograms of the optical radiated energy above the 50\% flux level ($E_{\rm rad, opt50}$) with and without redshift cuts at the median (0.071) and 84.1th percentile (0.166), representing more complete volume-limited subsamples (against Malmquist bias) close to the 97.7th percentile and median, respectively, of the peak $r$-band absolute magnitude (Figure~\ref{fig:Mr_z}). Two independent Gaussian distributions are preferred with the lowest BIC from GMM, with a similar mean value of each distribution for the samples with and without the redshift cuts.
} 
    \label{fig:Eradopt_BIC}
\end{figure}

\begin{figure}
    \centering
    \includegraphics[width=0.49\textwidth]{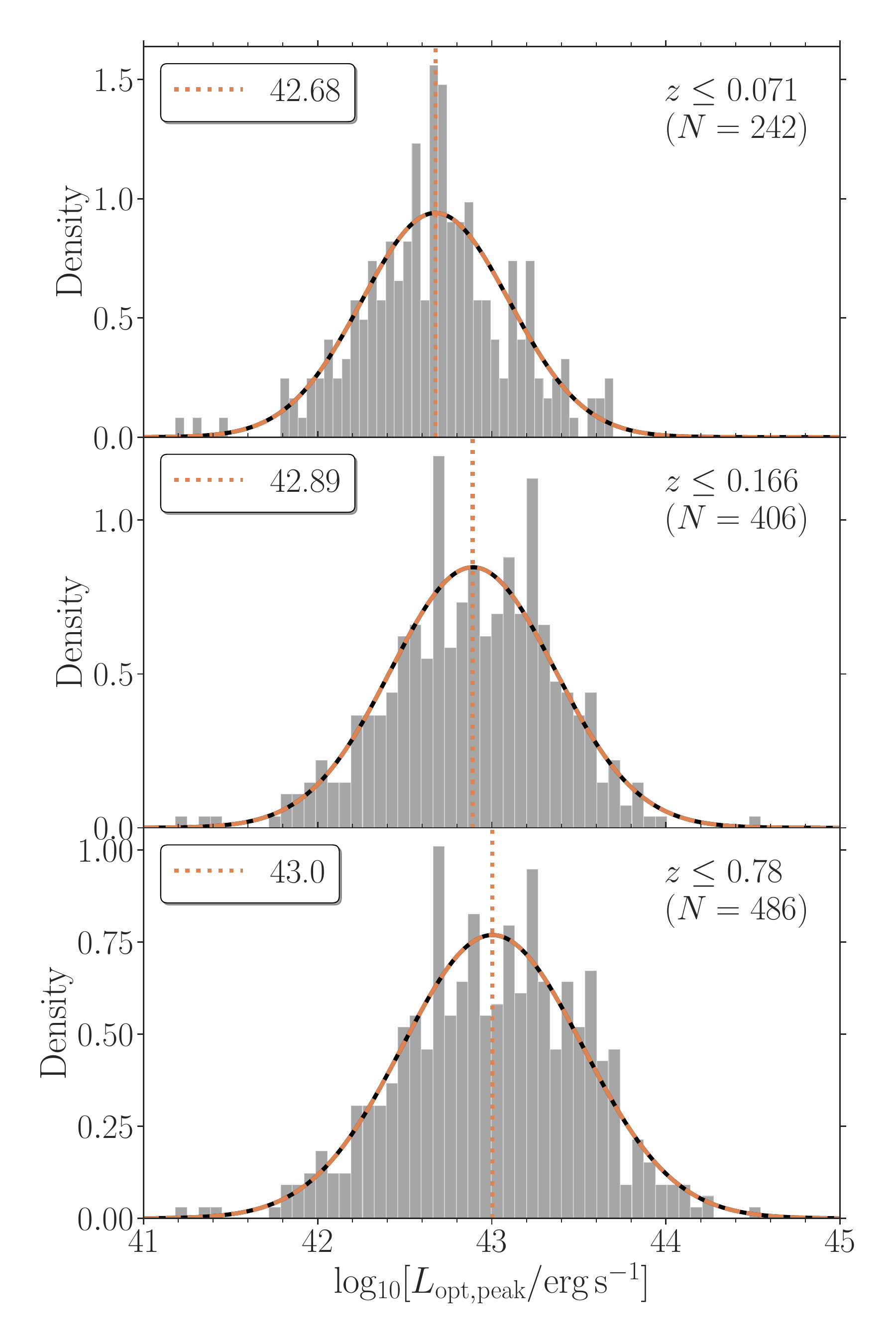}
\caption{Similar to Figure~\ref{fig:Eradopt_BIC}, but for $L_{\rm opt,peak}$, $t_{{\rm opt}50,{\rm rise}}$. Unlike $E_{\rm rad, opt50}$, no significant bimodality is found with GMM.
}
    \label{fig:BIC_Lp}
\end{figure}

\begin{figure}
    \centering
    \includegraphics[width=0.49\textwidth]{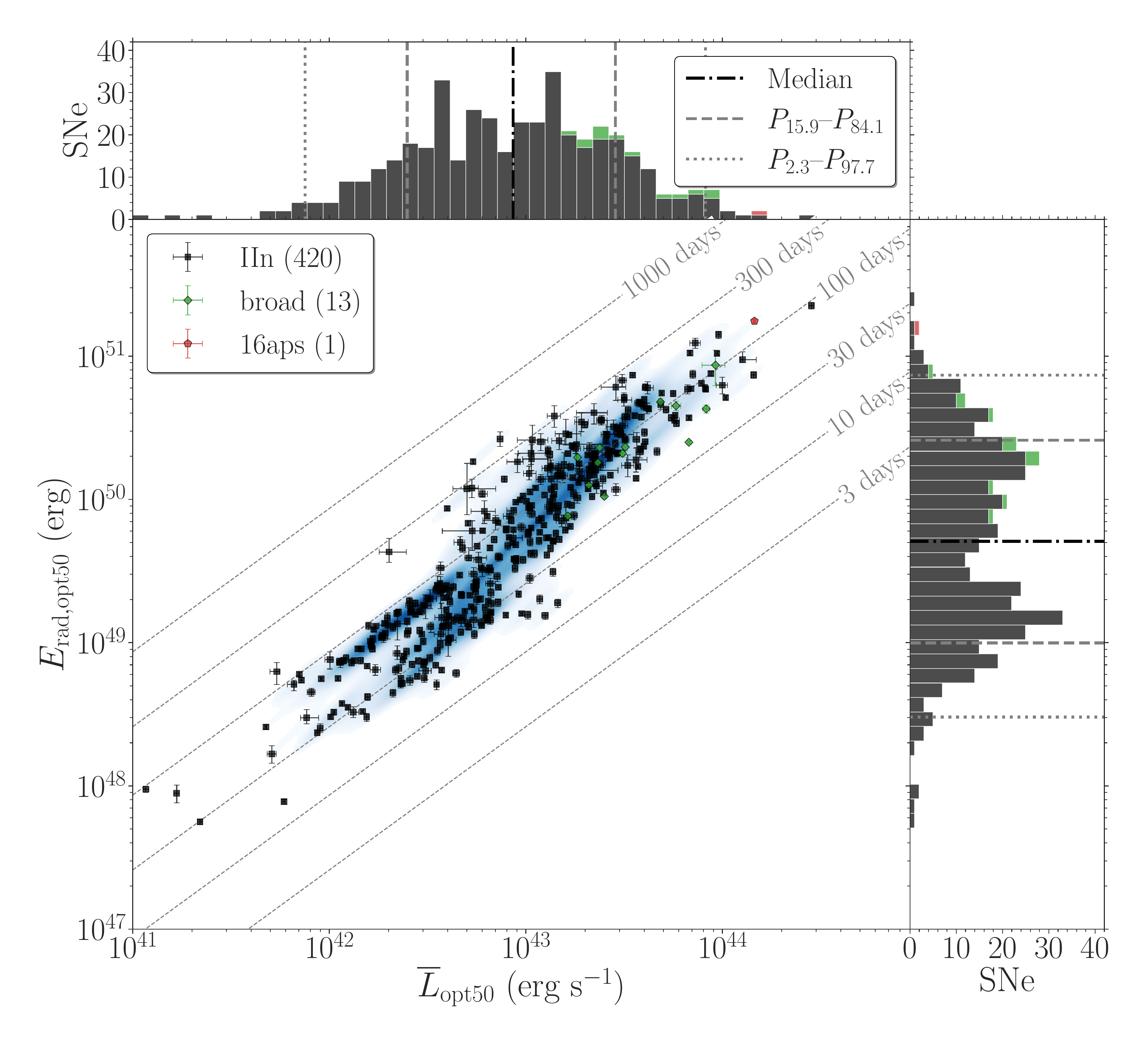}
\caption{Optical radiated energy ($E_{{\rm rad,opt}50}$) vs. mean luminosity ($\overline{L}_{{\rm opt}50}$) above the 50\% flux level. The error bars denote $1\sigma$ uncertainties, while the blue shaded regions represent the density maps. The diagonal dashed lines mark characteristic timescales, i.e., $E_{{\rm rad,opt}50}/\overline{L}_{{\rm opt}50}$. The events above the median radiated energy have a characteristic timescale of $\approx 100$ days (the luminous--slow group), while those on the faint end exhibit two potential branches with timescales of $\approx 40$ and $\approx 70$ days (faint--fast group).}
    \label{fig:Eradopt_mL50}
\end{figure}

We determine the radiated energy by integrating the bolometric and optical light curves over the total duration (rise plus decline) at the various fractional flux levels. In the left panels of Figure~\ref{fig:Eradopt}, we show the histograms of the optical radiated energy ($E_{\rm rad, opt}$) above each 10\% flux level. At the 50\% flux level, the median $\widetilde{E}_{{\rm rad, opt}50}$ is $(5.10^{+20.8}_{-4.10})\times10^{49}$\,erg ($P_{15.9}$--$P_{84.1}$ range), with the $P_{2.3}$--$P_{97.7}$ range of $(0.03-7.37)\times10^{50}$\,erg, a factor of $\approx250$. However, the $E_{\rm rad, opt}$ distributions are clearly bimodal around the median. 

To highlight this bimodality, in the right panels of Figure~\ref{fig:Eradopt}, we show the same histograms, but with the sample divided into two at the median $\widetilde{E}_{\rm rad, opt}$, where the median in each subsample roughly corresponds to the peak of its distribution. At the 50\% flux level, the low and high energy medians (i.e., peaks) are $(1.43^{+1.35}_{-0.79})\times10^{49}$ and $(1.74^{+2.52}_{-0.98})\times10^{50}$ erg, respectively ($P_{15.9}$--$P_{84.1}$ ranges). This is due to the correlation between the peak luminosity and timescale, as well as the bimodal distribution along the correlation (Figure~\ref{fig:LTrot}). The bimodality becomes more evident at the lower flux levels, where the two light-curve morphological groups become even more distinct.

To quantify the robustness of the bimodality, we apply a Gaussian mixture model (GMM) composed of $k$-independent Gaussian distributions (each with its own mean and variance) and find the best combination of $k$ Gaussians with the lowest Bayes information criterion (BIC) using \texttt{GaussianMixture} from \texttt{scikit-learn} \citep{Pedregosa2011JMLR...12.2825P}. Following \cite{Nyholm2020}, we consider $k=1,2,3$ and the model with $\Delta {\rm BIC}\geq6$ from the surrounding models to be significant. To compensate for the sample incompleteness due to Malmquist bias (\S\ref{sec:z}), we also apply redshift cuts at the median ($0.071$) and 84.1th percentile ($0.166$), yielding more complete volume-limited subsamples down to $-16.8$ and $-18.8$ mag at the   ZTF limiting magnitude, respectively, close to the 97.7th percentile ($-16.6$ mag) and median ($-19.2$ mag) of the peak $r$-band absolute magnitude (Figure~\ref{fig:Mr_z}). 
In Figure~\ref{fig:Eradopt_BIC}, we show the results of GMM for $E_{\rm rad, opt50}$ where $k=2$ is preferred with $\Delta {\rm BIC}\geq6$ over $k=1$ and $3$, regardless of the redshift cuts. The mean values of the two independent Gaussians are similar across the redshift cuts: log$_{10}(E_{\rm rad,opt50}/{\rm erg})\approx49.1$ and $50.2$, reinforcing the robustness of the bimodality to selection effects. 
In contrast, we find no significant bimodality when applying GMM for $L_{\rm opt,peak}$ with the same redshift cuts, as shown in Figure~\ref{fig:BIC_Lp}. This clearly demonstrates that the bimodality in the radiated energy distribution is not caused by peak luminosity.

In addition, we calculate the mean bolometric and optical luminosities ($\overline{L}=E_{\rm rad}/(t_{\rm rise}+t_{\rm decl})$) above each 10\% flux level. In Figure~\ref{fig:Eradopt_mL50}, we plot the optical $E_{{\rm rad,opt}50}$ versus $\overline{L}_{{\rm opt}50}$ at the 50\% flux level, along with characteristic timescales (i.e., $E_{{\rm rad,opt}50}/\overline{L}_{{\rm opt}50}$). The events at the luminous part of the distribution represent the luminous--slow group with a characteristic timescale of $\approx100$ days, while those on the faint end represent the faint--fast group with the two potential branches, with timescales of $\approx40$ and $\approx70$ days. This summarizes all the prior analysis on the peak brightness (\S\ref{sec:z}), timescale (\S\ref{sec:time}), and their correlations (\S\ref{sec:peaktime}), resulting in the clear bifurcation of the light-curve morphology (\S\ref{sec:lc}). 

Finally, we note that $E_{\rm rad, opt}$ represents a lower bound on the bolometric quantity $E_{\rm rad, bol}$, where a rough transformation can be obtained by the overall optical-bolometric correction of $\sim1.9$ (Figure~\ref{fig:bolP}), given their comparable timescales (Figures~\ref{fig:optT} and \ref{fig:bolT}). For completeness, we also show the histograms of $E_{\rm rad, bol}$ in Figure~\ref{fig:Eradbol}, where the same bimodality as in $E_{\rm rad, opt}$ can be seen.

\subsection{Light-curve Templates} 
\label{sec:temp}

\begin{figure}
    \centering
    \includegraphics[width=0.49\textwidth]{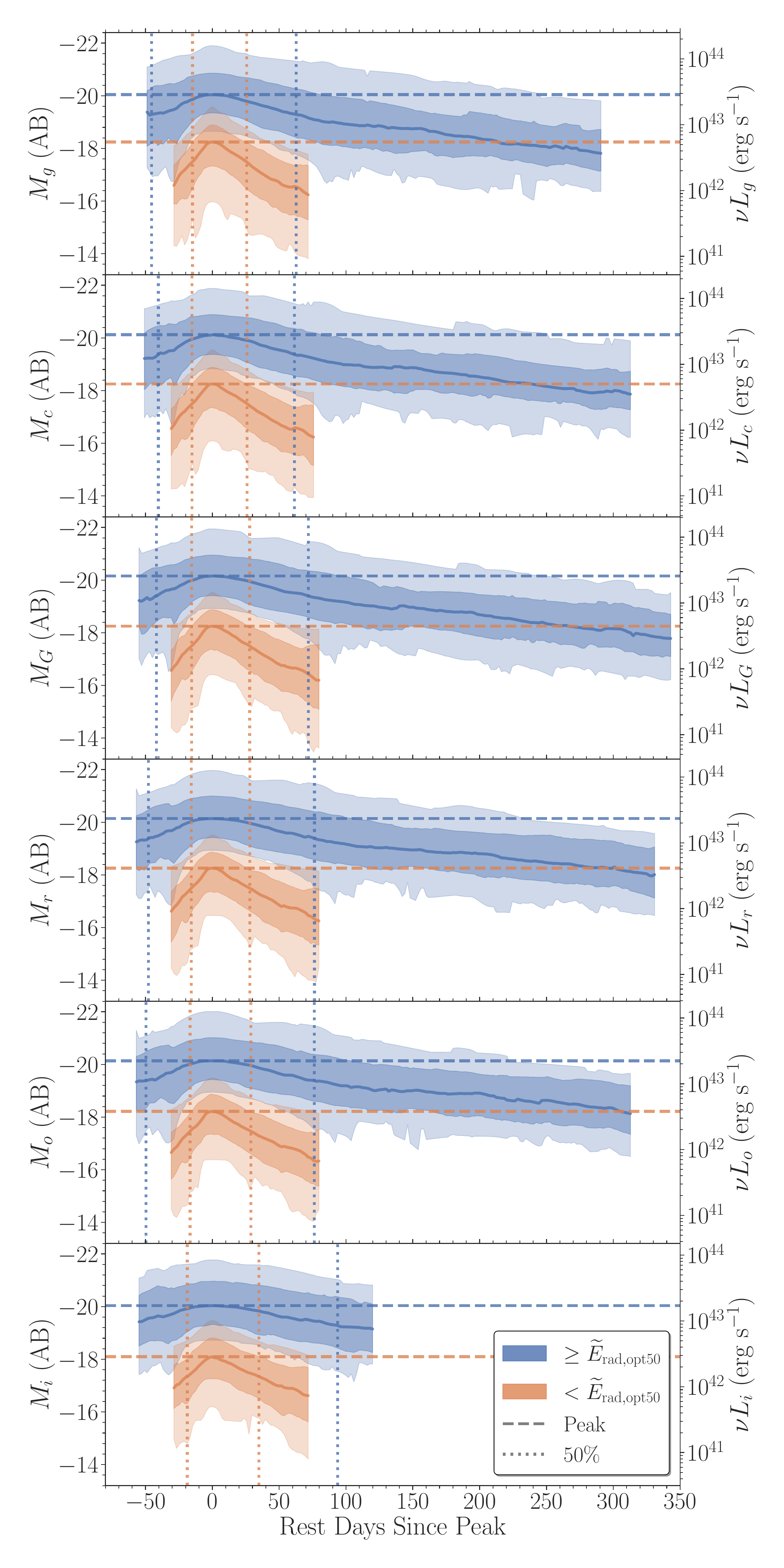}
\caption{Median light curves in each filter separated by the median optical radiated energy, $\widetilde{E}_{{\rm rad,opt}50}$ ($g$, $c$, $G$, $r$, $o$, and $i$ from \textit{top} to \textit{bottom}). The peak and 50\% measurements are marked by the horizontal dashed and vertical dotted lines, respectively. The shaded regions with different transparency represent $P_{15.9}$--$P_{84.1}$ and $P_{2.3}$--$P_{97.7}$ ranges. The two main groups of SNe~IIn (faint--fast and luminous--slow) are clear, with their templates overlapping only at the $P_{2.3}$--$P_{97.7}$ level. (The data used to create this figure will be available in machine-readable format upon publication.)
}
    \label{fig:LCtemp}
\end{figure}

\begin{figure}
    \centering
    \includegraphics[width=0.49\textwidth]{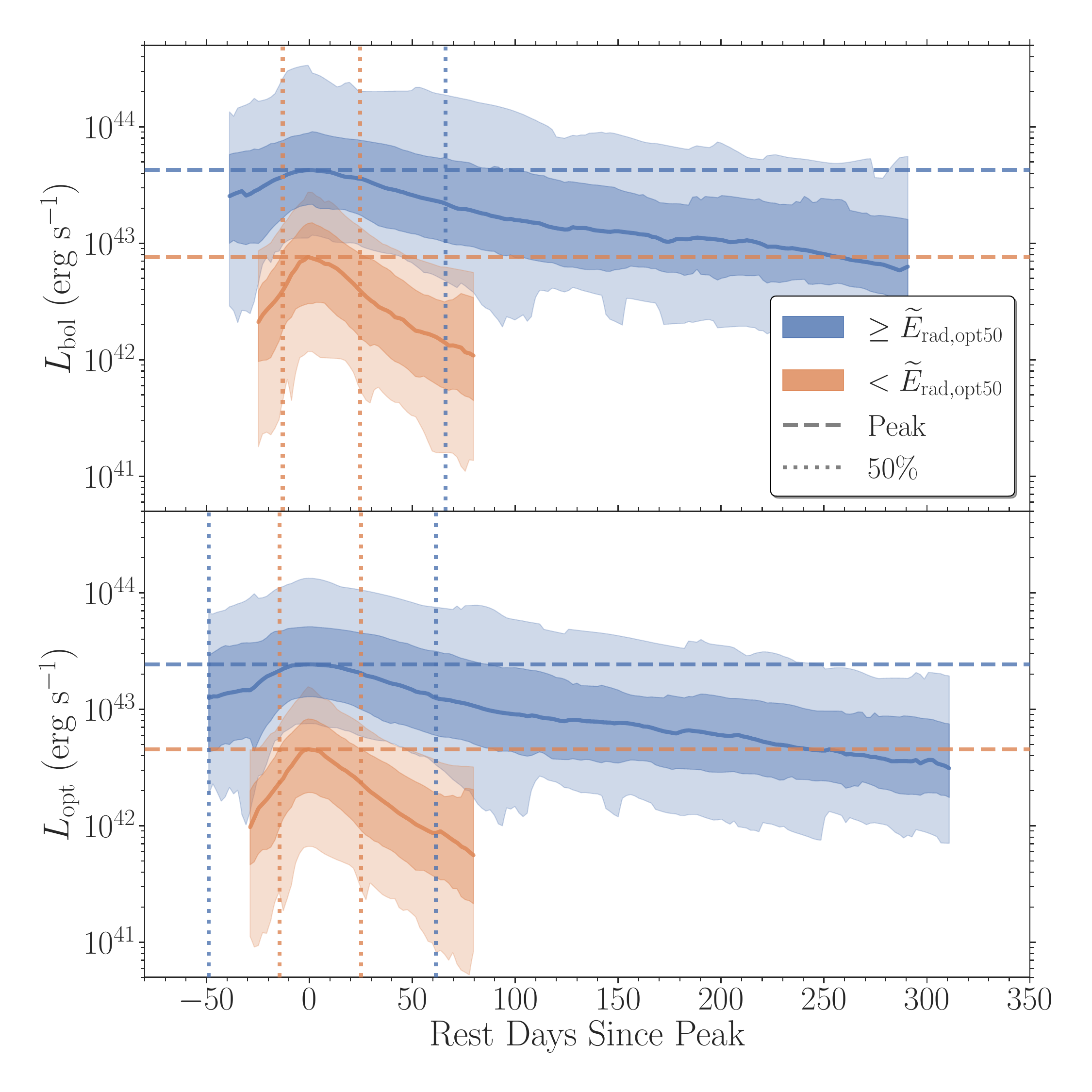}
\caption{Similar to Figure~\ref{fig:LCtemp}, but for the bolometric (\textit{top}) and optical (\textit{bottom}) luminosity. (The data used to create this figure will be available in machine-readable format upon publication.)
}
    \label{fig:Lboltemp}
\end{figure}

Motivated by the observed bifurcation of the light-curve morphology as well as the bimodality in the radiated energy distribution, we construct a median light curve of each morphological group, i.e., a light-curve template.  First, we divide the full sample into two at the median optical radiated energy, $\widetilde{E}_{{\rm rad,opt}50}$, which marks a clear dividing point in the bimodal distribution (Figure~\ref{fig:Eradopt}). For each subsample in each filter, we then combine all the GP light curves and calculate the median flux and its variance in 2 day bins if more than 100 SNe with significant flux measurements ($F/F_{\rm err}\geq3$) are available in each bin. %2000  
We perform the same procedure in the bolometric and optical light curves, as well. We also construct normalized templates at peak by performing the same procedure on the normalized band, bolometric, and optical light curves.

In Figures~\ref{fig:LCtemp} and \ref{fig:Lboltemp}, we plot the resulting light-curve templates in each filter, and in bolometric and optical luminosity, respectively. As expected, we find a clear separation into the faint--fast and luminous--slow groups, with the templates overlapping only at the $P_{2.3}$--$P_{97.7}$ level. 
The faint--fast and luminous--slow templates have peak absolute magnitudes of $\approx-18.2$ and $\approx-20.1$ mag across the filters, respectively, while they have peak luminosities of $\approx4.5\times10^{42}$ and $\approx2.4\times10^{43}$\,erg\,s$^{-1}$ in optical and $\approx7.6\times10^{42}$ and $\approx4.3\times10^{43}$\,erg\,s$^{-1}$ in bolometric. Their $r$-band 50\% rise and decline timescales are $\approx16$ and $\approx28$ days and $\approx48$ and $\approx76$ days, respectively, while those in optical luminosity are $\approx15$ and $\approx26$ days and $\approx48$ and $\approx62$ days, respectively. 
The optical total timescale of $\approx40$ days for the faint--fast template means that it is dominated by the below-median branch with shorter timescales (i.e., the left leg in Figure~\ref{fig:LTrot} left panel), rather than the one with longer timescales (i.e., more transitional events; see \S\ref{sec:peaktime}). 
As also noted in \S\ref{sec:lc}, the broad-lined SLSNe~II are photometrically identical to the luminous--slow template, despite their spectral differences.

\section{Summary and Discussion} 
\label{sec:sum}

Before discussing the implications of our findings, below we summarize the key observed properties of the SN~IIn sample (\S\ref{sec:ana}): 

\begin{itemize}
  \item Median peak absolute magnitudes of $\approx-19.1$ mag, with $P_{15.9}$--$P_{84.1}$ and $P_{2.3}$--$P_{97.7}$ spreads of $\approx2.6$ and $\approx5.0$ mag, respectively, in the $g$, $c$, $G$, $r$, $o$, and $i$ bands. %$\widetilde{M}_{\rm peak}$
  \item Median peak bolometric and optical luminosities of $\approx2.0\times10^{43}$ and $\approx1.1\times10^{43}$\,erg\,s$^{-1}$, respectively, with $P_{15.9}$--$P_{84.1}$ and $P_{2.3}$--$P_{97.7}$ spreads of about 1 and 2 orders of magnitude, respectively. %$\widetilde{L}_{\rm peak}$
  \item A bifurcation into two clear light-curve morphological groups: faint--fast and luminous--slow in both the single-band and bolometric/optical luminosities, separated around the median peak brightness.
  \item A median $r$-band rise (decline) timescale at the 50\% flux level of $\approx 26$ ($\approx 55$) days, with $P_{15.9}$--$P_{84.1}$ and $P_{2.3}$--$P_{97.7}$ ranges of $\approx13-49$ and $\approx7-89$ days ($\approx26-103$ and $\approx13-210$ days), respectively. %$t_{r50,{\rm rise}}$ ($t_{r50,{\rm decl}}$)
  \item Overall light-curve slope evolution from a steeper rise to shallower decline, with a more symmetric light-curve shape around the peak.
  \item Strong correlations between the peak absolute magnitude and both the rise and decline timescales, with multiple clusters along the overall correlations.
  \item A strong correlation between the peak optical luminosity and total (rise plus decline) timescale, with a clear bimodality around the medians and two potential branches below the medians.
  \item A median optical radiated energy of $5.1\times10^{49}$ erg at the 50\% flux level, with a strong bimodal distribution around the median, peaking at $\approx 1.4\times10^{49}$ and $\approx 1.7\times10^{50}$ erg.
  \item A characteristic timescale of $\approx100$ days for the events above the median luminosity, and two potential branches with characteristic timescales of $\approx40$ and $\approx70$ days for those below the median luminosity.
  \item A clear separation of light curves into the luminous--slow and faint--fast groups by the median optical radiated energy, with their templates overlapping only at the $P_{2.3}$--$P_{97.7}$ level.
\end{itemize}

\subsection{Comparison to Previous Studies}
\label{sec:comp}

The overall light-curve evolution (i.e., timescale and slope; Figure~\ref{fig:rT}) of the SN IIn sample is consistent with expectations for transients dominated by photon diffusion, in this case through the CSM (e.g., \citealt{Chevalier1982ApJ...258..790C,Moriya2013MNRAS.435.1520M,Khatami2024ApJ...972..140K}). More quantitatively, the decline versus rise time correlation we have uncovered, $t_{r50,{\rm decl}} \approx 2 \times t_{r50,{\rm rise}}$ (Figure~\ref{fig:rT205080}) matches well with the findings of 2D numerical simulations of CSM interaction (\citealt{Suzuki2019ApJ...887..249S}; see also Figures~\ref{fig:optT205080} and \ref{fig:bolT205080} for the similar correlations in the optical and bolometric light curves). Interestingly, in the numerical simulations, the correlation shows a weak dependence on the CSM geometry  (i.e., spherical versus disk). While these population properties are theoretically anticipated, they had not been previously characterized systematically in the observed sample.

\cite{Nyholm2020} studied a sample of
42 SNe~IIn from PTF with $-16.5 \gtrsim M_{\rm peak} \gtrsim -21.5$ (Table~\ref{tab:tot}), and found no significant correlation between $M_{\rm peak}$ and rise time (fitting the rise with $F(t) \propto t^{2}$), and only a moderate correlation between $M_{\rm peak}$ and decline rate (measured from the peak to +50 days). The lack of obvious correlations is most likely due to the small PTF sample size, as we recover the same correlations as for the full sample (in Figure~\ref{fig:MrT50}) even if we restrict the $M_{\rm peak}$ range as the PTF sample (leading to 424 SNe~IIn in our study, an order of magnitude increase). \cite{Nyholm2020} also hint at a bimodal distributions of the rise time and decline rate. While there is no clear bimodality in the rise timescale distribution in our larger sample, we identify a clear bimodality in the decline timescale/rate distributions (Figure~\ref{fig:rT}). However, we reiterate that the bimodality is two-dimensional along the correlations with peak brightness and timescale (Figure~\ref{fig:LTrot}).

\cite{Pessi2025A&A...695A.142P} recently carried out a sample study of 107 ``SLSNe-II'' from ZTF, defined as SNe~IIn with a cutoff of $M_{\rm peak}\leq -19.9$ mag. They found no significant correlations between $M_{\rm peak}$ and both fractional rise and decline timescales.  From our larger sample, covering the full range of SN IIn peak magnitudes, we conclude that their result is driven by the adopted arbitrary threshold for SLSNe-II. Specifically, using the same threshold of $\leq -19.9$ mag in our sample (131 events), we also miss the correlation.\footnote{With Pearson coefficient $r$ and log-p values of $\approx0.07$ and $\approx-0.4$, respectively, both on the rise and decline timescales.}  Critically, our large sample does not show any obvious transition in properties at $-19.9$ mag, or $L_{\rm opt, peak}\approx3\times10^{43}$\,erg\,s$^{-1}$ (Figure~\ref{fig:BIC_Lp}), and instead such events are merely the upper range in the luminous--slow group, which has a median peak absolute magnitude of $\approx-20.1$ (Figure~\ref{fig:LCtemp}).  Therefore, there is no justification for a separate class of ``SLSN-II'', specifically superluminous SNe~IIn (SLSN-IIn), and indeed this separation masks critical aspects of the SN IIn population. Instead, the classification of SNe IIn should be purely spectroscopic, not photometric, based on the presence of narrow Balmer series emission lines that reflect their emission mechanism, namely, interaction with CSM.

%``SLSN-II'' should be used only for broad-lined SLSNe~II (e.g., \citealt{Kangas2022MNRAS.516.1193K}).}

Finally, we note that \cite{Ransome2025ApJ...987...13R} recently presented semi-analytical model fits to 142 SNe~IIn using \texttt{MOSFiT} \citep{Guillochon2018ApJS..236....6G}. Using the best-fit model light curves, they find correlations between the peak absolute magnitude and the rise and decline times, as well as a bimodal distribution in the decline time. However, as these are model-dependent findings with simplifying assumptions (see e.g., \citealt{Zhang2026ApJ...999..186Z} for the discrepancies in peak luminosity and timescale when neglecting photon diffusion through optically thick CSM), direct comparisons with our work are not straightforward; we defer a more detailed discussion to a follow-up paper in this series focused on semi-analytical modeling.

\subsection{Possible Progenitor Channels}
\label{sec:prog}

SNe~IIn show narrow Balmer series emission lines in their spectra, indicating that the dominant power source is continued interaction between the SN ejecta and CSM, as the shock front propagates through the CSM. In this case, the conversion from kinetic energy to radiation is expected to be efficient, with a typical conversion efficiency of $\epsilon_{\rm rad}\gtrsim0.1$ (e.g., \citealt{Moriya2013MNRAS.435.1520M,Fransson2014ApJ...797..118F,Tsuna2019ApJ...884...87T,Khatami2024ApJ...972..140K}). The efficiency can increase up to order unity as the mass ratio of the CSM to SN ejecta becomes closer to (or even above) order unity. Given the luminosity-timescale correlation driven by diffusion of the radiation through the CSM (Figure~\ref{fig:LTrot}), higher efficiencies are generally expected for the luminous--slow group compared to the faint--fast group (Figure~\ref{fig:Lboltemp}). 

At the 20\% flux level, the median radiated energies of the faint--fast and luminous--slow SNe~IIn are $\sim2\times10^{49}$ and $\sim3\times10^{50}$ erg, respectively\footnote{These should be taken as lower limits, although the contributions from the phases below 20\% flux are marginal.} (Figures~\ref{fig:Eradopt} and \ref{fig:Eradbol}). 
Assuming $\epsilon_{\rm rad}\sim 0.1$ for the faint--fast SNe~IIn, the underlying SNe could be low-energy explosions, with $\sim {\rm few}\times 10^{50}$ erg.  The majority of these might correspond to explosions of low-mass red supergiant and/or super-asymptotic giant branch stars (e.g., \citealt{Kitaura2006A&A...450..345K,Janka2008A&A...485..199J,Smith2013MNRAS.434..102S,Moriya2014A&A...569A..57M,Woosley2015ApJ...810...34W,Muller2019MNRAS.484.3307M}), which would be consistent with the local environments of SNe~IIn that they weakly trace ongoing star formation compared to other core-collapse SN types (e.g., \citealt{Anderson2012MNRAS.424.1372A,Habergham2014MNRAS.441.2230H,Galbany2018ApJ...855..107G}). The potential branch in the low-luminosity group that has the longer timescale of $\sim 70$ days might instead arise from high-mass red supergiants with a denser CSM (e.g., \citealt{Smith2009AJ....137.3558S,Moriya2011MNRAS.415..199M,Burrows2024ApJ...964L..16B}); some of these high-mass explosions might be accompanied by black hole formation, resulting in lower ejecta kinetic energies (e.g., \citealt{Chan2018ApJ...852L..19C}).

%$\sim8-10\,M_\odot$; x
%$\sim17-25\,M_\odot$; 
%$\gtrsim 30\,M_\odot$; 
%$\gtrsim 20\,M_\odot$; 
%$\sim110-140\,\Msun$; 

On the other hand, with higher $\epsilon_{\rm rad}$ expected for the luminous--slow group, the underlying SNe have a higher typical explosion energy of $\sim 10^{51}$ erg or above. Luminous blue variables (e.g., \citealt{Humphreys1994PASP..106.1025H,Davidson1997ARA&A..35....1D,Vink2012ASSL..384..221V,Smith2014ARA&A..52..487S,Smith2017RSPTA.37560268S}) and/or massive interacting binaries (e.g., \citealt{Chevalier2012ApJ...752L...2C,Soker2013ApJ...764L...6S,Kashi2013MNRAS.436.2484K,Schroder2020ApJ...892...13S,Metzger2022ApJ...932...84M}) are more likely to give rise to these events. However, we note that the high radiated energy tail extends to $\gtrsim 10^{51}$ erg (up to $\sim 4\times10^{51}$ erg for SN~2016aps; see also \citealt{Nicholl2020NatAs...4..893N,Suzuki2021ApJ...908...99S}). For those events, even with the maximum $\epsilon_{\rm rad}=1$, higher explosion energies than those achievable with the neutrino-driven mechanism ($\sim2\times10^{51}$ erg; e.g., \citealt{Ertl2016ApJ...818..124E,Sukhbold2016ApJ...821...38S,Burrows2020MNRAS.491.2715B,Burrows2024ApJ...964L..16B}) are required. Potential enhancements can be achieved in pulsational pair-instability SNe (e.g.,  \citealt{Woosley2007Natur.450..390W,Blinnikov2010PAN....73..604B,Moriya2013MNRAS.428.1020M,Woosley2017ApJ...836..244W}) or jittering jet SNe (e.g., \citealt{Soker2010MNRAS.401.2793S,Papish2011MNRAS.416.1697P,Shishkin2023MNRAS.522..438S}).

We will quantitatively test these progenitor hypotheses in follow-up papers in this series focused on numerical and semi-analytical modeling as well as on local environments.

\subsection{Connection to Fast Transients}
\label{sec:fast}

\begin{figure*}
    \centering
    \includegraphics[width=0.49\textwidth]{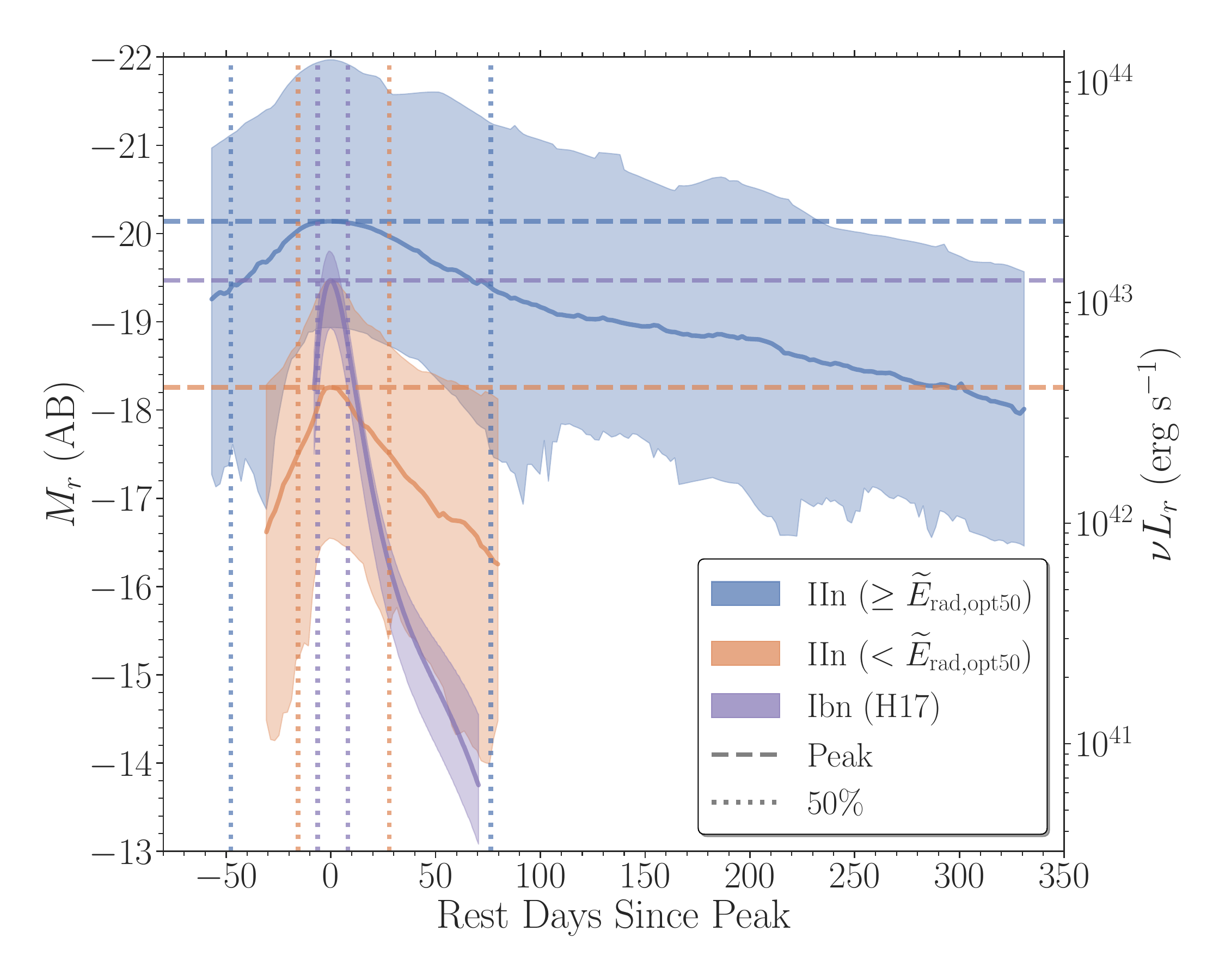}
    \includegraphics[width=0.49\textwidth]{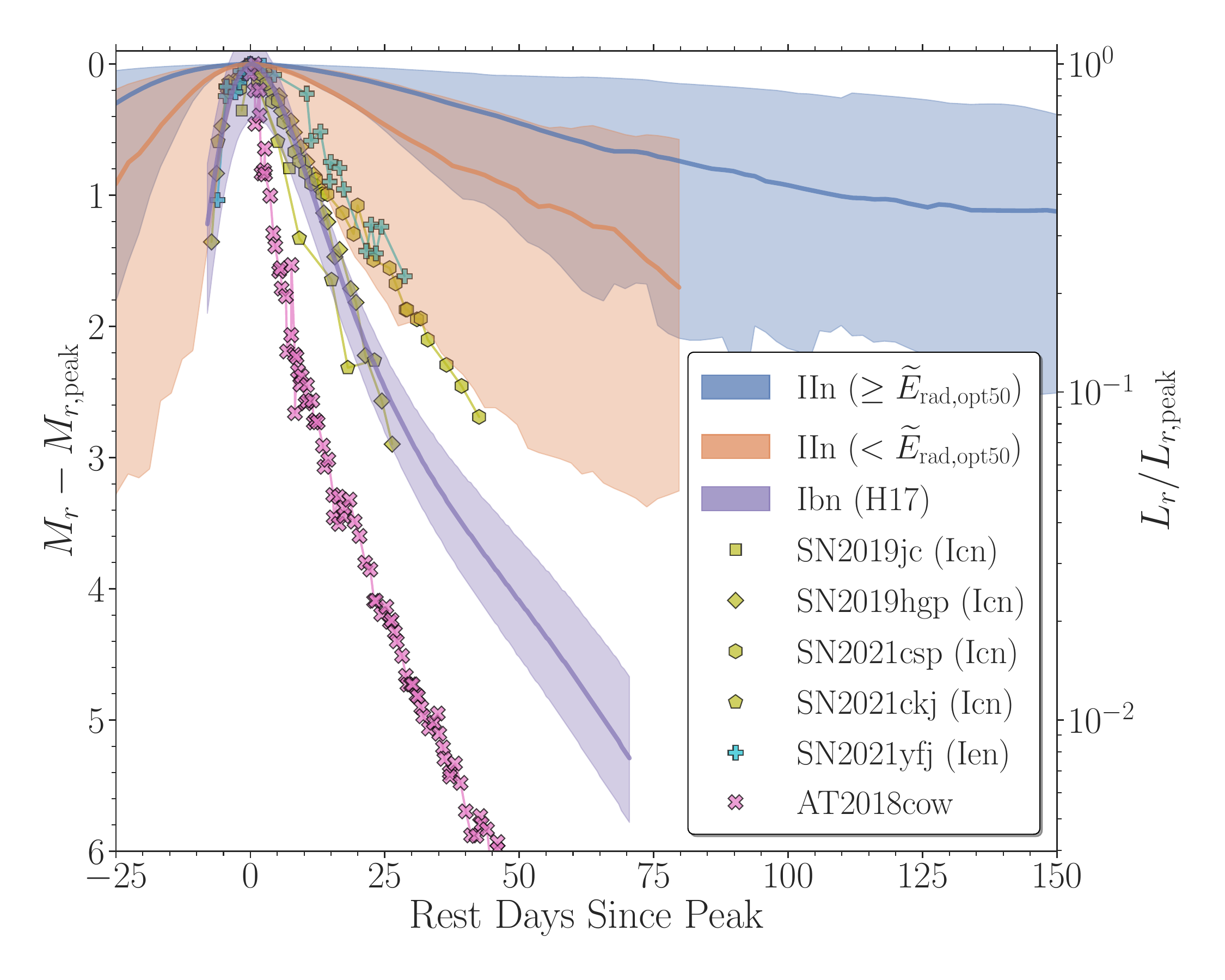}
\caption{\textit{Left}: similar to Figure~\ref{fig:LCtemp}, but with the SN~Ibn template from \citet[H17]{Hosseinzadeh2017ApJ...836..158H} in the $r/R$-band. The peak and 50\% measurements are marked by the horizontal dashed and vertical dotted lines, respectively. The shaded regions represent the $P_{2.5}-P_{97.5}$ (i.e., 95\%) range.
\textit{Right}: same template light curves, but normalized at peak and with additional interacting SESNe: SNe~Icn~2019jc, 2019hgp, 2021csp, and 2021ckj \citep{Gal-Yam2022Natur.601..201G,Perley2022ApJ...927..180P,Pellegrino2022ApJ...938...73P} and SN~Ien~2021yfj \citep{Schulze2025Natur.644..634S}, as well as the FBOT AT~2018cow \citep{Prentice2018ApJ...865L...3P,Xiang2021ApJ...910...42X}. The interacting SESNe and AT~2018cow fill in the unoccupied timescale region of SNe~IIn.
}
    \label{fig:Ibn}
\end{figure*}

The unoccupied region in the normalized light curves (Figure~\ref{fig:bandLC}) corresponds to the shortest timescale of SNe~IIn. At the 50\% flux level, the $P_{15.9}$ and $P_{2.3}$ ranges of the total timescale in the $r$ band ($t_{r50,{\rm rise}}+t_{r50,{\rm decl}}$) are $23$ and $15$ days, respectively (Figure~\ref{fig:rT}). Similarly, in the $g$ band, the $P_{15.9}$ and $P_{2.3}$ ranges are $21$ and $14$ days, respectively (Figure~\ref{fig:bandT50}). These $P_{2.3}$ ranges closely match the generally accepted timescale definition for ``fast transients,'' namely, $t_{50,{\rm rise}}+t_{50,{\rm decl}}\lesssim 12$ days (e.g., \citealt{Drout2014ApJ...794...23D,Pursiainen2018MNRAS.481..894P,Ho2019ApJ...887..169H,Ho2023ApJ...949..120H,Perley2022ApJ...927..180P,Pellegrino2022ApJ...938...73P}). This indicates that SNe~IIn contribute to a negligible fraction of fast transients.  
By contrast, the population of fast transients is partly composed of stripped-envelope SNe (SESNe) interacting with hydrogen-poor CSM (i.e., SNe~Ibn, Icn, Ien; e.g., \citealt{Pellegrino2022ApJ...926..125P, Gal-Yam2022Natur.601..201G, Perley2022ApJ...927..180P,Schulze2025Natur.644..634S}). Similar interaction signatures have also been observed in the fast blue optical transient (FBOT) AT~2018cow \citep{Fox2019MNRAS.488.3772F,Xiang2021ApJ...910...42X}. 

In the left panel of Figure~\ref{fig:Ibn}, we compare our $r$-band light-curve templates of SNe~IIn (Figure~\ref{fig:LCtemp}) with the template of SNe~Ibn from \citet{Hosseinzadeh2017ApJ...836..158H}.  We find that the SNe~Ibn have an intermediate peak luminosity between the faint--fast and luminous--slow groups of SNe~IIn, but have a shorter timescale than both SNe IIn groups. In the right panel of Figure~\ref{fig:Ibn}, we plot our normalized light-curve templates of the SN~IIn, in comparison to the normalized SN~Ibn template, as well as normalized SNe~Icn and Ien and AT~2018cow. Remarkably, the interacting SESNe and AT2018cow fill in the unoccupied timescale region of SNe~IIn.
This points to a continuum in timescales across different classes of interacting transients, from SNe~Ibc, Icn, Ien, and FBOTs to SNe~IIn. The difference can be attributed to the physical extent of the progenitor and CSM systems \citep{Khatami2024ApJ...972..140K}, where hydrogen-rich systems (e.g., red supergiants, luminous blue variables), which lead to SNe~IIn, have a larger physical extent than the hydrogen-poor systems that lead to interacting SESNe and FBOTs.

\section{Conclusions} 
\label{sec:conc}

We have presented the largest uniform study to date of SNe IIn, with a new sample of 427 events with high-quality light curves, supplemented by additional 63 events from past small sample studies in the literature.  In this paper, the first in a series that aims to delineate the population properties, we focus on the light-curve properties, analyzed in a uniform manner.  Our key findings and conclusions are:

\begin{itemize}
  \item SNe IIn span broad ranges of peak luminosity (about 2 orders of magnitude), timescale (about an order of magnitude), and radiated energy (about 3 orders of magnitude), but with strong correlations between the luminosity and timescale.
  \item SNe IIn exhibit a clear bimodal distribution along the luminosity-timescale correlations, separated around the median, resulting in a bifurcation of the light-curve morphology into luminous--slow and faint--fast groups; the below-median luminosity events also appear to separate into two branches. The same behavior is imprinted in a strong bimodality in the radiated energy distribution.
  \item The bimodality in radiated energies, and perhaps even more complex structure in the faint--fast group, may be indicative of at least two dominant progenitor channels that make up the overall SNe~IIn population.
  \item There is no evidence for a separate class of ``SLSNe-IIn'' ($M_{\rm peak}\leq -19.9$ mag) with distinctive properties; these events are simply the brighter half of the luminous--slow group.  
  \item There is a clear absence of SNe IIn with timescales of $\lesssim 15$ days, matching exactly the region occupied by fast transients with evidence for interaction with hydrogen-poor CSM (e.g., Ibn, Icn, Ien, FBOTs); this timescale differentiation is likely a reflection of the physical extent of the progenitor and CSM systems.
\end{itemize}

Our broad exploration of SNe~IIn will continue in subsequent papers in this series, which will cover a comparison to a large sample of model light curves based on numerical simulations; semi-analytical modeling of each SN in the sample to extract the explosion and CSM properties; a detailed study of the spectra and Balmer series line morphologies; and studies of the host galaxies and the locations of the SNe within them. We expect that correlations between the various photometric, spectroscopic, and environmental properties will shed more light on the progenitor systems leading to this diverse class of SNe IIn.

\section{acknowledgments}
We are grateful to Sho Fujibayashi, Hamid Hamidani, Kazumi Kashiyama, Shigeo S.~Kimura, Keiichi Maeda, Tatsuya Matsumoto, Lucy O.~McNeill, Takashi J.~Moriya, Toshikazu Shigeyama, Akihiro Suzuki, Masaomi Tanaka, Nozomu Tominaga, and Daichi Tsuna for useful discussions. 

D.H. is supported by STScI grants HST-GO-17770.002, JWST-GO-12468.001, and JWST-GO-09964.001.
The Berger Time-Domain research group at Harvard is supported by the NSF and NASA. 

TNS is supported by funding from the Weizmann Institute of Science, as well as grants from the Israeli Institute for Advanced Studies and the European Union via ERC grant No. 725161. 
We acknowledge WISeREP (\url{https://wiserep.org}).

This work has made use of data from the Zwicky Transient Facility (ZTF).  ZTF is supported by NSF grant No. AST-1440341 and a collaboration including Caltech, IPAC, the Weizmann Institute for Science, the Oskar Klein Center at Stockholm University, the University of Maryland, the University of Washington, Deutsches Elektronen-Synchrotron and Humboldt University, Los Alamos National Laboratories, the TANGO Consortium of Taiwan, the University of Wisconsin at Milwaukee, and Lawrence Berkeley National Laboratories. Operations are conducted by COO, IPAC, and UW. The ZTF forced-photometry service was funded under the Heising-Simons Foundation grant
No. 12540303 (PI: Graham).

This work has made use of data from the Asteroid Terrestrial-impact Last Alert System (ATLAS) project. ATLAS is primarily funded to search for near-Earth asteroids through NASA grant Nos. NN12AR55G, 80NSSC18K0284, and 80NSSC18K1575; byproducts of the NEO search include images and catalogs from the survey area. This work was partially funded by Kepler/K2 grant No. J1944/80NSSC19K0112 and HST grant No. GO-15889, and STFC grant Nos. ST/T000198/1 and ST/S006109/1. The ATLAS science products have been made possible through the contributions of the University of Hawaii Institute for Astronomy, the Queen’s University Belfast, the Space Telescope Science Institute, the South African Astronomical Observatory, and The Millennium Institute of Astrophysics (MAS), Chile.

We acknowledge ESA \textit{Gaia}, DPAC, and the Photometric Science Alerts Team (\url{http://gsaweb.ast.cam.ac.uk/alerts}).

This research has made use of the NASA Astrophysics Data System (ADS), the NASA/IPAC Extragalactic Database (NED), and NASA/IPAC Infrared Science Archive (IRSA, which is funded by NASA and operated by the California Institute of Technology).

\vspace{5mm}
\facilities{ADS, ATLAS, Gaia, IRSA, NED, TNS, WISeREP, ZTF}.

\software{
\texttt{astropy} \citep{Astropy2013A&A...558A..33A,Astropy2018AJ....156..123A,Astropy2022ApJ...935..167A}, 
\texttt{astroquery} \citep{Ginsburg2019AJ....157...98G},
\texttt{corner} \citep{Foreman-Mackey2016JOSS....1...24F}, 
\texttt{emcee} \citep{Foreman-Mackey2013PASP..125..306F}, 
\texttt{Matplotlib} \citep{Hunter2007CSE.....9...90H}, 
\texttt{NumPy} \citep{Oliphant2006}, 
\texttt{scikit-learn} \citep{Pedregosa2011JMLR...12.2825P}, 
\texttt{SciPy} \citep{SciPy2020NatMe..17..261V}, 
\texttt{seaborn} \citep{seaborn2020zndo...3629446W}.}

\clearpage
\appendix
\restartappendixnumbering

\section{Additional Sample Description}
\label{sec:extra}

We provide the details of the SN~IIn and broad-lined SLSN-II samples in Tables~\ref{tab:sample} and \ref{tab:broad}, respectively. In Table~\ref{tab:exc}, we summarize the excluded sample with the reasoning.

\startlongtable
\begin{deluxetable*}{cccccccccc}
\tabletypesize{\scriptsize}
\tablecaption{SN~IIn Sample \label{tab:sample}}
\tablehead{
\colhead{SN} & \colhead{R.A.} & \colhead{Decl.} & \colhead{Redshift} & \colhead{$dL$} & \colhead{$A_{V,{\rm MW}}$} & \colhead{TNS Classification} & \colhead{WISeREP} & \colhead{Discovery} & \colhead{Classfication} \\[-6pt]
\colhead{} & \colhead{(deg)} & \colhead{(deg)} & \colhead{($z$)} & \colhead{(Mpc)} & \colhead{(mag)} & \colhead{Classification} & \colhead{Classification} & \colhead{Report} & \colhead{Report}
}
\startdata
\multicolumn{10}{c}{This Work} \\
\hline
2023abtj & 21.75575 & 33.954931 & 0.014 & 59.8 & 0.131 & IIn & IIn & \cite{2023TNSTR3371....1T} & \cite{2024TNSCR..13....1L} \\
2023acmb & 233.487917 & 4.3237 & 0.038 & 165.1 & 0.127 & IIn & IIn & \cite{2024TNSTR.197....1F} & \cite{2024TNSCR2218....1P} \\
2023zjw & 116.495071 & -3.279069 & 0.047 & 205.6 & 0.205 & IIn & IIn & \cite{2023TNSTR3189....1F} & \cite{2024TNSCR.520....1P} \\
2023zmy & 104.251717 & 30.981869 & 0.056 & 246.5 & 0.234 & IIn & IIn & \cite{2023TNSTR3193....1M} & \cite{2024TNSCR3192....1R} \\
2023zaz & 59.730925 & -26.057442 & 0.0625 & 276.4 & 0.084 & IIn & IIn & \cite{2023TNSTR3156....1J} & \cite{2023TNSCR3364....1D} \\
2023yxs & 96.519325 & -31.257056 & 0.112 & 512.3 & 0.141 & IIn & IIn & \cite{2023TNSTR3131....1T} & \cite{2023TNSCR3299....1P} \\
2023yqw & 4.409683 & 22.73215 & 0.086 & 386.6 & 0.258 & IIn & IIn & \cite{2023TNSTR3116....1T} & \cite{2023TNSCR3299....1P} \\
\enddata 
%\tablecomments{Light curves and measurements of each sample will be published in its entirety in machine-readable form.}
\tablenotetext{a}{In YSE DR1 \citep{Aleo2023ApJS..266....9A}.}
\tablenotetext{b}{Classified by C. Angus and K. Auchettl, but the official report is not available on ADS.}
\tablenotetext{c}{See also \cite{Bruch2023ApJ...952..119B}.}
\tablenotetext{d}{See also \cite{Strotjohann2021ApJ...907...99S}.}
\tablenotetext{e}{See also \cite{Nicholl2020NatAs...4..893N}.}
\tablenotetext{f}{From \cite{Taddia2013}.}
\tablenotetext{g}{From \cite{Kiewe2012ApJ...744...10K}.}
\tablenotetext{h}{From \cite{Sako2018PASP..130f4002S}.}
\tablenotetext{i}{From \cite{Villar2020ApJ...905...94V}.}
\tablenotetext{j}{From \cite{Nyholm2020}.}
\end{deluxetable*}
\startlongtable
\begin{deluxetable*}{cccccccccc}
\tabletypesize{\scriptsize}
\tablecaption{Broad-lined SLSN-II Sample \label{tab:broad}}
\tablehead{
\colhead{SN} & \colhead{R.A.} & \colhead{Decl.} & \colhead{Redshift} & \colhead{$dL$} & \colhead{$A_{V,{\rm MW}}$} & \colhead{TNS} & \colhead{WISeREP} & \colhead{Discovery} & \colhead{Classfication} \\[-6pt]
\colhead{} & \colhead{(deg)} & \colhead{(deg)} & \colhead{(z)} & \colhead{(Mpc)} & \colhead{(mag)} & \colhead{Classification} & \colhead{Classification} & \colhead{Report} & \colhead{Report}
}
\startdata
2023acft & 78.771542 & -34.401111 & 0.101 & 458.7 & 0.084 & SLSN-II & SLSN-II & \cite{2023TNSTR3393....1T} & \cite{2024TNSCR.672....1C} \\
2023gpw & 195.577917 & -5.853042 & 0.08 & 358.1 & 0.094 & SLSN-II & SLSN-II & \cite{2023TNSTR.887....1F} & \cite{2023TNSCR1334....1K} \\
2020hgr\tablenotemark{a} & 214.110442 & 70.413561 & 0.1258 & 580.7 & 0.04 & SLSN-II & SLSN-II & \cite{2020TNSTR1039....1N} & \cite{2022TNSCR.662....1K} \\
2020bfe\tablenotemark{a} & 269.461208 & 33.796739 & 0.099 & 449.0 & 0.116 & IIn & SLSN-II & \cite{2020TNSTR.301....1N} & \cite{2020TNSCR1738....1Y} \\
2019uba\tablenotemark{a} & 35.869758 & -1.983033 & 0.3035 & 1551.5 & 0.073 & SLSN-II & SLSN-II & \cite{2019TNSTR2257....1N} & \cite{2022TNSCR.661....1K} \\
2019xfs\tablenotemark{a} & 287.520437 & 32.344842 & 0.116 & 532.0 & 0.49 & SLSN-II & SLSN-II & \cite{2019TNSTR2655....1F} & \cite{2020TNSCR.868....1D} \\
2019cqc\tablenotemark{a} & 275.429383 & 30.992636 & 0.1173 & 538.5 & 0.314 & SLSN-II & SLSN-II & \cite{2019TNSCR.514....1T} & \cite{2019TNSCR2814....1F} \\
2019kwr\tablenotemark{a} & 200.551458 & 49.914639 & 0.2022 & 978.1 & 0.034 & SLSN-II & SLSN-II & \cite{2019TNSTR1265....1P} & \cite{2019TNSCR1186....1P} \\
\enddata
%\tablecomments{Light curves and measurements of each sample will be published in its entirety in machine-readable form.}
\tablenotetext{a}{In the \cite{Kangas2022MNRAS.516.1193K} sample.}
\tablenotetext{b}{See \cite{Yao2023ApJ...955L...6Y}.}
\end{deluxetable*}
\startlongtable
\begin{deluxetable*}{cccccc}
\tabletypesize{\scriptsize}
\tablecaption{Excluded Sample \label{tab:exc}}
\tablehead{
\colhead{SN} & \colhead{TNS} & \colhead{WISeREP} & \colhead{Discovery} & \colhead{Classification} & \colhead{Reason} \\[-6pt]
\colhead{} & \colhead{Classification} & \colhead{Classification} & \colhead{Report} & \colhead{Report} & \colhead{} 
}
\startdata
\multicolumn{6}{c}{Spectroscopic Cuts} \\
\hline
\hline
2023acft & SLSN-II & SLSN-II & \cite{2023TNSTR3393....1T} & \cite{2024TNSCR.672....1C} & Broad-lined SLSN-II (Table~\ref{tab:broad}) \\
2023vto & SLSN-II & SLSN-II & \cite{2023TNSTR2714....1F} & \cite{2023TNSCR3050....1P} & TDE \citep{Kumar2024ApJ...974L..36K} \\
2023mgr & IIn & IIn & \cite{2023TNSTR1550....1T} & \cite{2023TNSCR1570....1H} & SN~II with flash features \\
2023gpw & SLSN-II & SLSN-II & \cite{2023TNSTR.887....1F} & \cite{2023TNSCR1334....1K} & Broad-lined SLSN-II (Table~\ref{tab:broad}) \\ %\citealt{Kangas2026A\&A...705A..52K}; 
2023jo & IIn & IIn & \cite{2023TNSTR..76....1T} & \cite{2023TNSCR.128....1P} & Spectrum dominated by galaxy \\
2023alo & IIn & IIn & \cite{2023TNSTR.182....1F} & \cite{2023TNSCR.907....1P} & Nuclear transient? \\
2022rze & IIn & IIn & \cite{2022TNSTR2428....1H} & \cite{2022TNSCR3514....1H} & Changing-look AGN \citep{Pessi2025MNRAS.542.3354P} \\
\enddata
\tablenotetext{a}{SN~IIn In YSE DR1 \citep{Aleo2023ApJS..266....9A}.}
\end{deluxetable*}

%\clearpage
\section{Timescale and Radiated Energy in Luminosity Space}
\label{sec:Ltime}

\begin{figure*}
    \centering
    \includegraphics[width=0.98\textwidth]{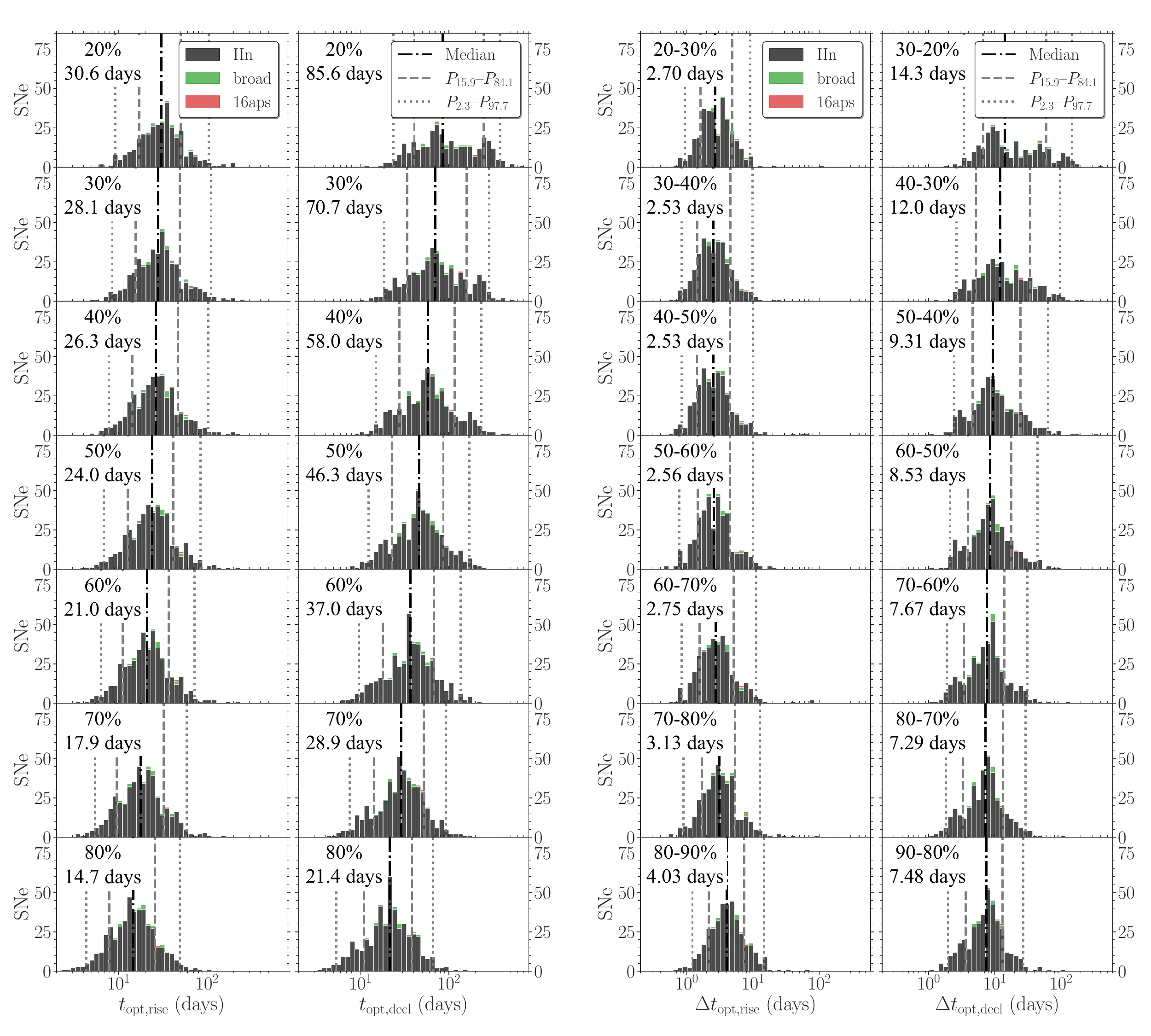}
\caption{Similar to Fig.~\ref{fig:rT}, but for optical luminosity.
}
    \label{fig:optT}
\end{figure*}

\begin{figure*}
    \centering
    \includegraphics[width=0.98\textwidth]{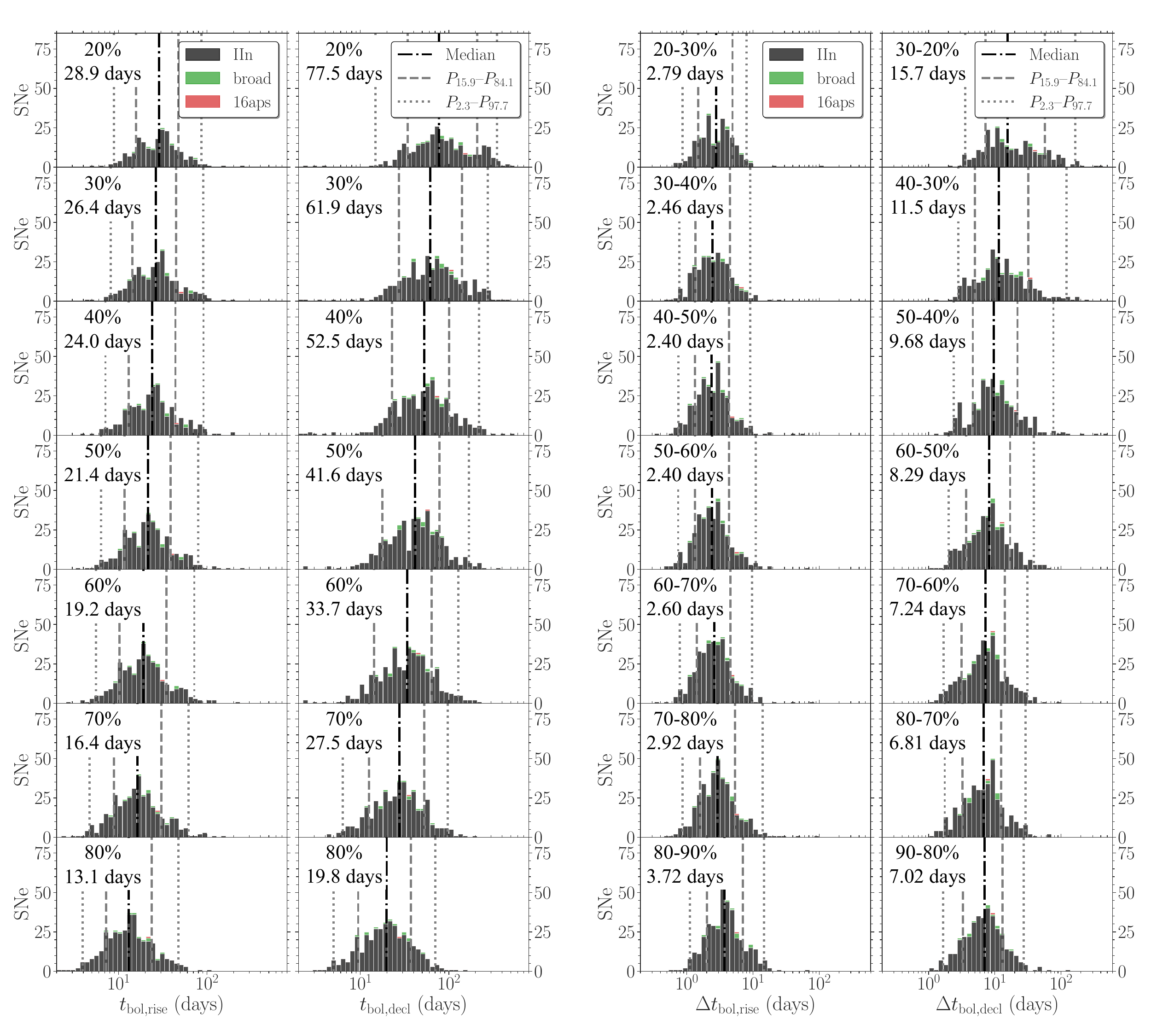}
\caption{Similar to Fig.~\ref{fig:rT}, but for bolometric luminosity.
}
    \label{fig:bolT}
\end{figure*}

\begin{figure}
    \centering
    \includegraphics[width=0.44\textwidth]{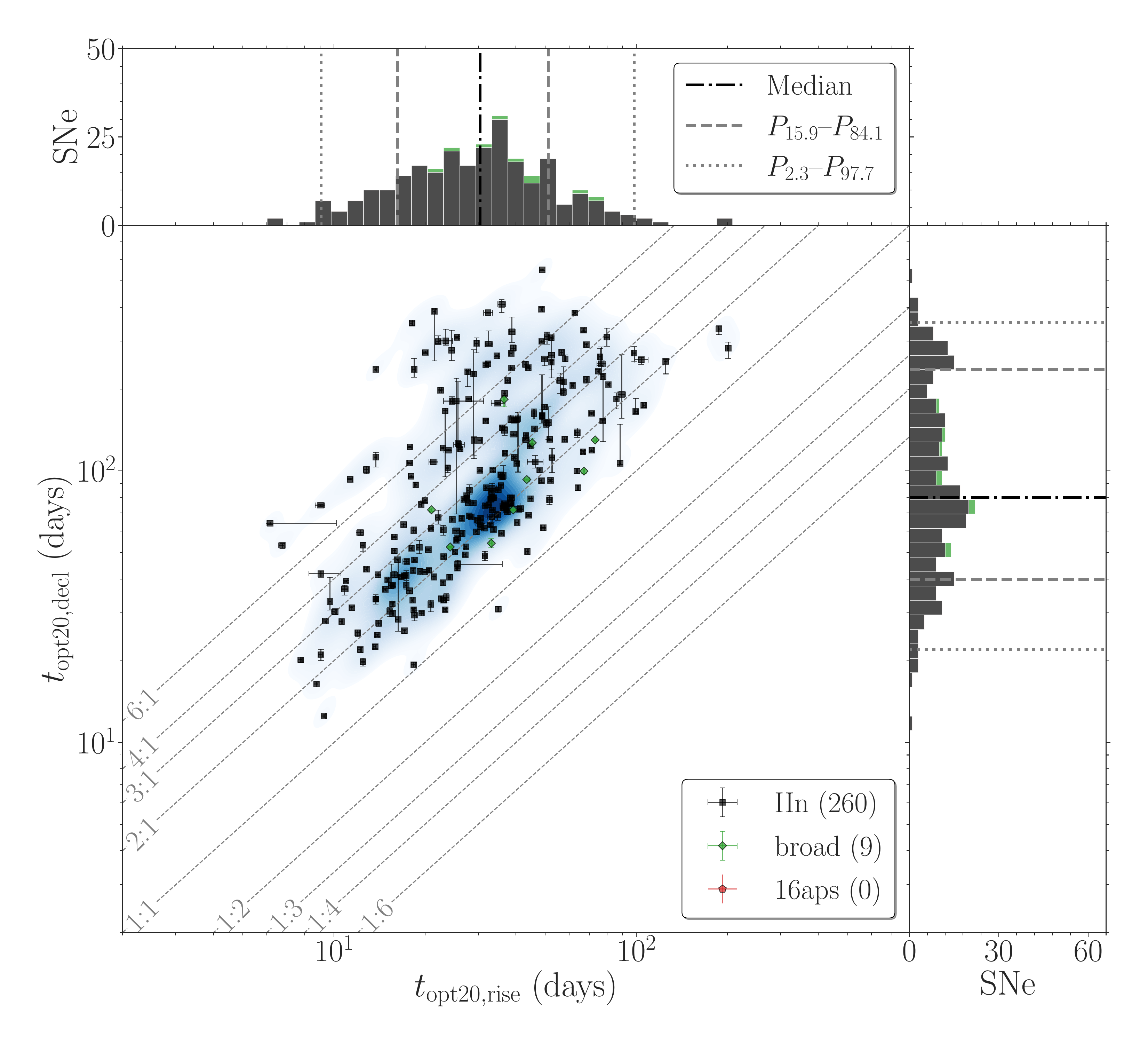}
    \includegraphics[width=0.44\textwidth]{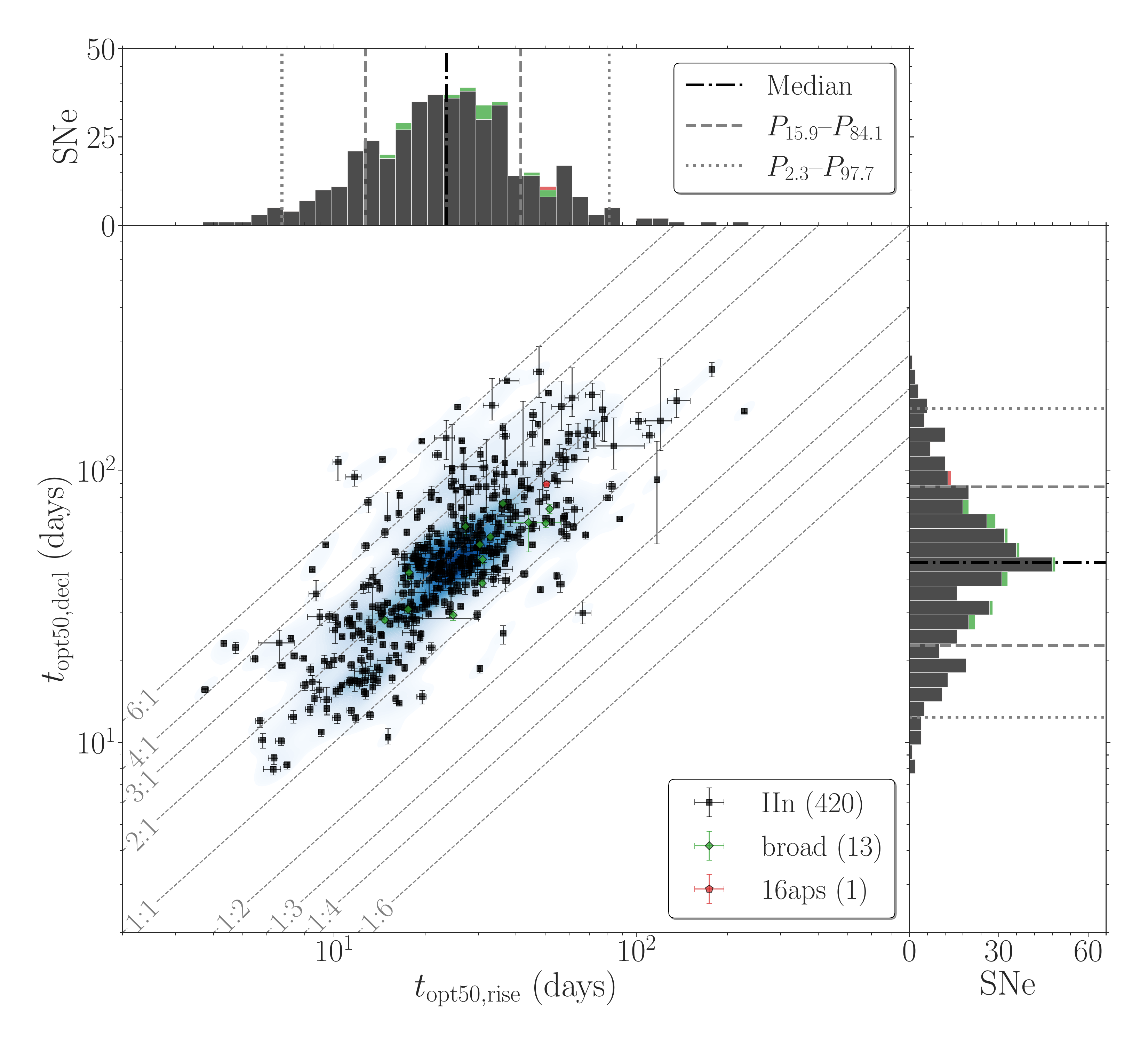}
    \includegraphics[width=0.44\textwidth]{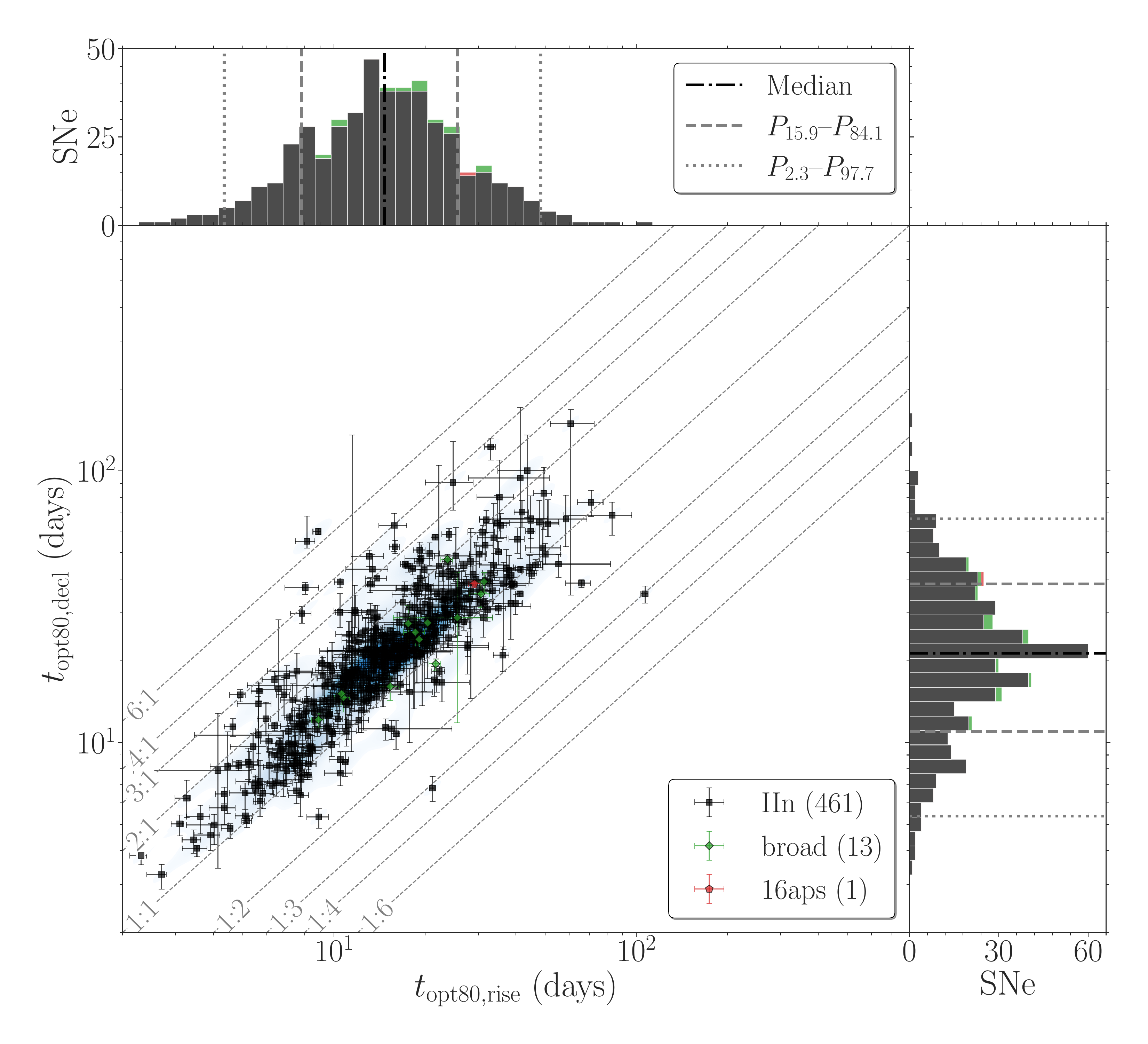} 
\caption{Similar to Figure~\ref{fig:rT205080}, but for optical luminosity.
}
    \label{fig:optT205080}
\end{figure}

\begin{figure}
    \centering
    \includegraphics[width=0.44\textwidth]{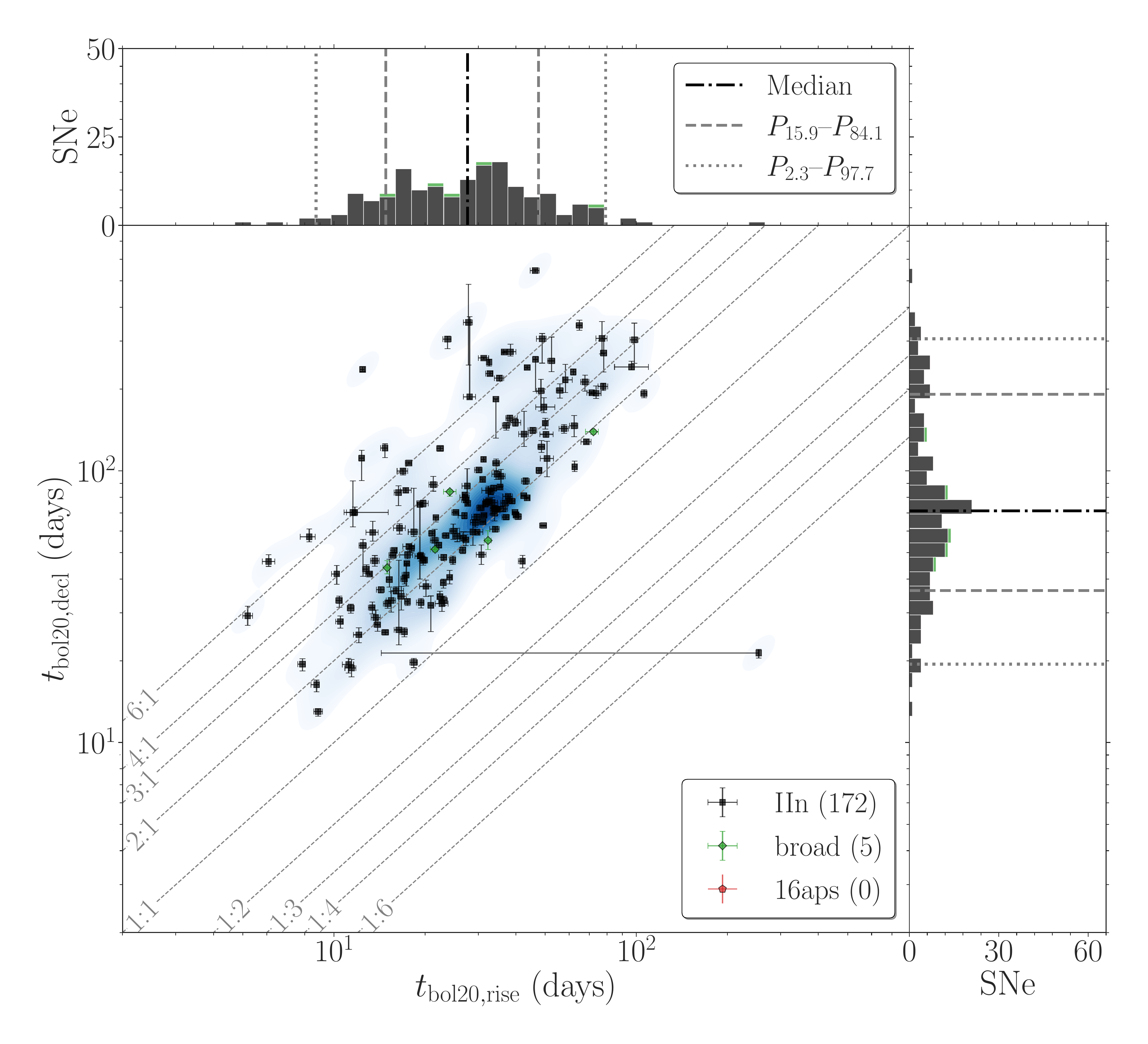}
    \includegraphics[width=0.44\textwidth]{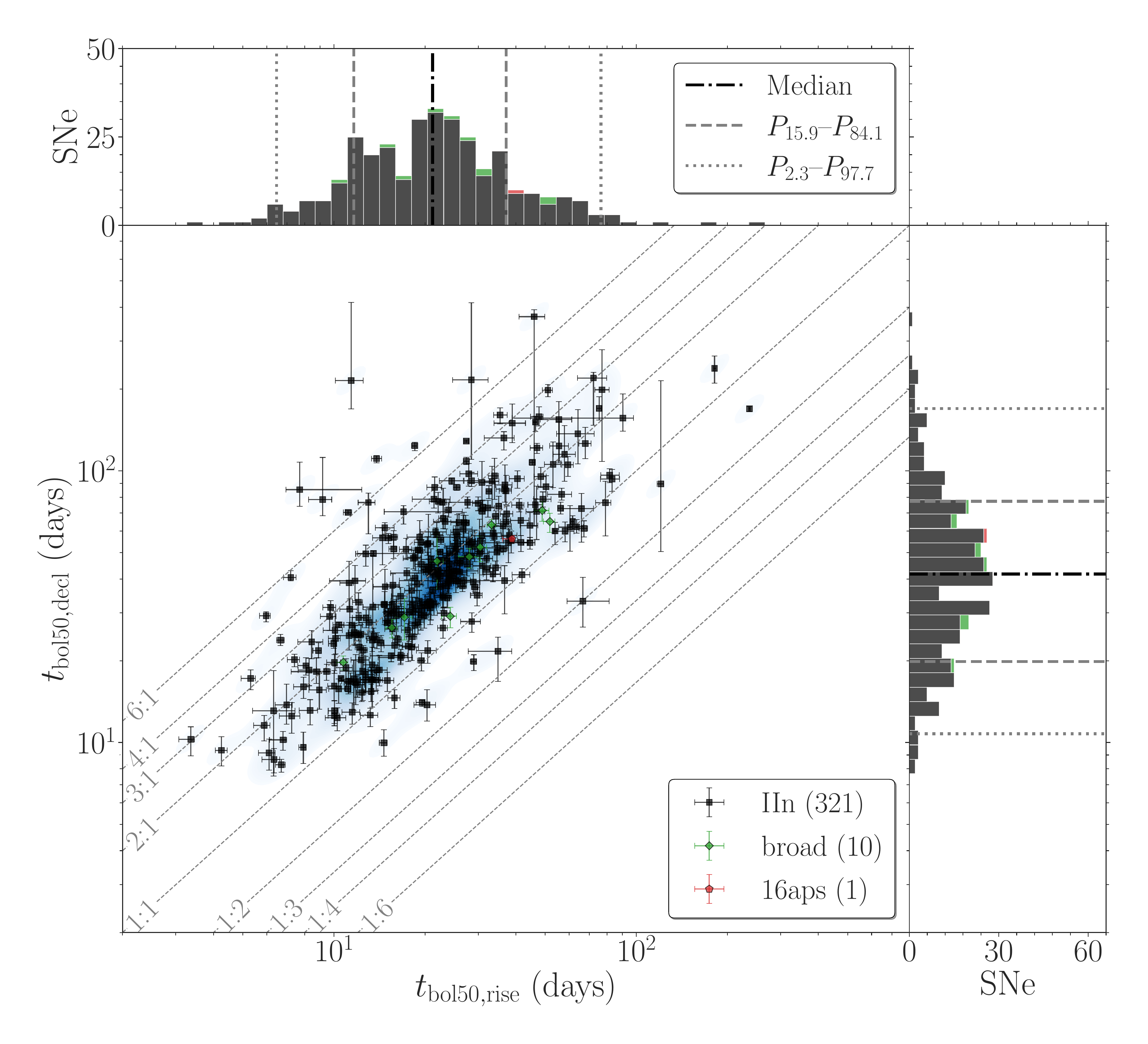}
    \includegraphics[width=0.44\textwidth]{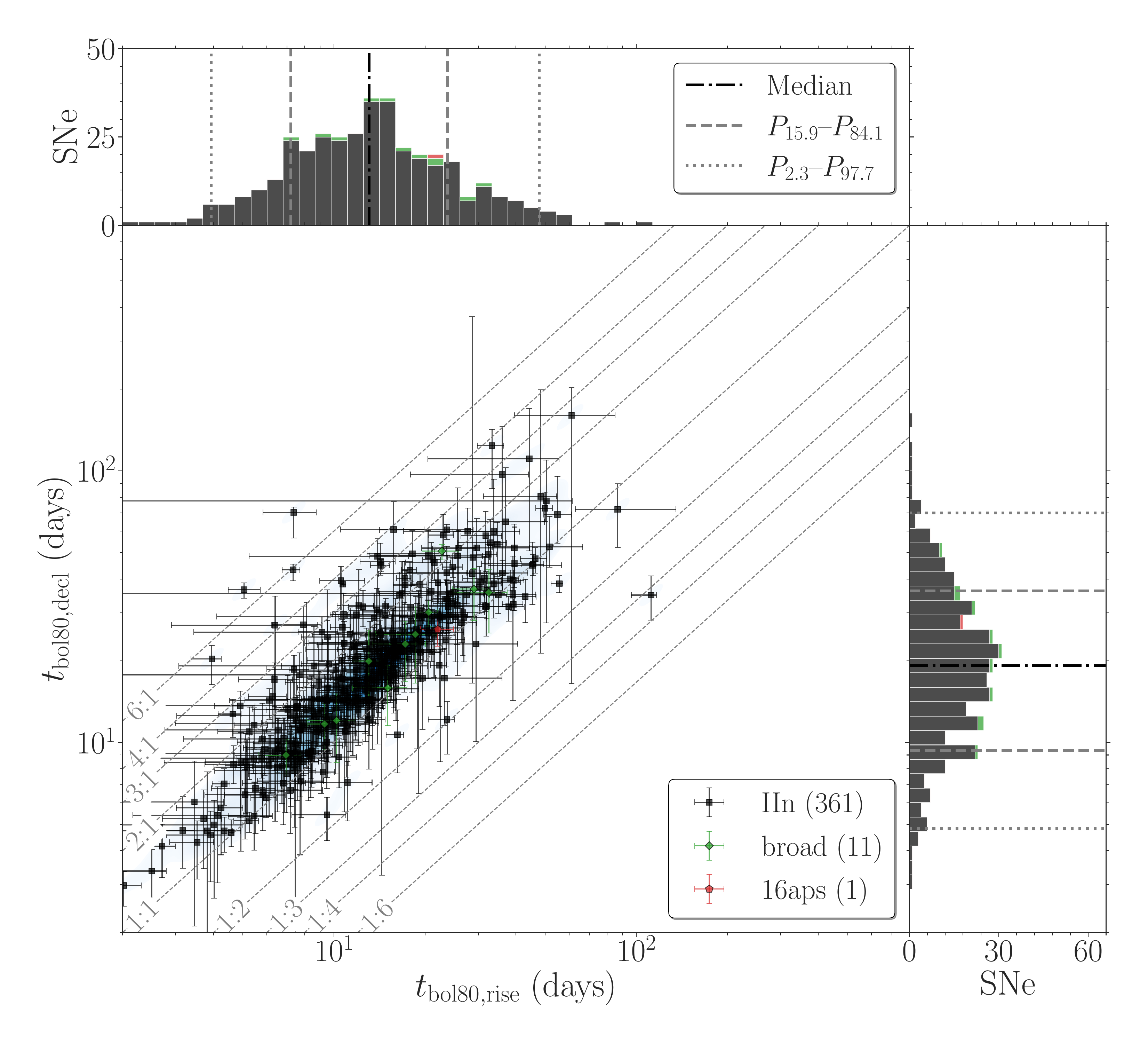} 
\caption{Similar to Figure~\ref{fig:rT205080}, but for bolometric luminosity.
}
    \label{fig:bolT205080}
\end{figure}

\begin{figure*}
    \centering
    \includegraphics[width=0.98\textwidth]{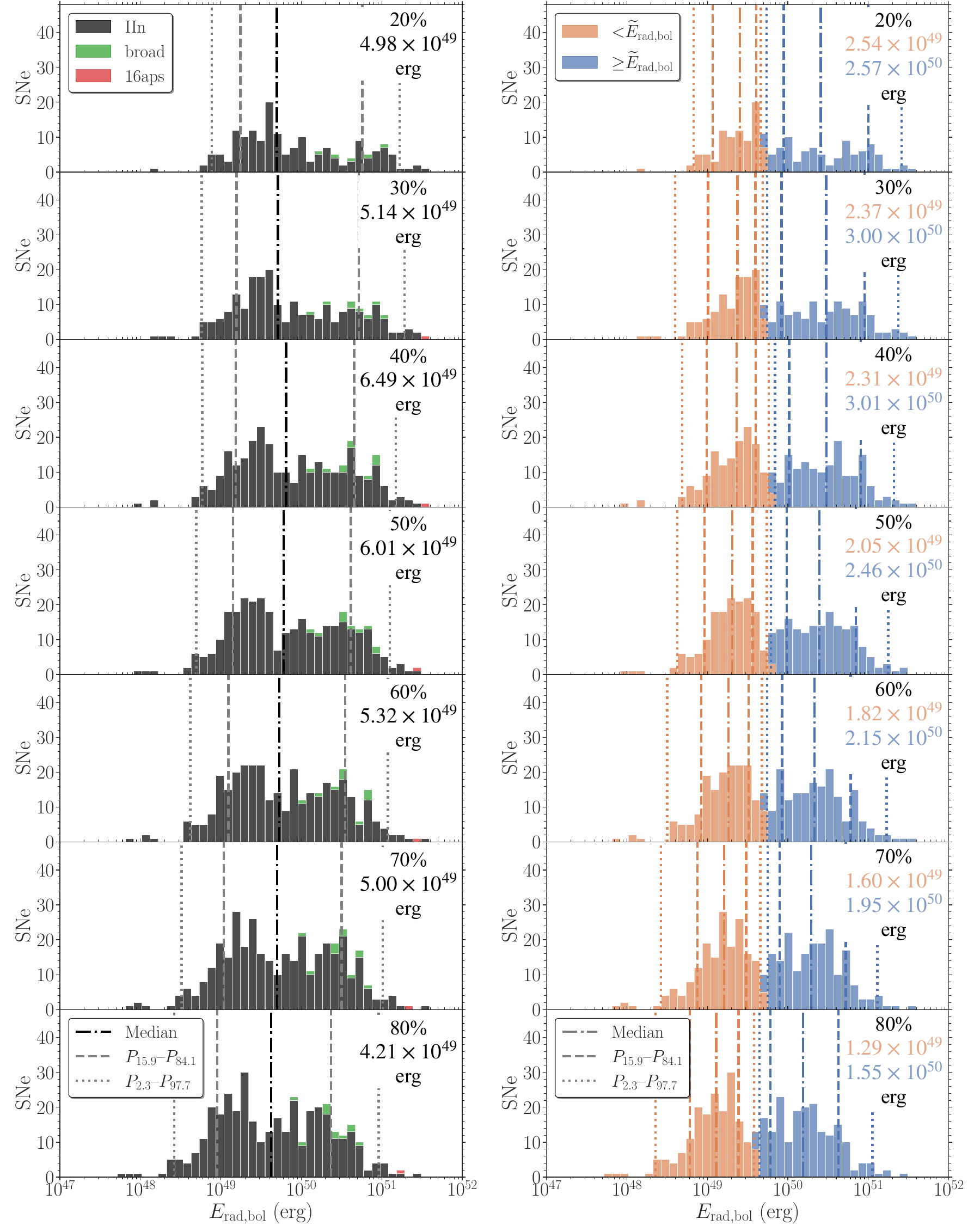}
\caption{Similar to Figure~\ref{fig:Eradopt}, but for bolometric luminosity. 
}
    \label{fig:Eradbol}
\end{figure*}

For completeness, we show the timescale histograms and correlations in bolometric and pseudo-bolometric (optical: 3250--8900$\,\Angstrom$) luminosity in Figures~\ref{fig:optT}, \ref{fig:bolT}, \ref{fig:optT205080}, and \ref{fig:bolT205080}. In Figure~\ref{fig:Eradbol}, we show the histograms of the bolometric radiated energy.

\clearpage
\bibliography{main,TNS_disc,TNS_class}

%\end{CJK*}
\end{document}